\documentclass[notitlepage,english,aps,prd,longbibliography]{revtex4-1}
\usepackage[T1]{fontenc}
\usepackage[utf8]{inputenc} 
\usepackage{color}
\usepackage{bm}
\usepackage[linesnumbered,ruled,vlined]{algorithm2e}
\usepackage{xcolor}
\usepackage{amsmath}
\usepackage{mathtools}
\usepackage{amssymb}
\usepackage{graphicx}
\usepackage{natbib}
\usepackage{braket}
\usepackage{psfrag}
\usepackage{tikz}
\usepackage[normalem]{ulem}
\usepackage{qcircuit}
\usepackage{siunitx}
\usepackage[colorlinks=true,urlcolor=blue,citecolor=blue,linkcolor=blue]{hyperref}

\DeclarePairedDelimiter\ceil{\lceil}{\rceil}
\DeclarePairedDelimiter\floor{\lfloor}{\rfloor}

\begin{document}
 

\title{Foundational Patterns for Efficient Quantum Computing}
 
\author{Austin Gilliam}
\affiliation{%
JPMorgan Chase}%

\author{Charlene Venci}
\affiliation{%
JPMorgan Chase}%

\author{Sreraman Muralidharan}
\affiliation{%
JPMorgan Chase}%

\author{Vitaliy Dorum}
\affiliation{%
JPMorgan Chase}%

\author{Eric May}
\affiliation{%
JPMorgan Chase}%

\author{Rajesh Narasimhan}
\affiliation{%
JPMorgan Chase}%
 
\author{Constantin Gonciulea}
\affiliation{%
JPMorgan Chase}%
 
 
\date{\today}
 
\begin{abstract}
We present a number of quantum computing patterns that build on top of fundamental algorithms, that can be applied to solving concrete, NP-hard problems.
In particular, we introduce the concept of a quantum dictionary as a summation of multiple patterns and algorithms, and show how it can be applied in the context of Quadratic Unconstrained Binary Optimization (QUBO) problems.
We start by presenting a visual approach to quantum computing, which avoids a heavy-reliance on quantum mechanics, linear algebra, or complex mathematical notation, and favors geometrical intuition and computing paradigms.
We also provide insights on the fundamental quantum computing algorithms (Fourier Transforms, Phase Estimation, Grover, Quantum Counting, and Amplitude Estimation). 
\end{abstract}
 
\maketitle
 
 
\section{\label{sec:introduction} Introduction}
 
If asked to calculate the squares of integers from 0 to 99, one might think they had been given a trivial task.
However, 70 years ago this was a difficult problem for classical computers. Under the supervision of Sir Maurice Wilkes, professor at the University of Cambridge (and eventual Turing Award winner~\cite{TuringAward}), a team of researchers spent two years building the “Electronic Delay Storage Automatic Calculator” (EDSAC), noteworthy for marking the transition from "test to tool" in civilian computing.
On May 6, 1949, the EDSAC ran its first stored program, which calculated the squares of integers from 0 to 99 in 2 minutes and 35 seconds~\cite{EDSAC}. 

Quantum computing is at a stage similar to that of classical computing 70 years ago.
Even though classical computers have the capability to perform complex calculations and analyze large data sets, solving certain classes of problems is difficult or even impossible for them.
Quantum computers will be able to solve such problems in the near future. While quantum computers cannot replace classical computers entirely, they will complement them when the ability to perform exponentially-hard calculations is required. 

While there are multiple quantum computing primers~\cite{Nannicini2017,Chatterjee2003,Landsberg2018,Yanofsky2007,Zalka1998,Rieffel2000,Coles2018,Kopczyk2018}, this paper specifically focuses on introducing efficient patterns.
Patterns are not algorithms – as stated by Christopher Alexander, “Each pattern describes a problem which occurs over and over again in our environment, and then describes the core of the solution to that problem, in such a way that you can use this solution a million times over, without ever doing it the same way twice”~\cite{Alexander1977}.
We approach quantum computing from a computer scientist’s perspective, avoiding linear algebra and quantum mechanics whenever possible, and stick to the geometrical intuition of how quantum computing works.

To the best of our knowledge, this is a list of singular contributions presented in this paper:
\begin{itemize}
    \item Quantum Dictionary Pattern
    \begin{itemize}
        \item Entangled Key-Value Registers
        \item Lookup by Key
        \item Value Counting (Equality and Inequality-Based Value Matching)
    \end{itemize}
    \item Number Representation using Phase Estimation
    \begin{itemize}
        \item Handling Addition and Comparison
        \item Encoding Discrete Probability Spaces
        \item Function and Probability Distribution Encoding
    \end{itemize}
    \item Quantum State Visualization using Complex Histograms
    \item Alternative Oracle Implementation (Multiplying Amplitudes of Selected States by -1)
    \item Identification of the Probability Distribution in Phase Estimation as the Fejer Distribution
    \item Novel, Illustrative Examples for Quantum State, Operators, Oracles, and Patterns
    \item New Insights on Phase Estimation, Quantum Counting, and Amplitude Estimation
\end{itemize}

This paper has three parts.
Section~\ref{sec:a-visual-approach-to-quantum-computing} is an elementary introduction to quantum state and quantum systems.
Section~\ref{sec:insights-on-quantum-algorithms} is an overview of the fundamental quantum algorithms~\cite{Shor1997,Nielsen2011,Grover1996,Brassard1998} with some unique insights.
Section~\ref{sec:quantum-dictionary} describes an original pattern applied to solving computer science problems – the quantum dictionary.
We show how one can use a quantum dictionary to encode problems in a quantum state, and utilize some of the algorithms discussed in Section~\ref{sec:insights-on-quantum-algorithms} to create a common framework for formulating and solving problems.

\section{\label{sec:a-visual-approach-to-quantum-computing}A Visual Approach to Quantum Computing}

\subsection{\label{sec:classical-random-quantum-bits} Classical, Random, \& Quantum Bits}

\subsubsection{\label{subsec:complex-numbers-and-vectors}Complex Numbers \& Vectors}

For those who need a refresher on complex numbers, vectors, roots of unity, or inner products, we will give a quick overview. Otherwise, this subsection can be skipped.

\begin{figure}[htb]
\includegraphics[width=10cm]{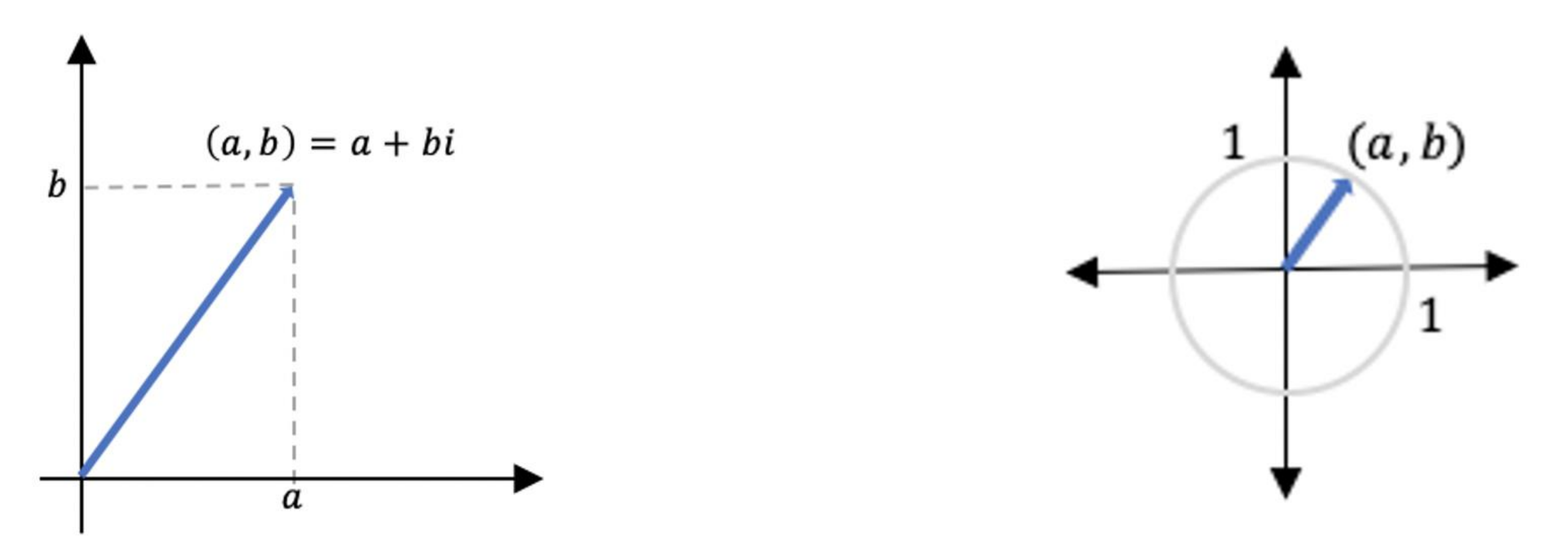}
\caption{\label{fig:1}A visual representation of (left) a complex number, and (right) its corresponding representation on the unit circle.}
\end{figure}

A \textbf{complex} number is composed of a real part $a$ and an imaginary part $b$, and can be written in the form $a+b*i$. 
We can use an arrow with \textbf{magnitude} and \textbf{direction} to visualize a complex number (as in Fig.~\ref{fig:1}).
This arrow starts from the origin of the real and imaginary axes and goes to point $(a,b)$.
The magnitude of an arrow is given by $\sqrt{a^2+b^2}$. 

\begin{figure}[htb]
\includegraphics[width=10.5cm]{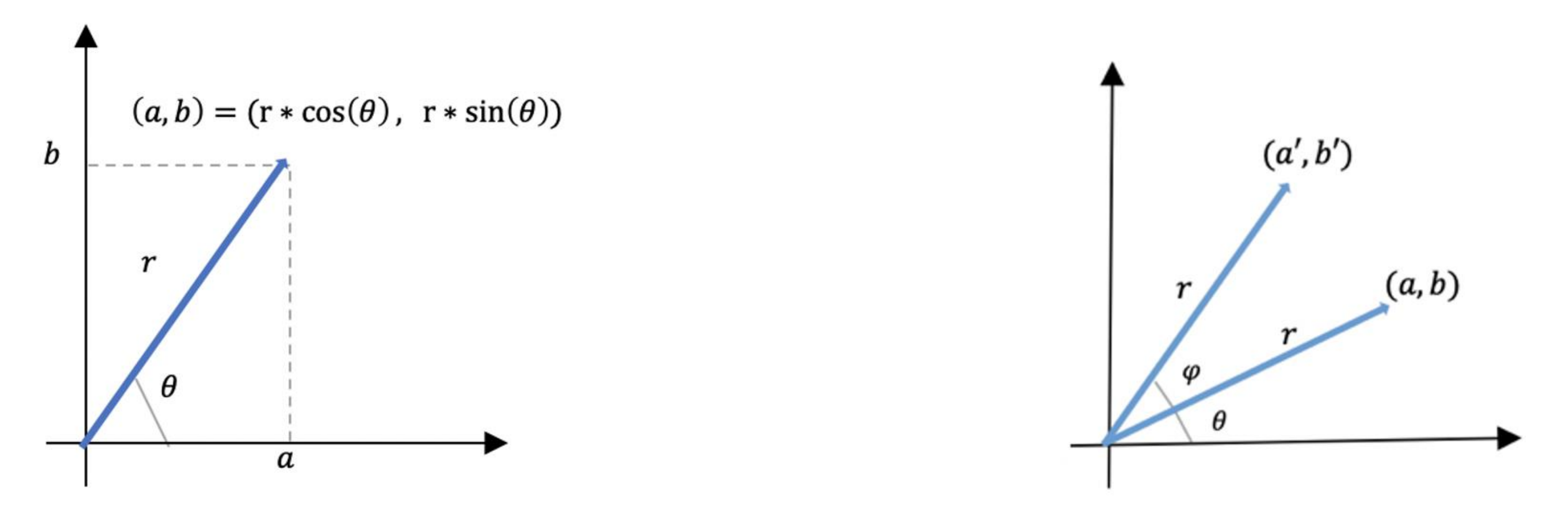}
\caption{\label{fig:2}(left) An alternate form for describing a complex number. (right) Rotation of a complex number by an angle $\phi$.}
\end{figure}

The \textbf{unit circle} is a circle with radius equal to $1$ (Fig.~\ref{fig:1}), centered at the origin $(0,0)$.
We can plot the set of complex numbers with magnitude $1$ on the unit circle, and an arrow on this circle is a \textbf{unit vector}.
A complex number can also be written in polar form $r(\cos{\theta}+i\sin{\theta})$, where $r$ is the length of the arrow and $\theta$ is the angle between the arrow and the x-axis, also called the \textbf{phase} of the complex number.
The coordinates of the arrowhead are expressed as $a = r\cos{\theta}$ and $b = r\sin{\theta}$.
If we rotate the arrow $r(\cos{\theta}+i*\sin{\theta})$ in the counterclockwise direction by $\phi$, we obtain a new arrow $r(\cos{(\theta + \phi)}+i\sin{(\theta + \phi)})$ (Fig.~\ref{fig:2}).

\begin{figure}[htb]
\includegraphics[width=5cm]{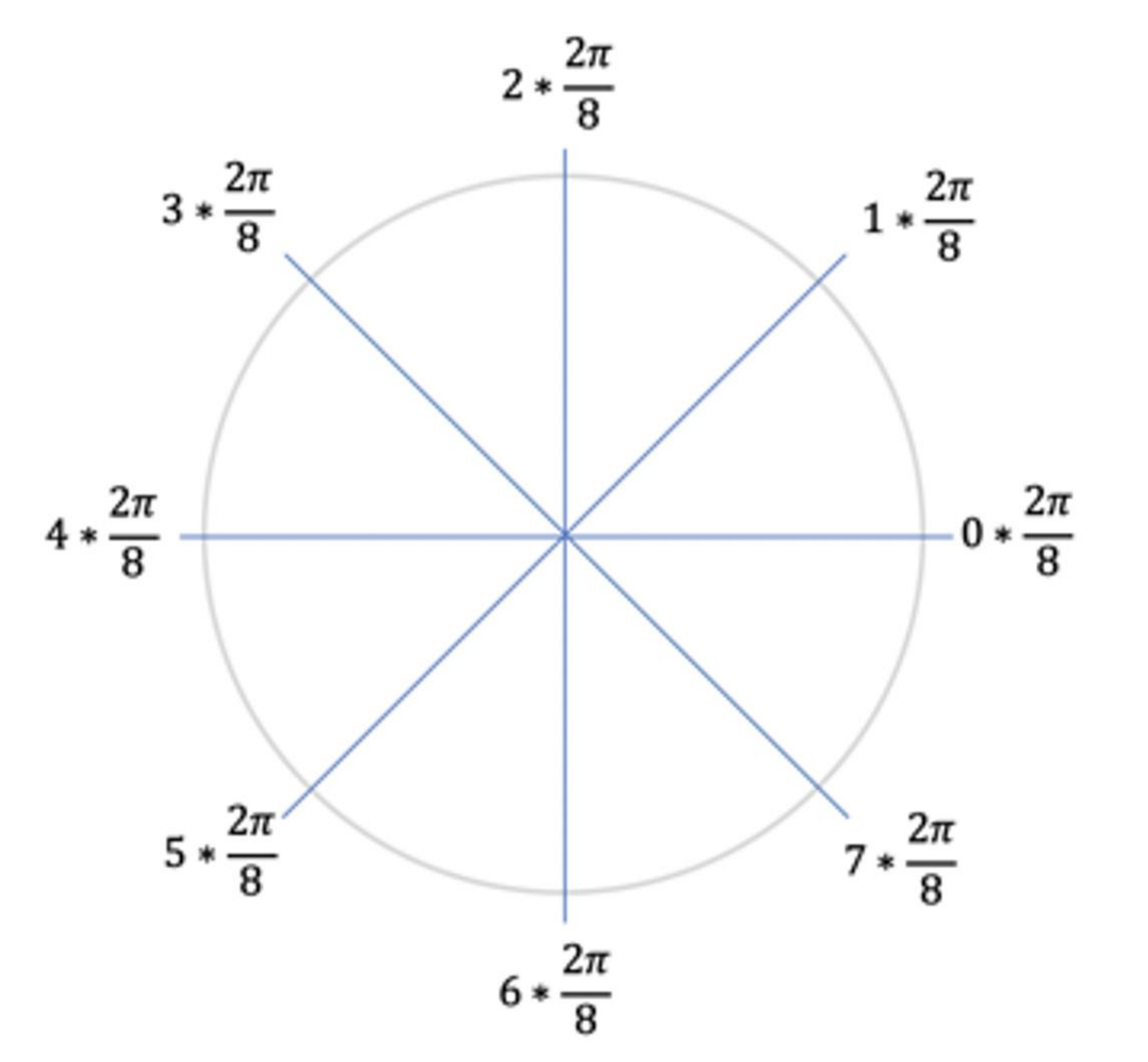}
\caption{\label{fig:3}The 8th roots of unity, labelled by the respective values of $\theta$ for the polar form $r(\cos\theta+i\sin\theta)$.}
\end{figure}

A \textbf{root of unity} is a complex number $c$ such that $c^n=1$ for a positive integer $n$. 
There are $n$ solutions to this equation (called the $n$th roots of unity), which can be mapped to the complex plane in the polar form $\cos{\frac{2k}{n}}+i\sin{\frac{2k}{n}}$, where $k \in \{0, 1, ..., n-1\}$ (Fig.~\ref{fig:3}).

\begin{figure}[htb]
\includegraphics[width=10cm]{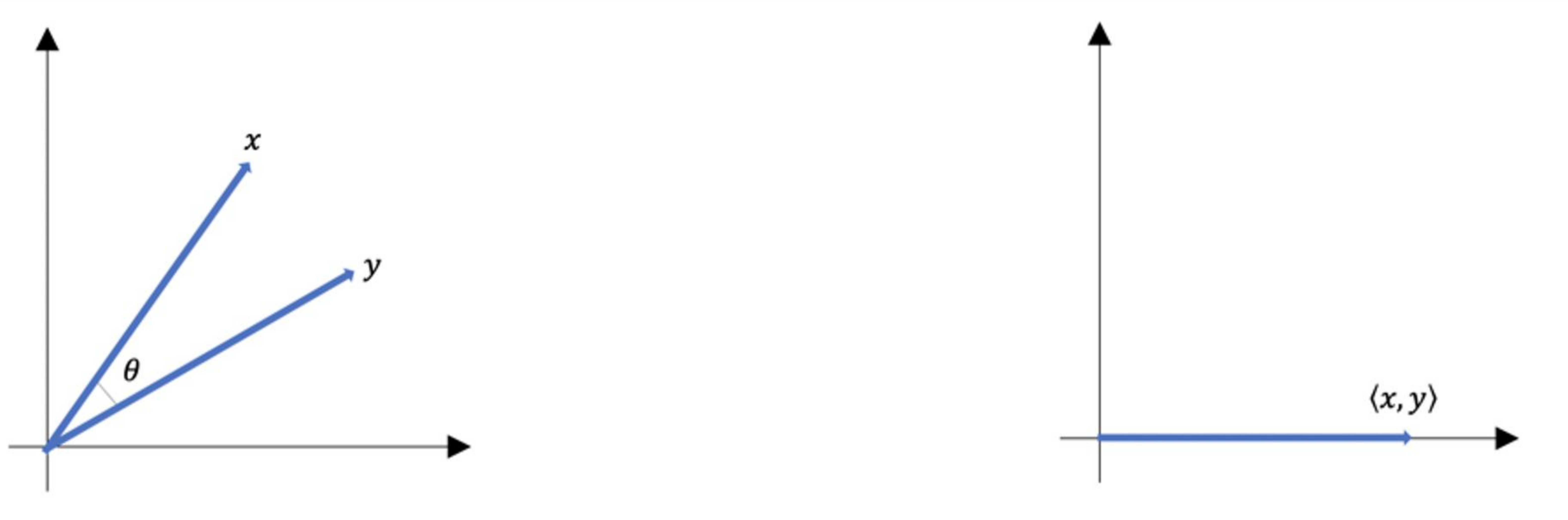}
\caption{\label{fig:4}The inner product of two complex numbers $x$ and $y$.}
\end{figure}

Given two complex numbers $x$ and $y$, their \textbf{inner product} is defined as $\langle x, y \rangle = x\overline{y}$.
An alternate form is $\langle x, y \rangle = |x| |y| \cos{\theta}$, where $\theta$ is the angle between $x$ and $y$.
An inner product of two sequences of complex numbers $[x_0,x_1,...,x_n]$ and $[y_0,y_1,...,y_n]$ is defined as $x_0\overline{y_0} + x_1\overline{y_1} + ... + x_n\overline{y_n}$.
Note that the closer the directions of $x$ and $y$, the greater the magnitude of the inner product, which can be seen as a measurement of how close or similar they are (Fig.~\ref{fig:4}).

\subsubsection{\label{subsec:random-and-classical-bits}Random \& Classical Bits}

\begin{figure}[htb]
\includegraphics[width=7cm]{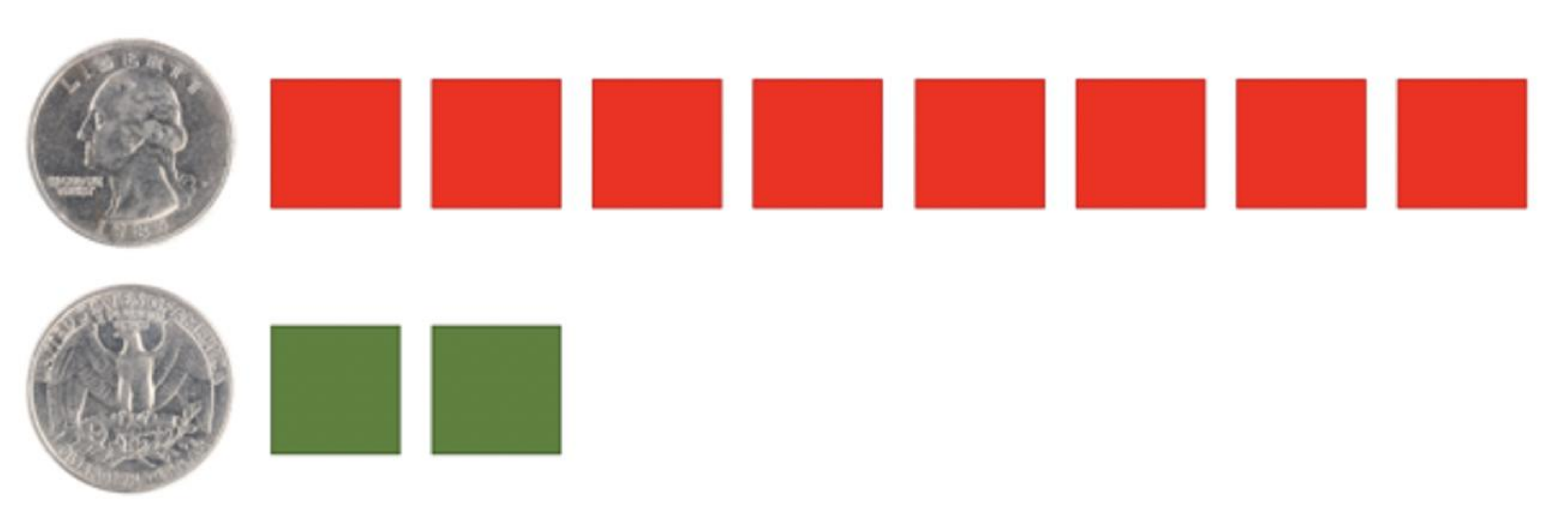}
\caption{\label{fig:5}Given a new coin with an unknown bias towards heads, we toss the coin $10$ times.}
\end{figure}

Let’s look at the process of tossing a fair coin.
We can regard this process as random with two possible outcomes: heads or tails.
When one of the outcomes has a greater probability than the other, we call the coin biased.
The \textbf{bias} is an unknown factor which can skew the probability from what we would expect from a fair coin. 
This unknown bias cannot be determined with only a single coin flip, so the coin would need to be flipped multiple times to get a more precise estimation of what the bias is.
The frequency of the outcomes from the coin flips will provide information as a set of probabilities, which must add up to $1$.
For example, if we were to flip a coin with an unknown bias towards heads $10$ times, we may see that we get tails only twice (Fig.~\ref{fig:5}).

We can think of a biased coin as a \textbf{probabilistic bit} – i.e. there is some randomness to the result, as opposed to deterministically measuring $0$ or $1$.
We can interpret a classical bit as an extremely-biased probabilistic bit, with one of the outcomes having probability $1$, and the other having probability $0$.
The probabilities associated with each outcome can be thought of as the parameters of the random process before sampling it, similar to a spinning coin before landing on one of its sides.
When the number of possible outcomes is greater than one, we refer to the state as being in \textbf{superposition} of those outcomes.
A state is said to be in \textbf{equal superposition} if all outcomes are equally likely.
While this term is often used in quantum computing, this idea is not unique to the field, and exists for any probabilistic system.

\subsubsection{\label{subsec:the-quantum-bit}The Quantum Bit}

\begin{figure}[htb]
\includegraphics[width=6cm]{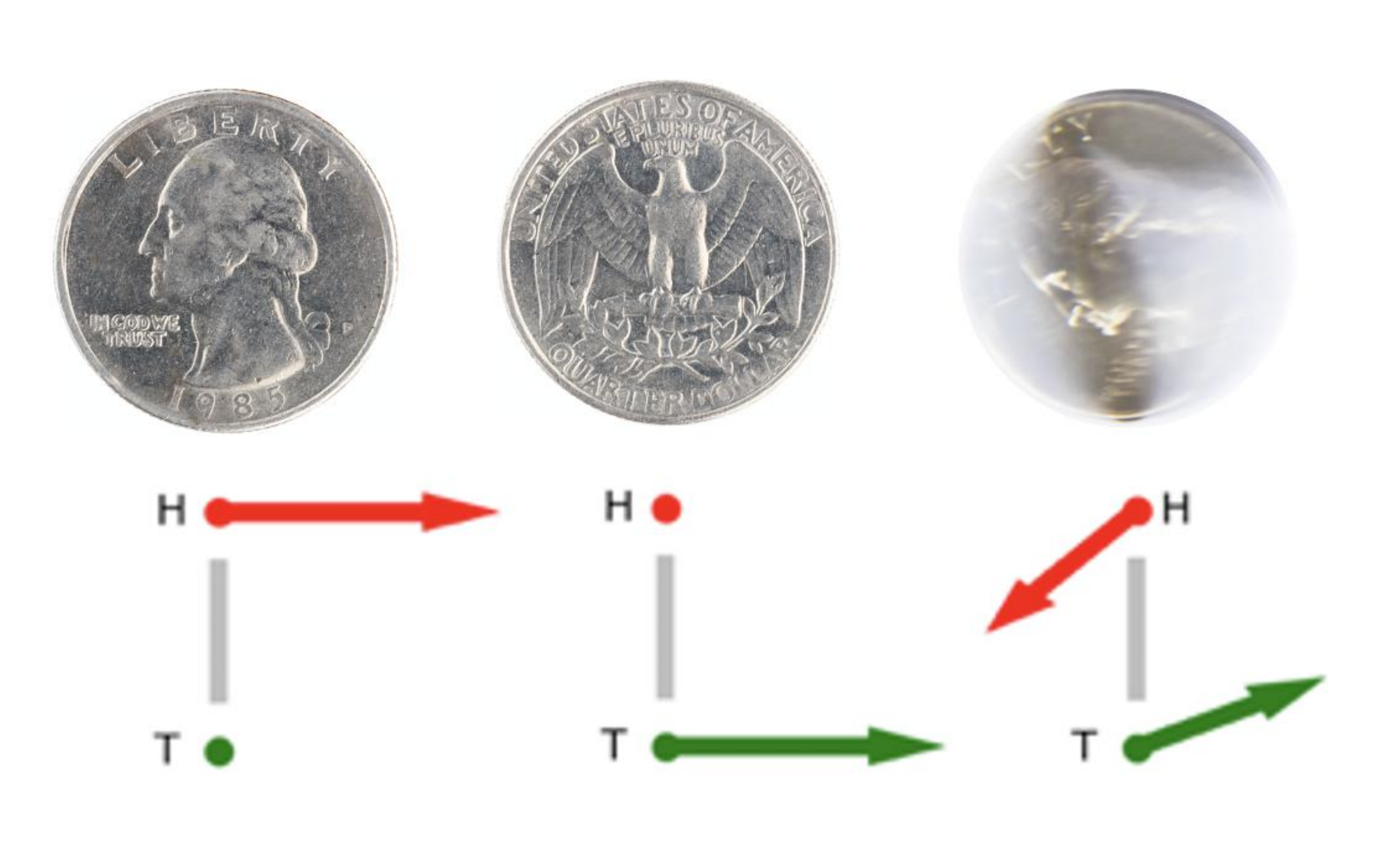}
\caption{\label{fig:6}When the coin is heads or tails the amplitude for that side is $1$, and the other is $0$. When in superposition, we see that the arrow notation can represent complex amplitudes.}
\end{figure}

A \textbf{quantum bit} (abbreviated qubit~\cite{Schumacher1995} or qbit~\cite{Mermin2007}) is a generalized version of the probabilistic bit.
Instead of associating probabilities with each outcome (as we do in the probabilistic bit), we associate 2-dimensional vectors or arrows~\cite{Feynman2006}, which are called \textbf{amplitudes}.
The probabilities of the outcomes are correlated to the magnitudes of their corresponding amplitudes. More precisely, the probability of an outcome is the square of the length of its corresponding arrow.
Therefore, we can use arrows to represent each outcome of a quantum bit, as seen in Fig.~\ref{fig:6}.

A qubit starts in the \textbf{default state}, where the probability of measuring $0$ is $1$ (for example, the coin would always turn up heads).
The default state is one of the \textbf{basis states}, which correlate directly with a possible outcome of the quantum state.
There are two basis states for a single qubit: $0$ and $1$.
We define a \textbf{quantum state} by its amplitudes and their corresponding basis states, much like a dictionary of key-value pairs.
The arrow notation can therefore be spoken of interchangeably with the quantum state itself.

\begin{figure}[htb]
\includegraphics[width=7cm]{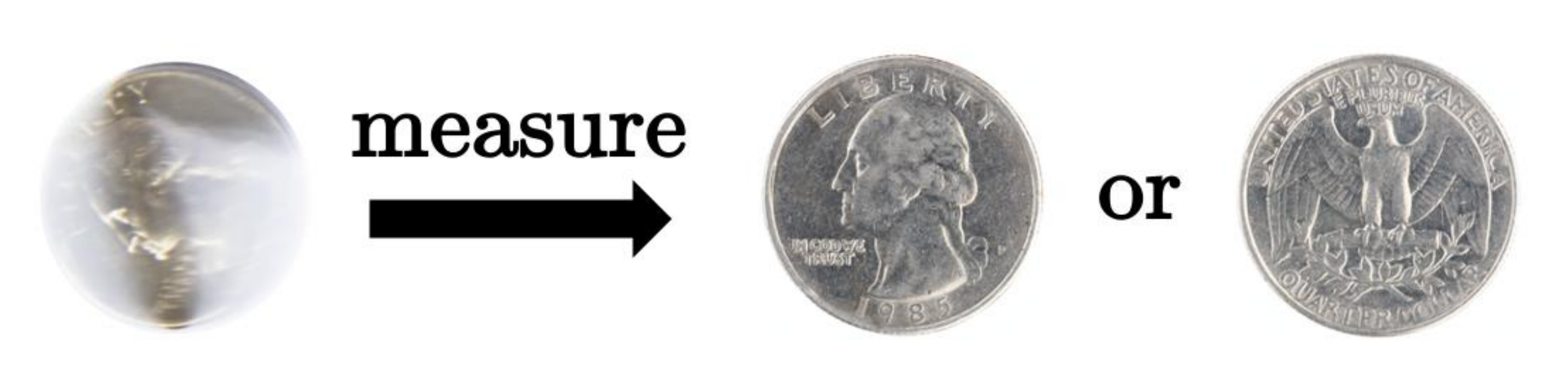}
\caption{\label{fig:7}When a qubit is measured, we observe one of the two possible measurement outcomes – heads ($0$) or tails ($1$).}
\end{figure}

With a classical system we can observe its current state at any time without affecting it.
But the state of a quantum system randomly \textbf{collapses} to one of its possible outcomes when measured (Fig.~\ref{fig:7}).
We can never observe the quantum state directly – measurement returns a binary string, and the state after observation is the basis state that corresponds to that binary string (the state before measuring is destroyed).
Given this fact, we will refer to \textbf{output states}, basis states, and measurement outcomes interchangeably.

\begin{figure}[htb]
\includegraphics[width=6cm]{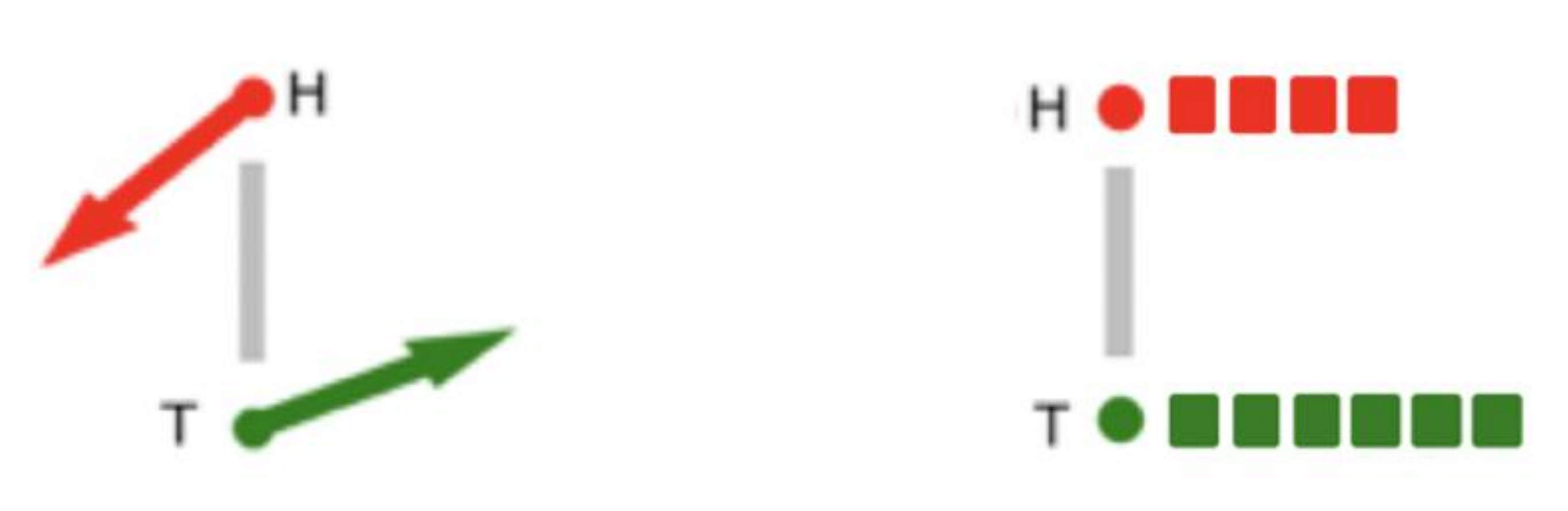}
\caption{\label{fig:8}(left) The quantum state at the beginning of the computation and (right) the resulting histogram after measuring.}
\end{figure}

In order to retrieve any meaning from the state of a quantum system, its state has to be recreated and measured multiple times, and the random results have to be interpreted as a pattern in the context of the given problem.
Fig.~\ref{fig:8} shows the result of repeating the same computation $10$ times, and measuring the state at the end of each iteration.
When we need to keep track of the cumulative result we can use a classical system to generate a histogram, giving us an insight about the amplitudes.

The challenge is to construct a quantum state that gives the desired answer in the least amount of measurements, which manifests as a balance between accuracy and computation time.

\subsection{\label{sec:quantum-systems}Quantum Systems}

The definition of a quantum system is based on the Postulates of Quantum Computing~\cite{Nielsen2011}, which are formally stated in Appendix~\ref{sec:posulates}.

\subsubsection{\label{subsec:composition-of-a-quantum-system}Composition of a Quantum System}

\begin{figure}[htb]
\includegraphics[width=6cm]{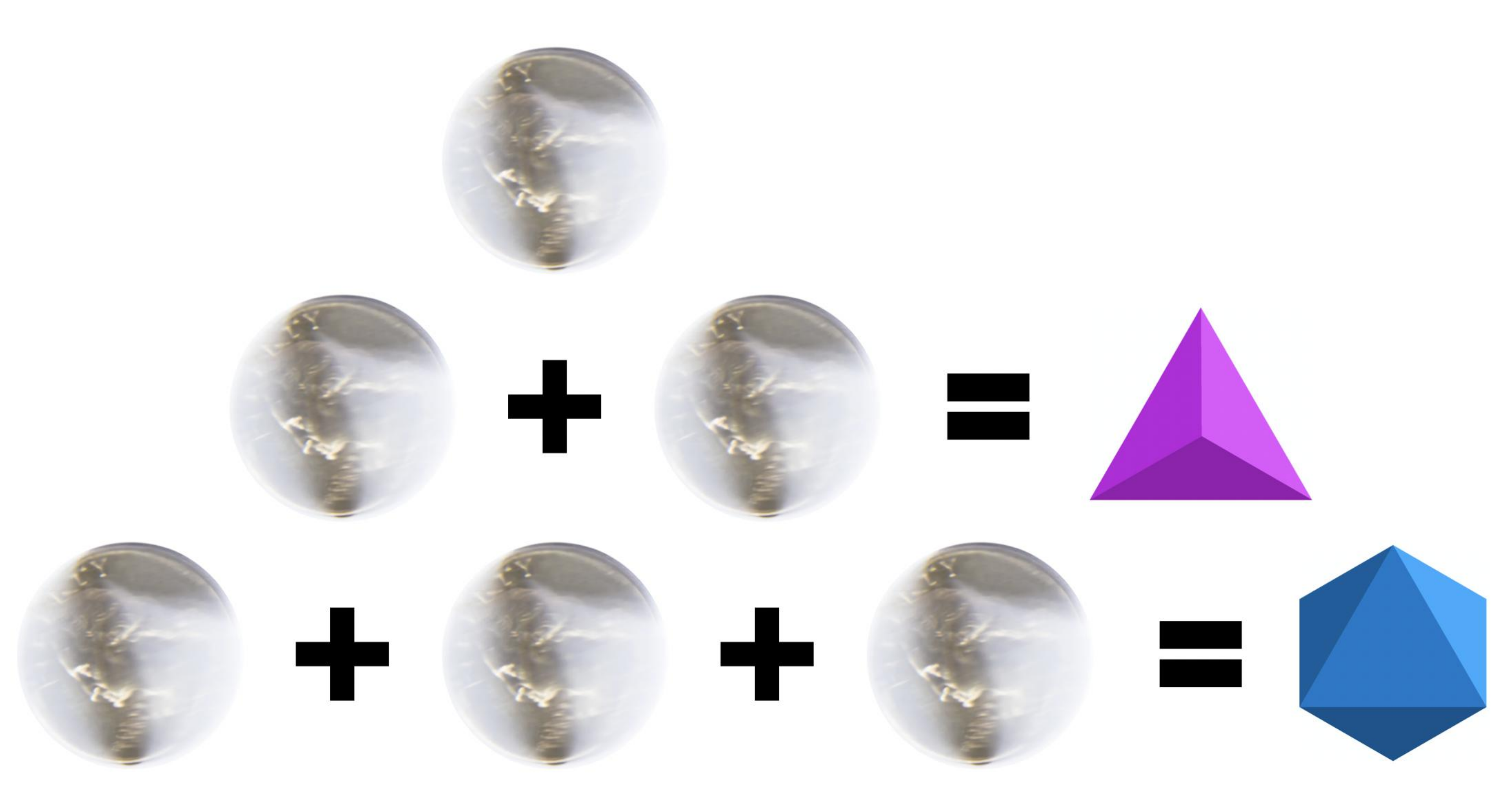}
\caption{\label{fig:9}As we increase the number of “quantum coins”, we can no longer think of them as their individual parts, and instead we treat them as “quantum dice”. The die has $2^n$ sides, where $n$ is the number of coins (qubits).}
\end{figure}

A \textbf{quantum computing system} is comprised of multiple qubits and has an associated state, which consists of one amplitude for each possible measurement outcome.
For a quantum system with n qubits, there are $N=2^n$ outcomes.
The simplest system contains only a single qubit, and is the building block with which larger systems are composed. It is important not to think of the quantum system in terms of the state of the individual qubits, but instead to consider the state as a whole. If we continue with the example using a coin, we can think of the composition of the quantum system as quantum coins combining into a die (Fig.~\ref{fig:9}).

\begin{figure}[htb]
\includegraphics[width=10cm]{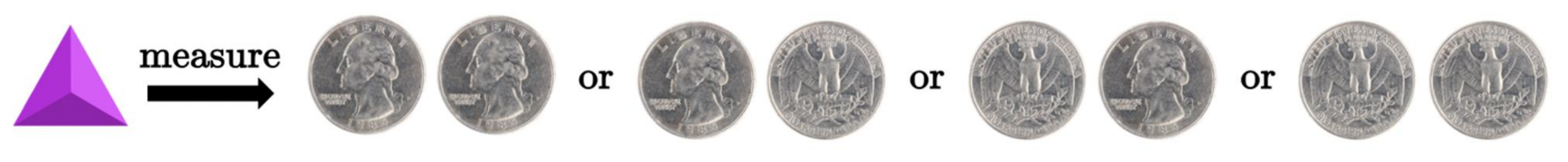}
\caption{\label{fig:10}Once the quantum state is measured, the quantum die will land on one side, which is one of the possible combinations of two coin flips.}
\end{figure}

We don’t discuss the outcomes of each individual coin (or qubit) separately. 
Each face corresponds to a possible combination of heads and tails (Fig.~\ref{fig:10}).

\begin{figure}[htb]
\includegraphics[width=2.5cm]{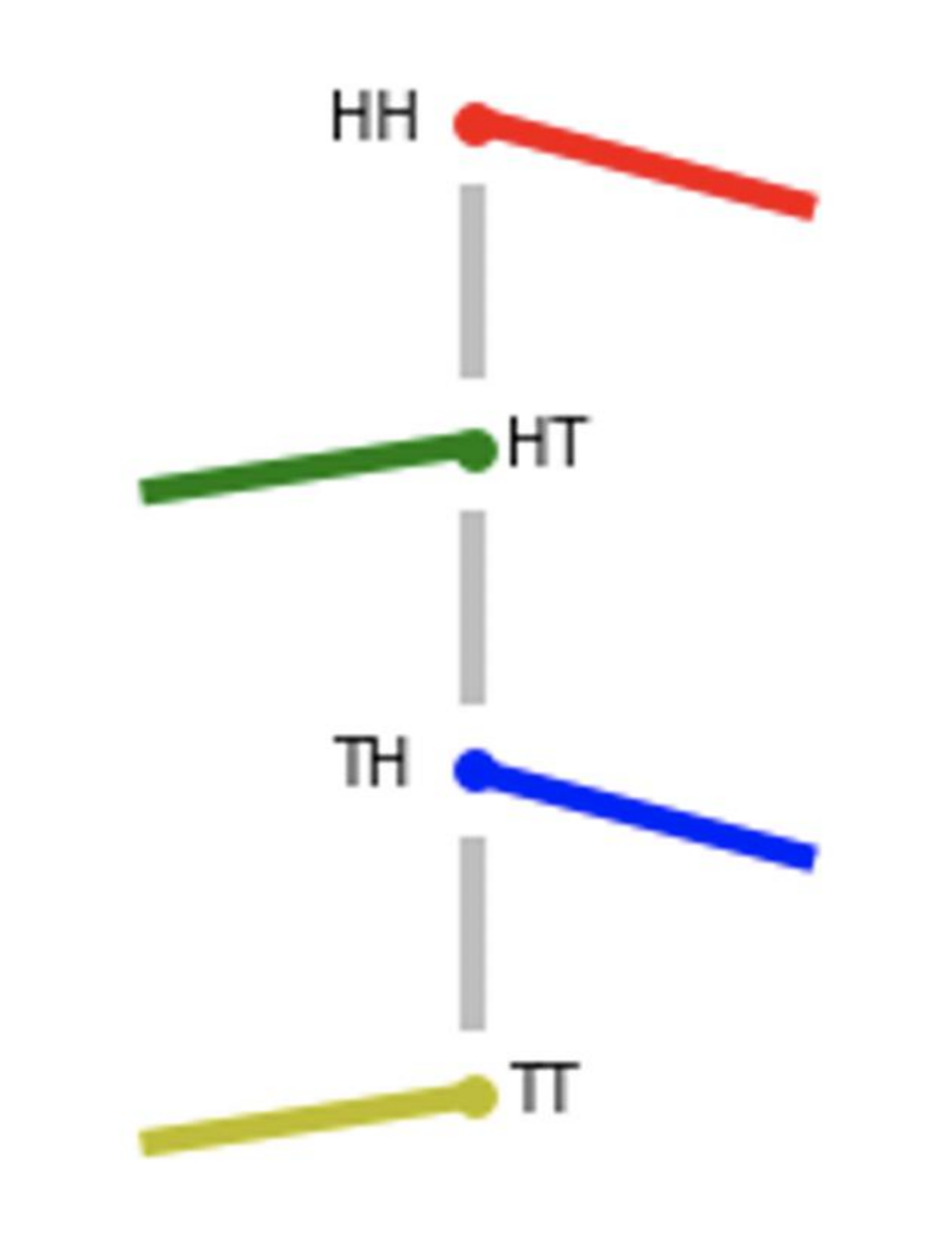}
\caption{\label{fig:11}An arrow representation of a 2-coin quantum system. Note that the arrow itself is not necessary.}
\end{figure}

In the case of a 2-coin system (seen as a four-sided die) each face has its own arrow, and a respective probability of being measured (Fig.~\ref{fig:11}).

In general we will label each outcome using its binary string representation.
As a reminder, binary strings are indexed from right to left, which correlates with the powers of 2.
The decimal value of a binary string is the sum of the powers of 2 that have 1 in their corresponding index.
Any integer can be represented uniquely as a sum of powers of two, and the powers correspond to the 1 digits in its binary representation.
For example, the binary representation of $5=2^2+2^0=1*2^2+0*2^1+1*2^0$ is $101$.

\subsubsection{\label{subsec:composition-example}An Example of How Quantum Computations Work}

\begin{figure}[htb]
\includegraphics[width=2.5cm]{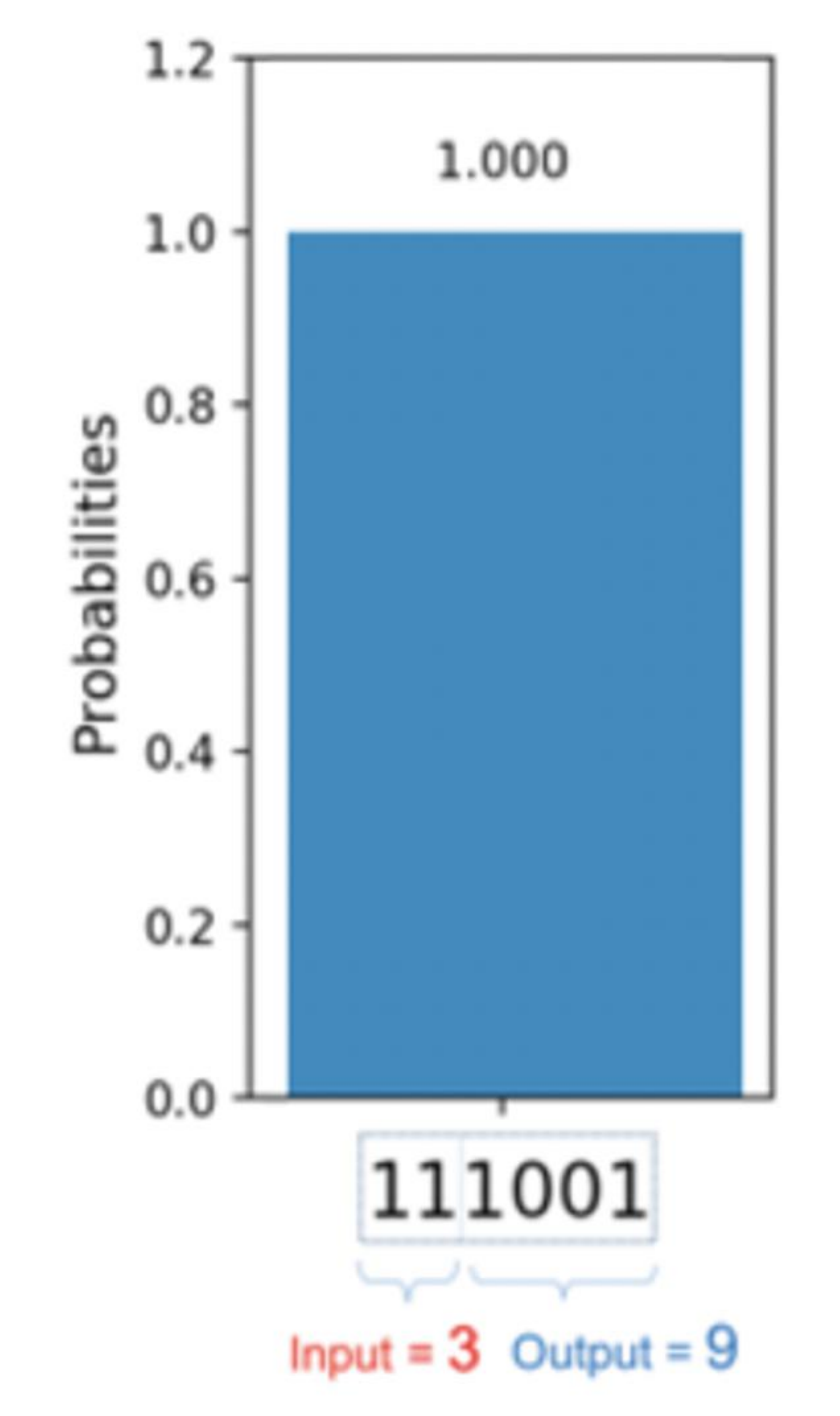}
\caption{\label{fig:12}The measured outcome of the computation with one input.}
\end{figure}

Suppose we want to implement a quantum version of the first program run on the EDSAC, calculating squares.
Given an input, we can design a computation that encodes the square of that input.
The details of the implementation will be described in Sec.~\ref{sec:quantum-dictionary}.
For now, all we need to know is that an outcome will consist of two parts, the input and the output (Fig.~\ref{fig:12}).

\begin{figure}[htb]
\includegraphics[width=6.5cm]{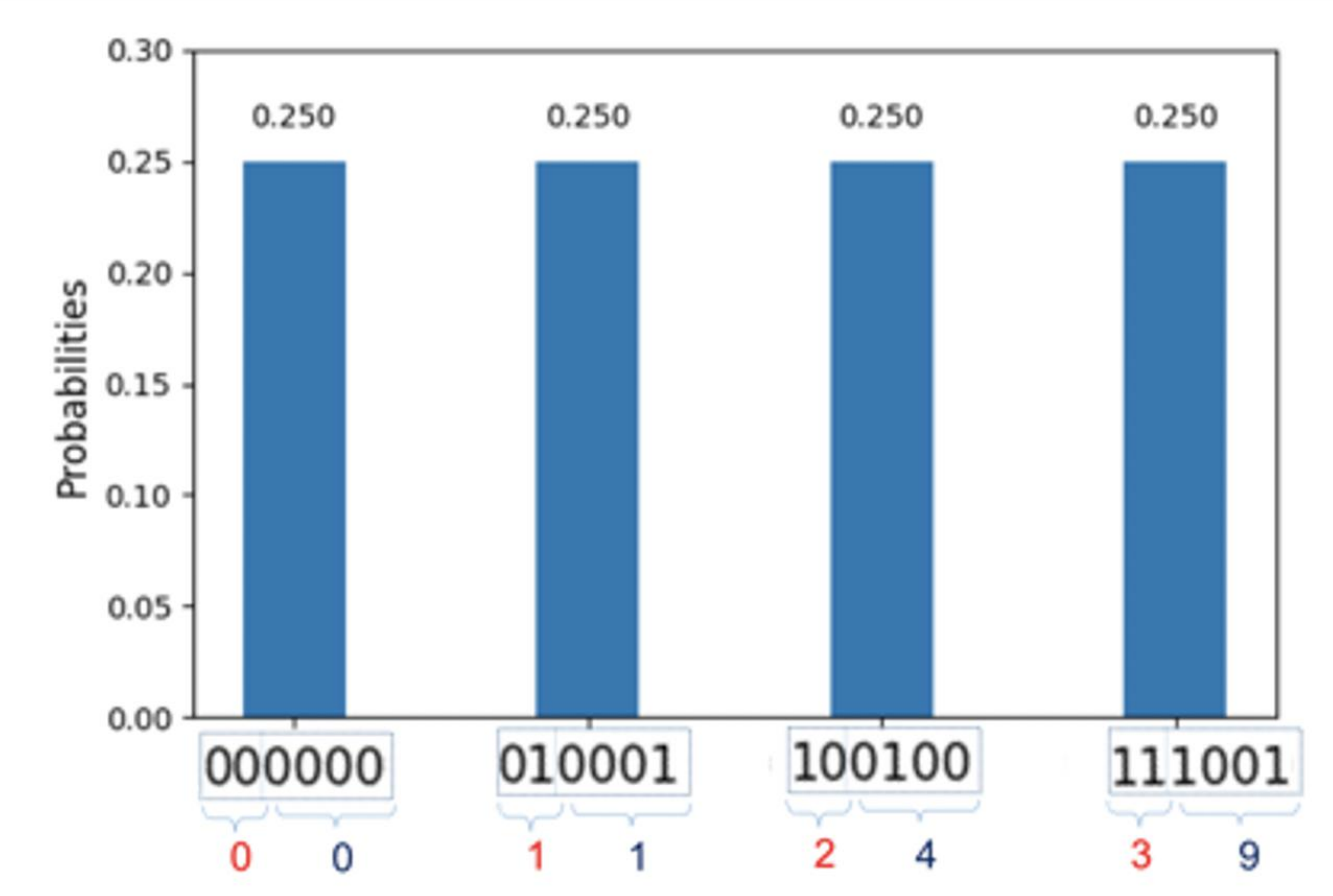}
\caption{\label{fig:13}All possible outcomes of the computation that simultaneously includes all inputs.}
\end{figure}

However, this doesn’t provide any advantage over the classical version. 
In order to pursue a quantum advantage, we can simultaneously encode all inputs and their squares in a quantum state (Fig.~\ref{fig:13}).

While this is a step towards a quantum advantage, we aren’t there yet because measurement reveals a random output - which means that we have to repeat the computation at least 4 times to return the squared values of all of the inputs. 
Some people may find useful an analogy between quantum computations and slot machines, with only two symbols (0 and 1) for each slot and various probabilities for each combination. In order to arrive at a winning combination, one needs to pull the lever many times.
As we will see later, there are techniques to increase the likelihood of receiving a specific outcome. In summary, there are many problems that make pursuing quantum computing worthwhile, but the challenge is in the implementation. Preparing a specific quantum state is both an art and a science.

\subsection{\label{sec:quantum-gates}Quantum Gates}

Similar to an assembly language, there are a limited number of instructions that change the state of a physical quantum system - for example, IBM’s current quantum computer only has four~\cite{Mckay2018}.
The quantum state evolves as we apply \textbf{quantum gates}, which are state transformations built from the instructions.

\begin{figure}[htb]
\includegraphics[width=5cm]{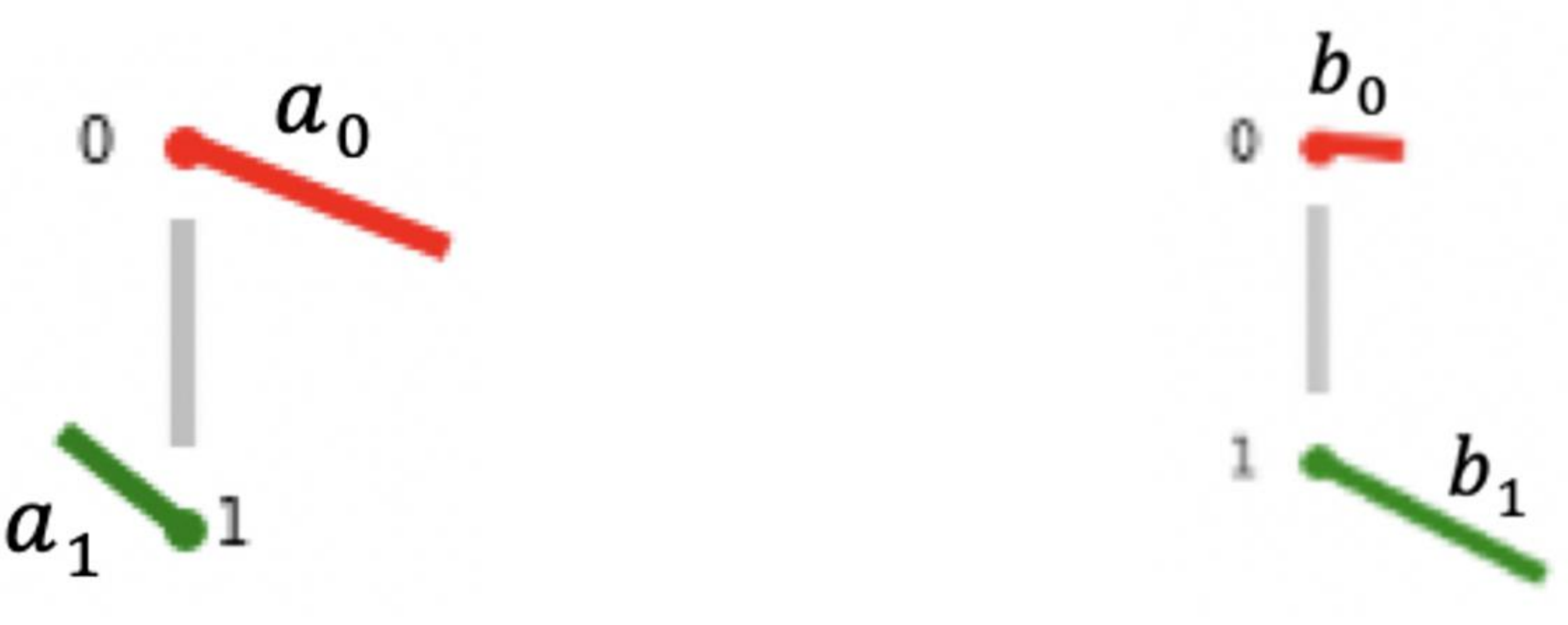}
\caption{\label{fig:14}An arrow representation of (left) an arbitrary, single-qubit quantum state and (right) the quantum state after applying a gate.}
\end{figure} 

Recall that we can think of a qubit as a spinning coin, with two amplitudes that correspond to the probability of observing each side.
We will denote these amplitudes as $a_0$ and $a_1$ (Fig.~\ref{fig:14}).
A transformation of a quantum state leads to new amplitudes $b_0$ and $b_1$, which are a linear combination of the initial amplitudes ($a_0$ and $a_1$) using well-defined formulae, which take the form of weighted averages with complex weights:
$$b_0=c_{00}*a_0  +  c_{01}*a_1$$
$$b_1=c_{10}*a_0  + c_{11}*a_1$$
We will review the descriptions of some of the most common elementary gates, which can be combined into more complex quantum gates.

\subsubsection{\label{subsec:elementary-gates}The Elementary Single-Qubit Gates}

\begin{center}
    \textbf{$X$ gate and $R_X$ gate}
\end{center}

The X gate swaps the two amplitudes (of the outcomes) of a single-qubit system:
$$b_0=a_1$$
$$b_1=a_0$$
The $R_X$ gate can be thought of as a sort of generalization of the $X$ gate, parameterized by an angle.
The formulae for $R_X(\theta)$ are as follows:
$$b_0=\cos\frac{\theta}{2}a_0-i\sin\frac{\theta}{2}a_1$$
$$b_1=-i\sin\frac{\theta}{2}a_0+\cos\frac{\theta}{2}a_1$$
Note that $X=i*R_X(\pi)$, taking into account that $\sin\frac{\pi}{2}=1$ and $\cos\frac{\pi}{2}=0$.

\begin{center}
    \textbf{$Z$ gate and $R_Z$ gate}
\end{center}

The $Z$ gate changes the sign of $a_1$ and does not affect $a_0$:
$$b_0=a_0$$
$$b_1=-a_1$$
An $R_Z(\theta)$ gate performs the following transformation:
$$b_0=(\cos\frac{\theta}{2}- i*\sin\frac{\theta}{2}*a_0$$
$$b_1=(\cos\frac{\theta}{2}+ i*\sin\frac{\theta}{2}*a_1$$
Note that $Z=i*R_Z(\pi)$.
\newpage
\begin{center}
    \textbf{Hadamard $H$ gate}
\end{center}

\begin{figure}[htb]
\includegraphics[width=13cm]{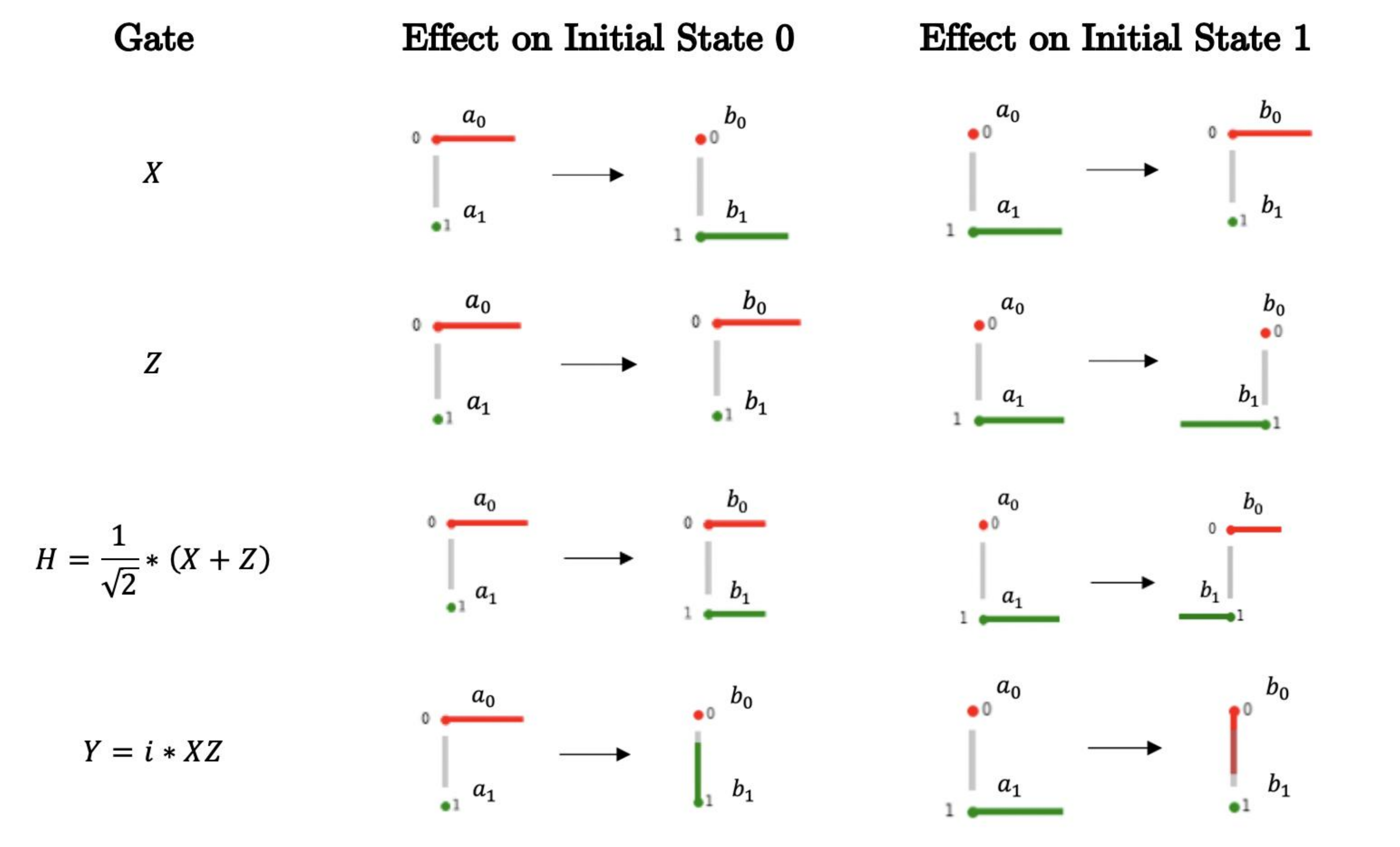}
\caption{\label{fig:16}The transformation effects of the elementary gates on the basis states.}
\end{figure}

The Hadamard gate acts on amplitudes to produce their normalized sum and the difference.
$$b_0=\frac{1}{\sqrt{2}}(a_0+a_1)$$
$$b_1=\frac{1}{\sqrt{2}}(a_0-a_1)$$
Note that $H=\frac{1}{\sqrt{2}}(X+Z)$.
The practical use of the Hadamard gate comes from its ability to create a superposition of basis states, shown in (Fig.~\ref{fig:16}).
If you start with either $0$ or $1$ states, we see that the resulting state is an equal superposition, and both outcomes become equally likely to be observed.

\begin{center}
    \textbf{$Y$ gate and $R_Y$ gate}
\end{center}

The $Y$ gate first swaps the $a_0$ and $a_1$ amplitudes of a single-qubit system, and then rotates $a_0$ by $\frac{\pi}{2}$ and $a_1$ by $-\frac{\pi}{2}$:
$$b_0=-ia_1$$
$$b_1=ia_0$$
An $R_Y(\theta)$ gate performs the following transformation: 
$$b_0=\cos\frac{\theta}{2}a_0-\sin\frac{\theta}{2}a_1$$
$$b_1=\sin\frac{\theta}{2}a_0+\cos\frac{\theta}{2}a_1$$

Note that $Y=iR_Y(\pi)$ and $Y=iXZ$. 

\begin{figure}[htb]
\includegraphics[width=13cm]{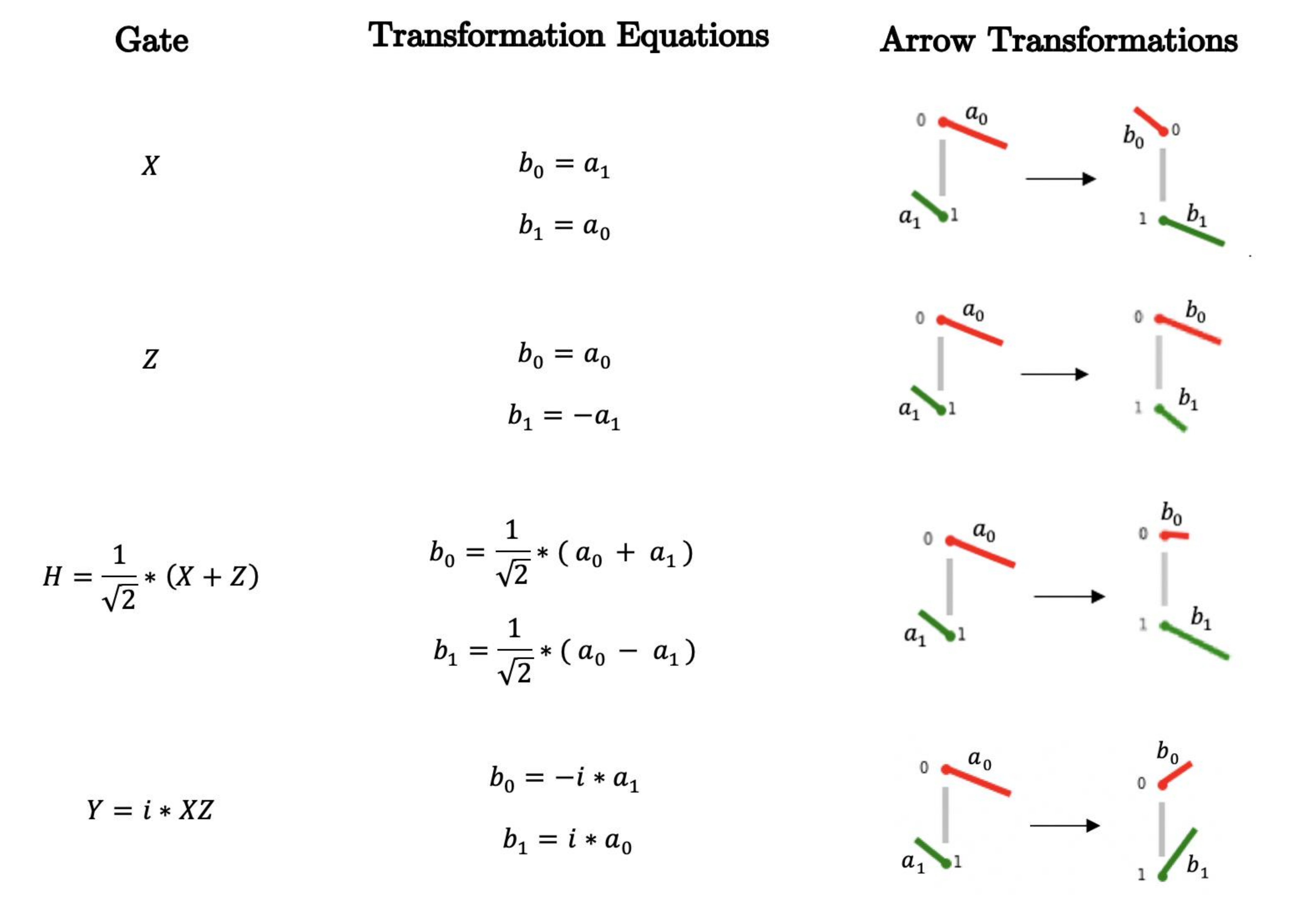}
\caption{\label{fig:15}The transformation effects of the elementary gates in algebraic and geometric forms.}
\end{figure} 

\clearpage

\subsubsection{\label{subsec:example-single-qubit-gates}Examples of Applying Single-Qubit Gates}

\begin{figure}[htb]
\includegraphics[width=5cm]{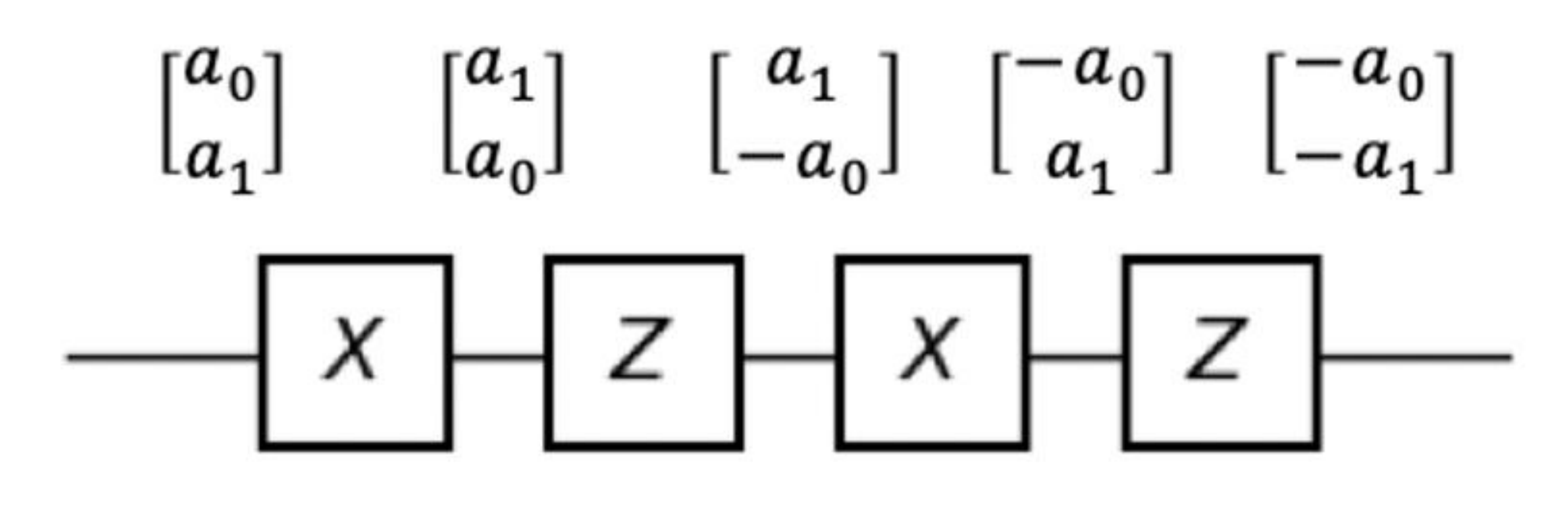}
\caption{\label{fig:17}A simple example of applying multiple single-qubit gates, read as $ZXZX$ which multiplies the amplitudes of a quantum state by $-1$.}
\end{figure}

We can apply a sequence of multiple gates to a given qubit, represented as shown in Fig.~\ref{fig:17}.
In analogy with classical circuits, the line on which the gates are shown is sometimes called a \textbf{wire}.
Each quantum gate is applied one after the other corresponding to different time steps, represented by their sequential order (circuits are read left to right).

The way to interpret the composition sequence $ZXZX$ is as follows: First, we need to specify the initial state (in this case, an arbitrary $a_0$ and $a_1$).
Second, gates are applied sequentially, transforming the state as shown in Fig.~\ref{fig:17}.
This computation is actually an implementation of an important pattern in algorithms, multiplying the amplitudes of the quantum state by $-1$.
This pattern can be applied to any state, and is a straightforward way to negate amplitudes. 
We will explore this pattern (and its alternatives) in a future section.

\begin{figure}[htb]
\includegraphics[width=4cm]{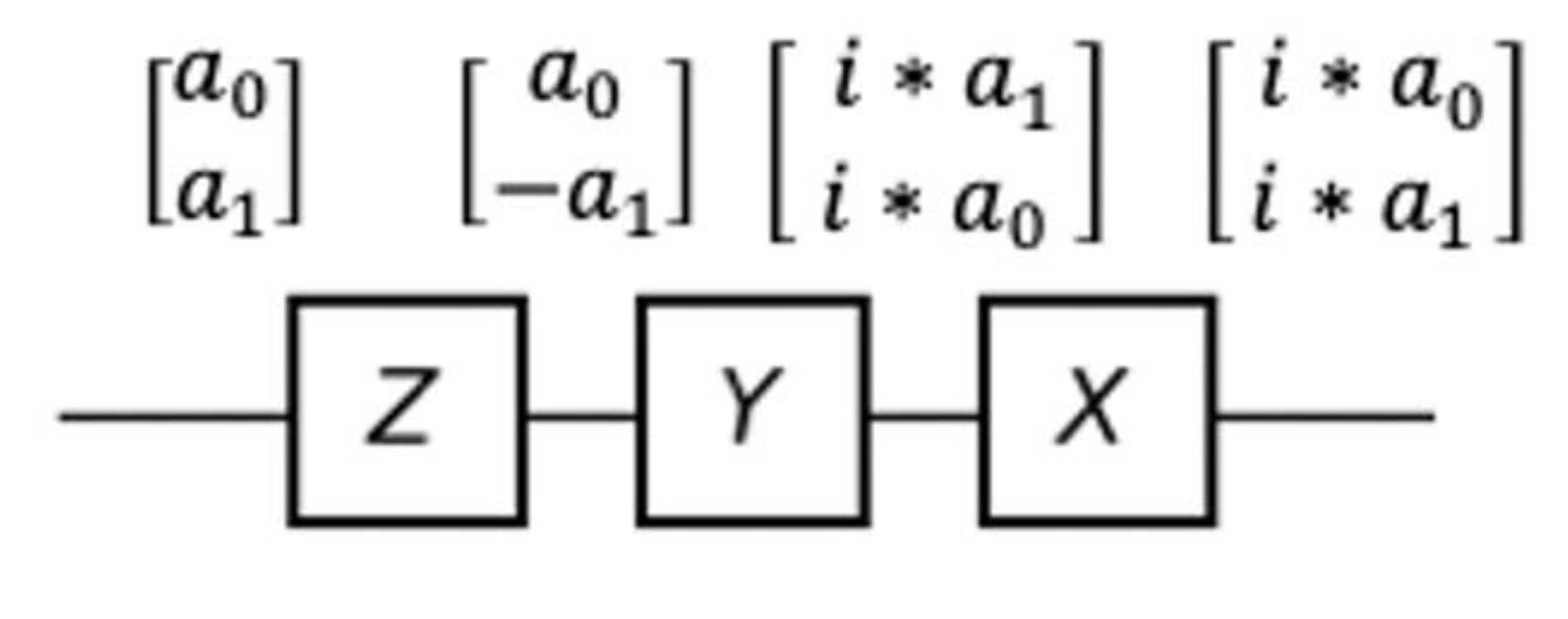}
\caption{\label{fig:18}An $XYZ$ circuit, which multiplies the amplitudes of a quantum state by $i$.}
\end{figure}

A similar pattern is multiplying both amplitudes by $i$, seen in Fig.~\ref{fig:18}. 

\subsubsection{\label{subsec:multi-qubit-circuits}Multi-Qubit Circuits}

\begin{figure}[htb]
\includegraphics[width=3cm]{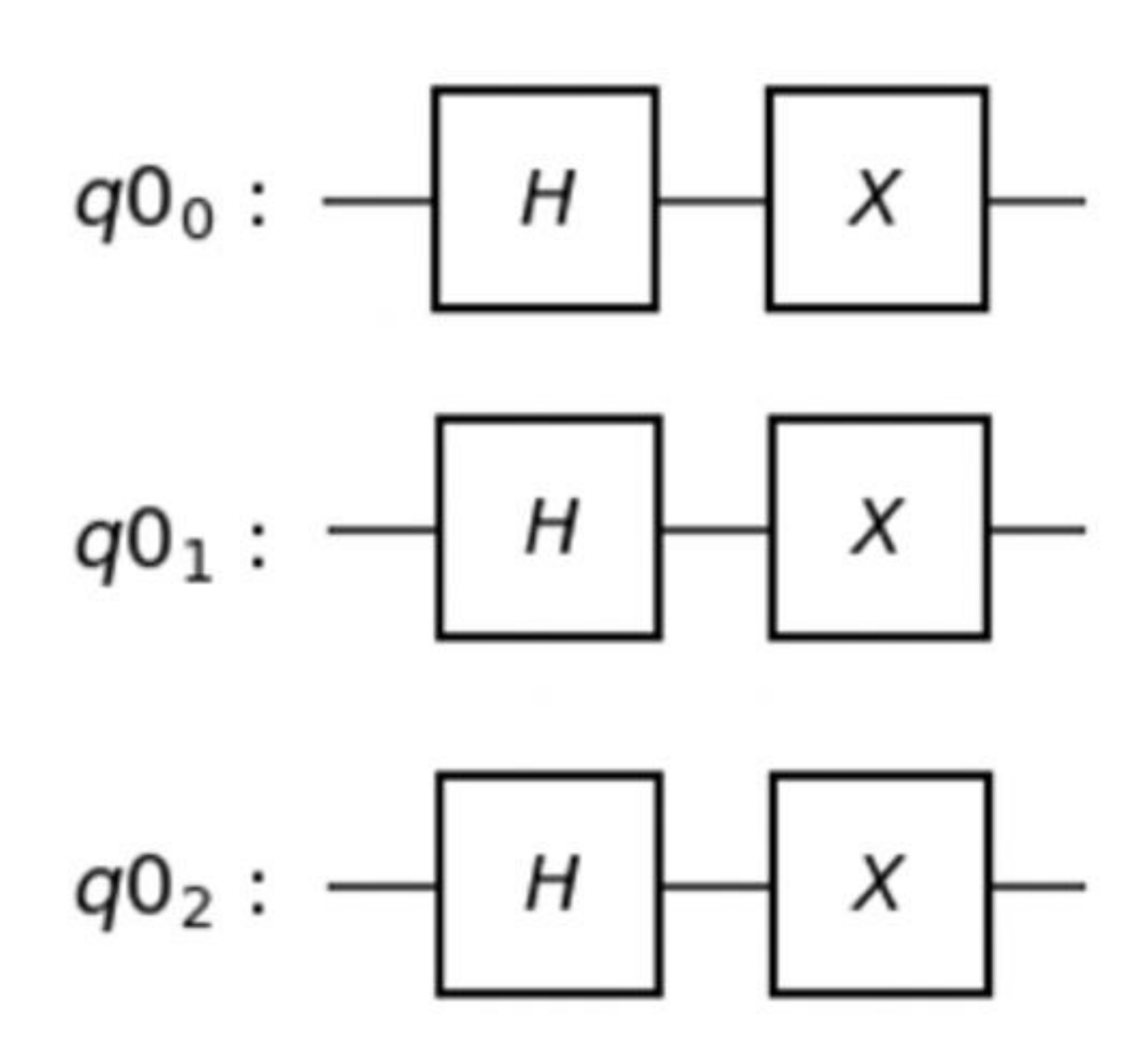}
\caption{\label{fig:19}An example of a multi-qubit quantum circuit.}
\end{figure}

A quantum circuit consists of the composition of multiple qubits, as discussed in Sec.~\ref{subsec:composition-of-a-quantum-system}.
When building circuits we can logically group qubits into registers, which is useful when separating input, output, and ancillary qubits.
Single-qubit gates can be applied in sequence to any of the qubits in the system, as shown in Fig.~\ref{fig:19}.
In this paper, the bit at index $0$ in a binary string is the top-most wire, unless otherwise specified.

\begin{figure}[htb]
\includegraphics[width=12cm]{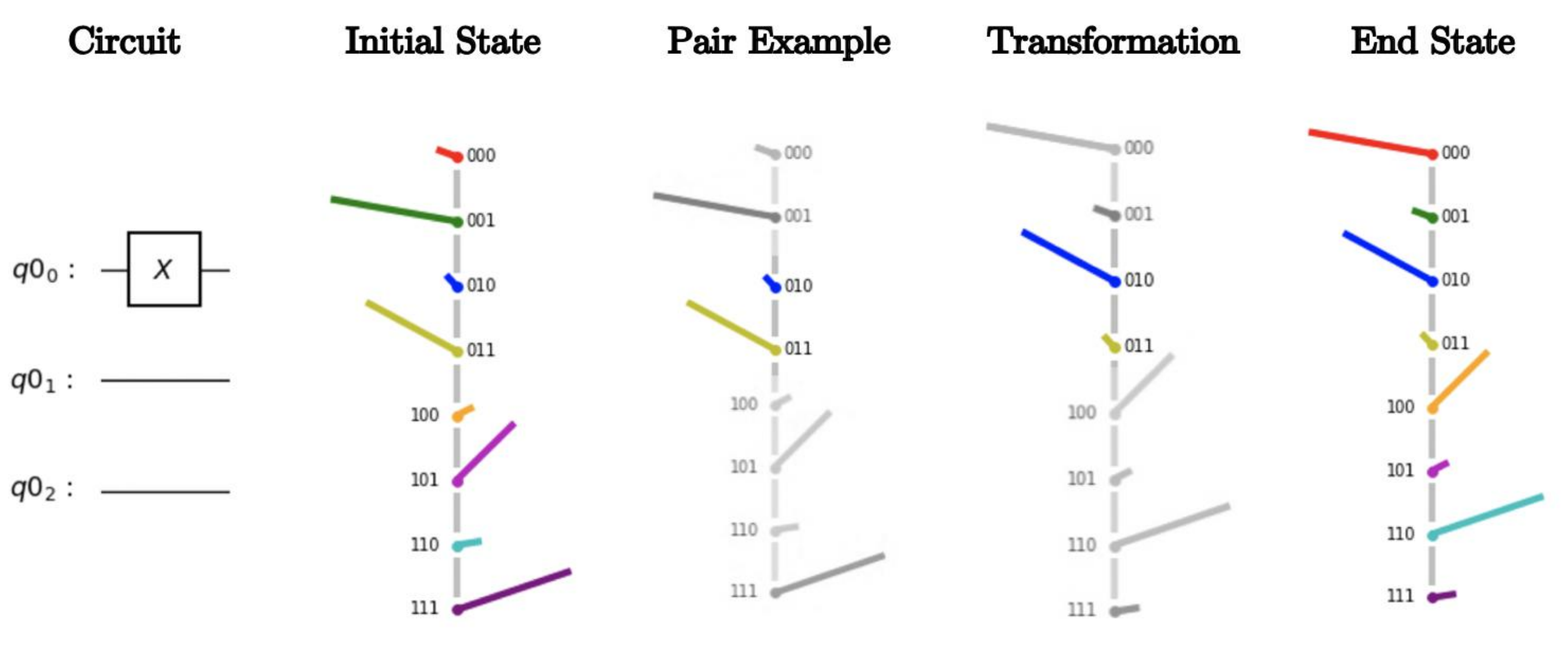}
\caption{\label{fig:20}A single-qubit gate will potentially transform all the amplitudes in the state of a multi-qubit quantum system.}
\end{figure}

If we consider the quantum state as a dictionary of key-value pairs, with the label (a binary string) as the key and the amplitude as the value.
When applying a gate to a qubit, we are indicating an index $j$ such that $0 \le j < n$, where $n$ is the number of bits in the label.
We then pair two keys, such that they differ only in position $j$.
The pairs are then transformed according to the definition of the gate.
Suppose we start with an arbitrary state of three qubits (Fig.~\ref{fig:20}).
In this case $j=0$, and thus $010$ and $011$ are pairs.
We simultaneously identify the three other pairs in the state (these are greyed out in Fig.~\ref{fig:20} for illustration purposes).
The amplitudes of these pairs are swapped by the $X$ gate, highlighted in the transformation step, resulting in the end state.

\subsubsection{\label{subsec:conditional gates}Conditional Gates}

\begin{figure}[htb]
\includegraphics[width=12cm]{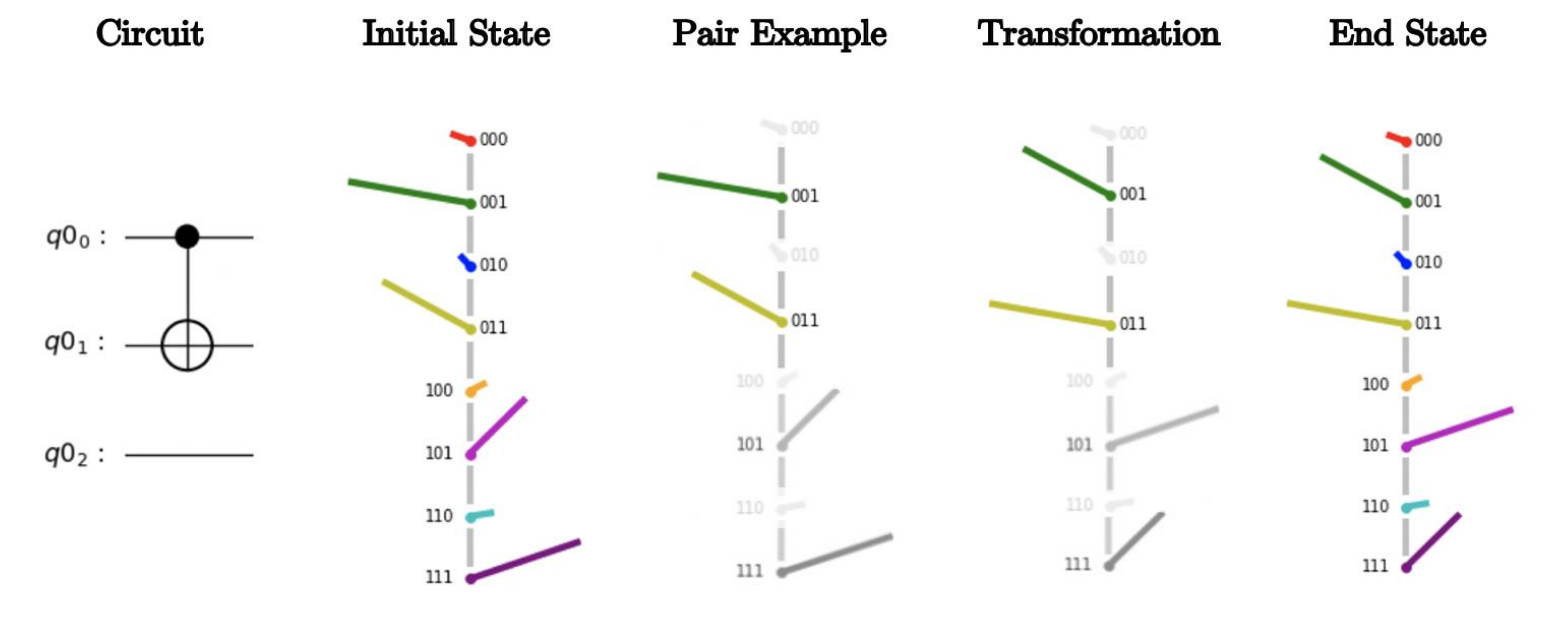}
\caption{\label{fig:21}A single-qubit controlled gate will potentially transform at most half of the amplitudes in the state of a multi-qubit quantum system.}
\end{figure}

A \textbf{conditional gate} is a gate that is applied only to a subset of the amplitudes---those that have $1$ in the control bits.
Let us consider the $cX$ gate, where an $X$ gate is performed on the \textbf{target qubit} when the \textbf{control qubit} is $1$ (Fig.~\ref{fig:21}). 

In Fig.~\ref{fig:21} we also show how conditional gates perform the transformation of pairs.
As described in the previous subsection, when applying a gate to a qubit, we are indicating an index $j$ such that $0 \le j < n$, where $n$ is the number of bits in the label.
Half the states are $1$ in position $j=0$ and only those states are considered for the transformation (the unaffected outputs are faded out in Fig.~\ref{fig:21}).
We then select the pairs on which the $X$ gate is applied.

\subsubsection{\label{subsec:entanglement-example}Example of Entanglement - A Die as a Coin}

\begin{figure}[htb]
\includegraphics[width=5.5cm]{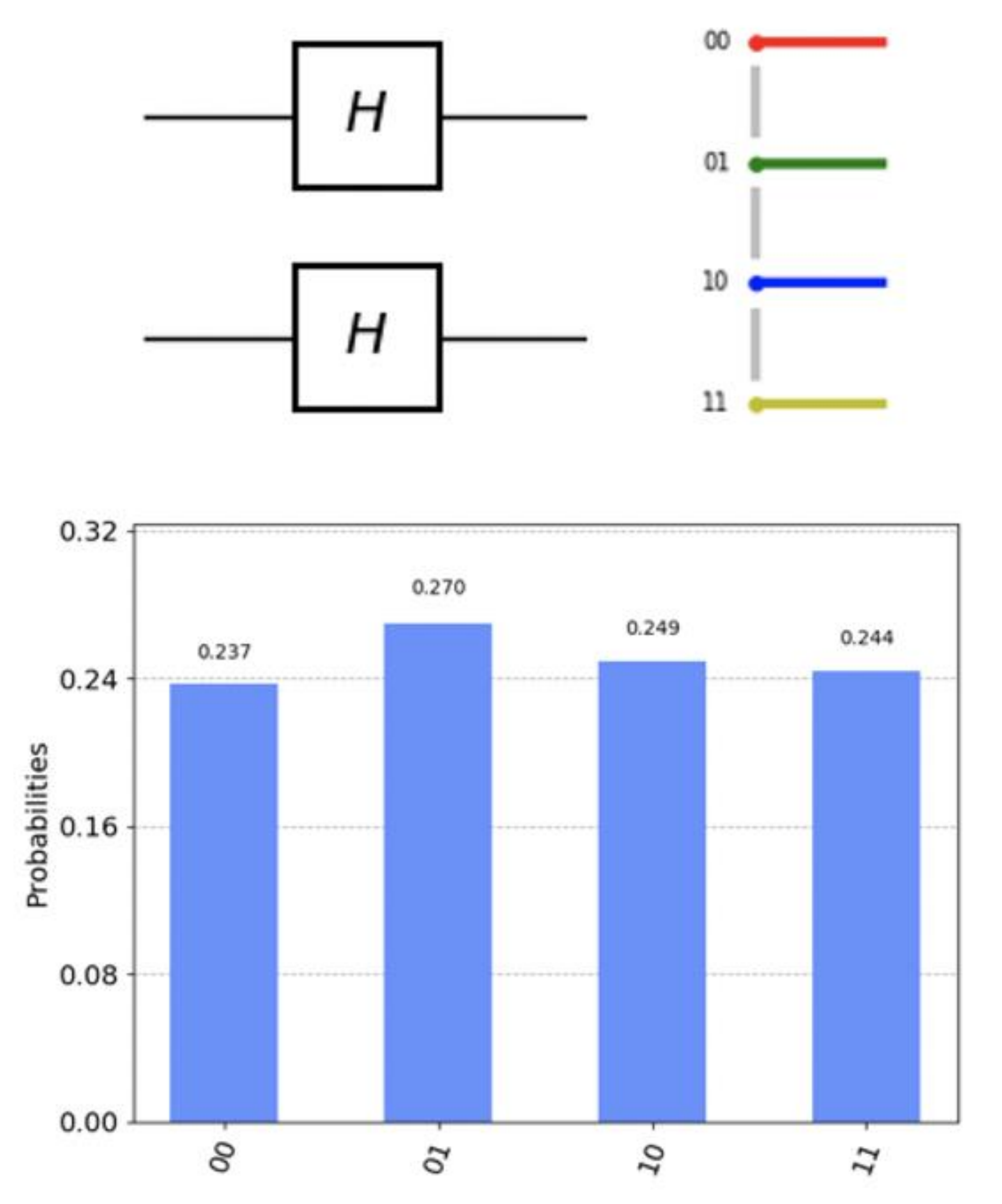}
\caption{\label{fig:22}The resulting probabilities of a two-qubit state generated from (top-right) a state of equal superposition made by (top-left) a circuit of Hadamard gates.}
\end{figure}

Let’s take the example of a $2$-qubit system, seen as a four-sided die with sides labelled $0$ through $3$.
We roll the die $1000$ times, resulting in the cumulative histogram seen in Fig.~\ref{fig:22}.

\begin{figure}[htb]
\includegraphics[width=5.5cm]{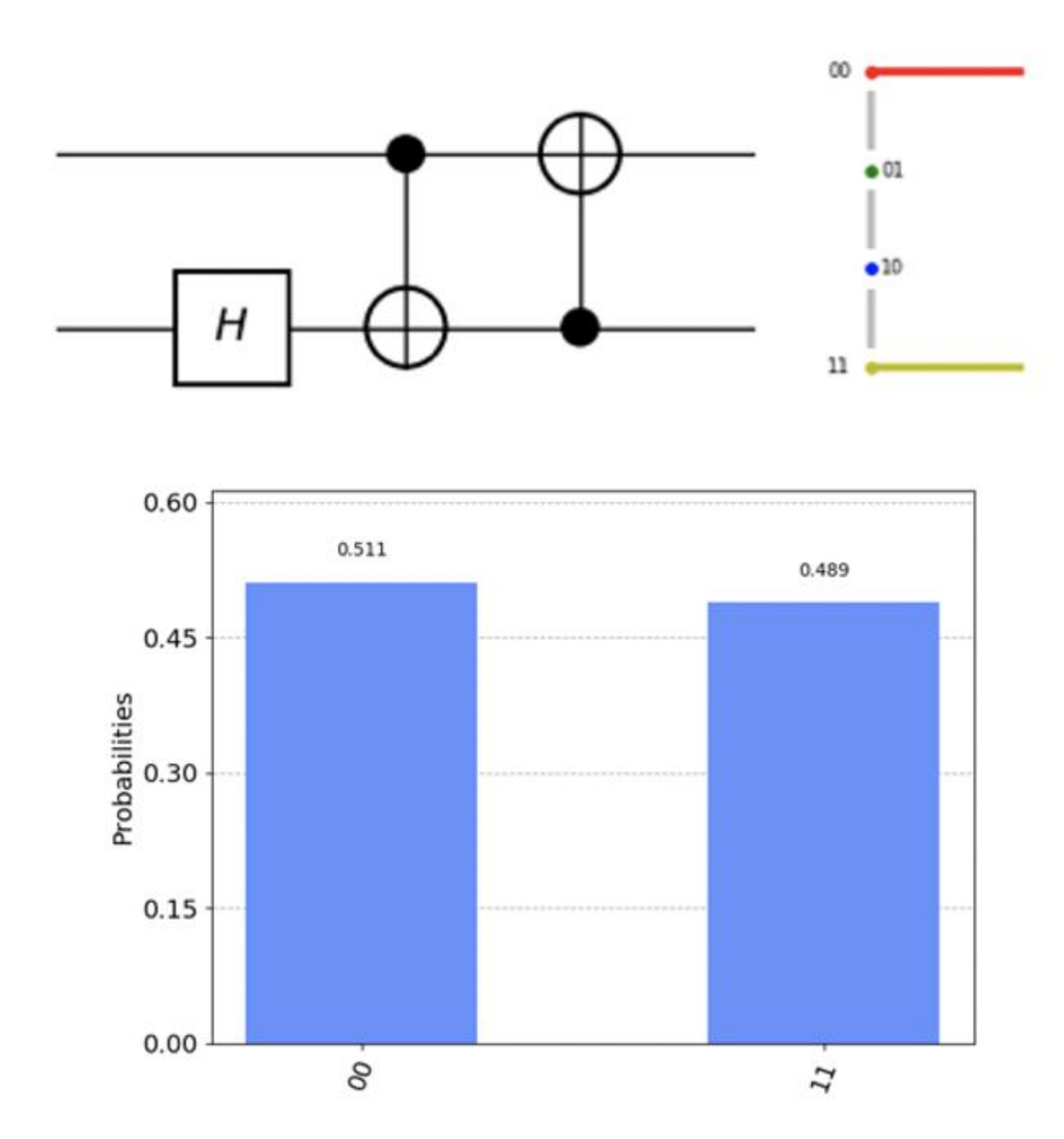}
\caption{\label{fig:23}(bottom) The resulting probabilities of (top-right) the entangled, two-qubit state created by (top-left) a circuit control gates and a Hadamard.}
\end{figure}

We want to force the die to only return one of two results---$00$ or $11$.
In a quantum system, this can be done with control gates to create entanglement between the two qubits.
Specifically, we force one qubit’s measurement to match the other, eliminating the outcomes $01$ and $10$, which can be done as seen in Fig.~\ref{fig:23}.

To achieve the resulting histogram, we repeat the computation $1000$ times, and see that we are approximating a fair coin toss using a four-sided die.

\subsubsection{\label{subsec:fair-results-example}Example - Fair Results from a Biased Coin}

In 1951, John von Neumann proposed a procedure for getting fair results from a biased coin~\cite{vonNeumann1951}:
\begin{enumerate}
    \item Toss the coin twice.
    \item If the results match, discard both results and start over.
    \item If the results differ, use the first result, and forget the second.
\end{enumerate}

\begin{figure}[htb]
\includegraphics[width=5.5cm]{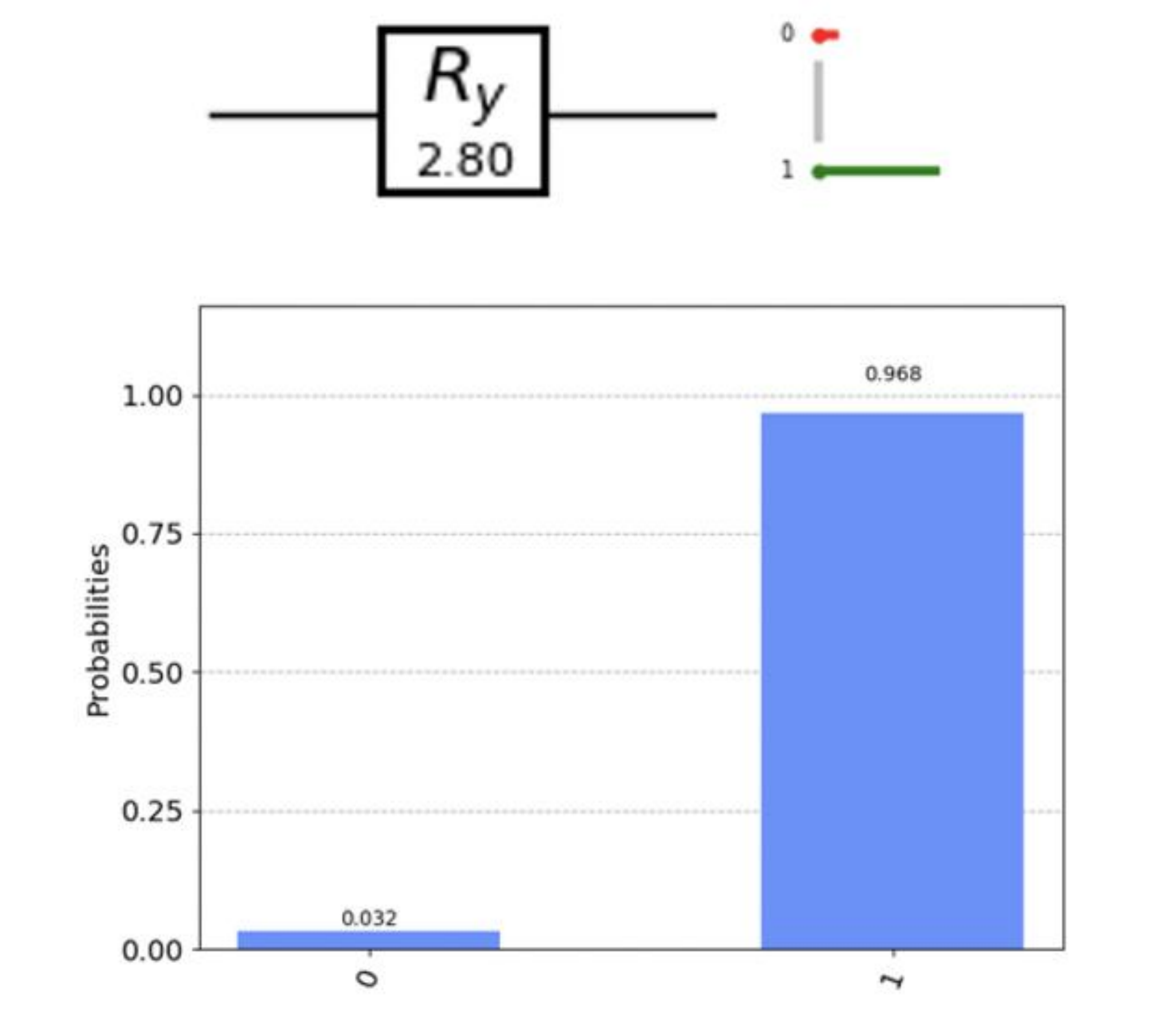}
\caption{\label{fig:24}A quantum representation of a biased coin using an $R_Y$ gate and 1000 shots.}
\end{figure}

We can simulate tossing a biased coin using a quantum system.
As discussed in Sec.~\ref{sec:classical-random-quantum-bits}, a qubit can be thought of as a quantum coin, with the bias defined as the probability of measuring $1$.
For an angle $\theta \in [0, 2\pi)$ we can use $R_Y(\theta)$ on such a qubit to represent a coin toss.
We can get an estimate of the bias by repeating the computation a large number of times (e.g. $1000$ repetitions) and counting the number of times each output occurs (Fig.~\ref{fig:24}).
In probability theory, this is called a Bernoulli random process~\cite{Klenke2014}.

\begin{figure}[htb]
\includegraphics[width=5.5cm]{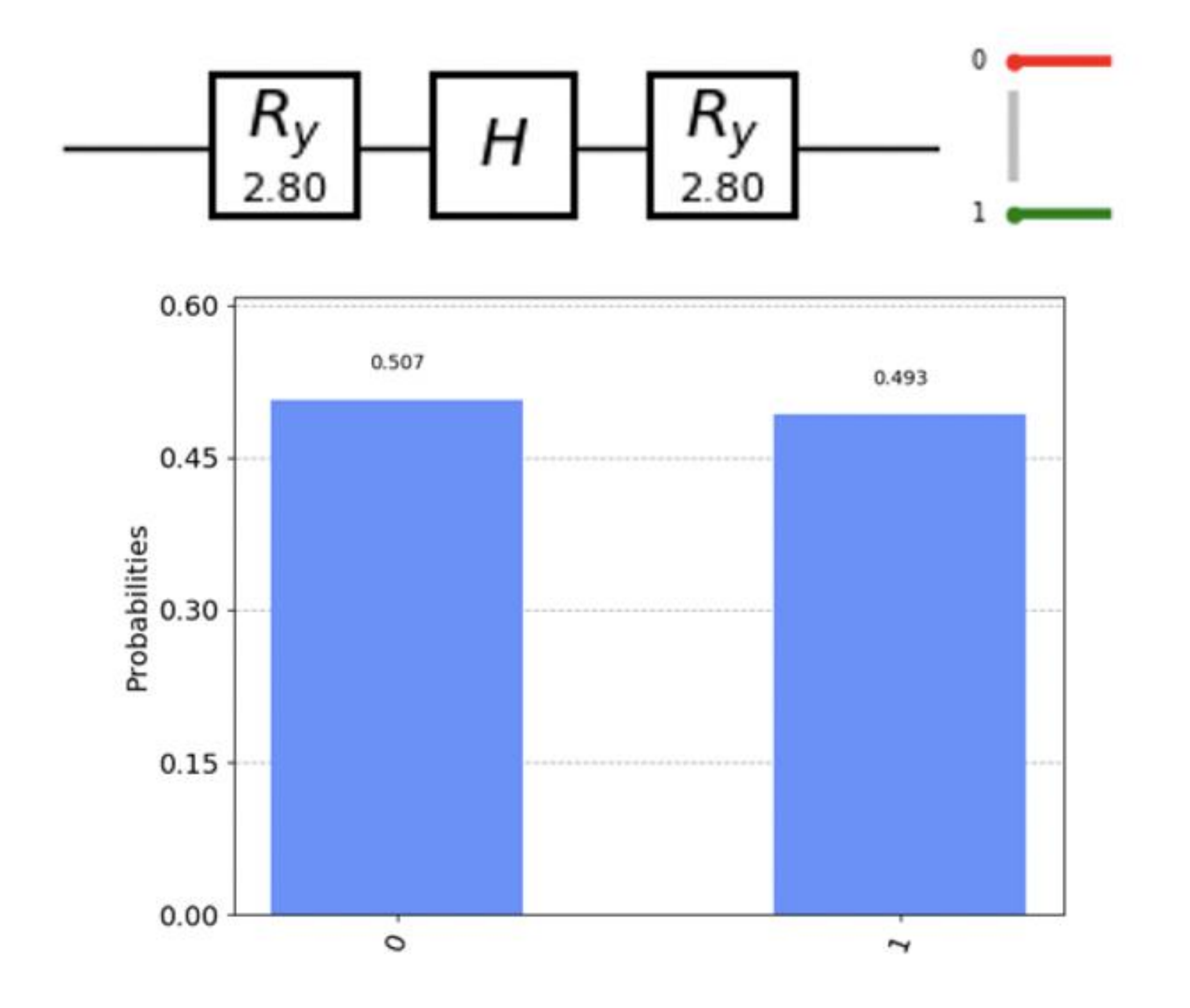}
\caption{\label{fig:25}(top) A quantum circuit representing two biased coin flips, made “fair” by a Hadamard gate in the middle. This is verified by (bottom) the probabilities.}
\end{figure}

\begin{figure}[htb]
\includegraphics[width=5cm]{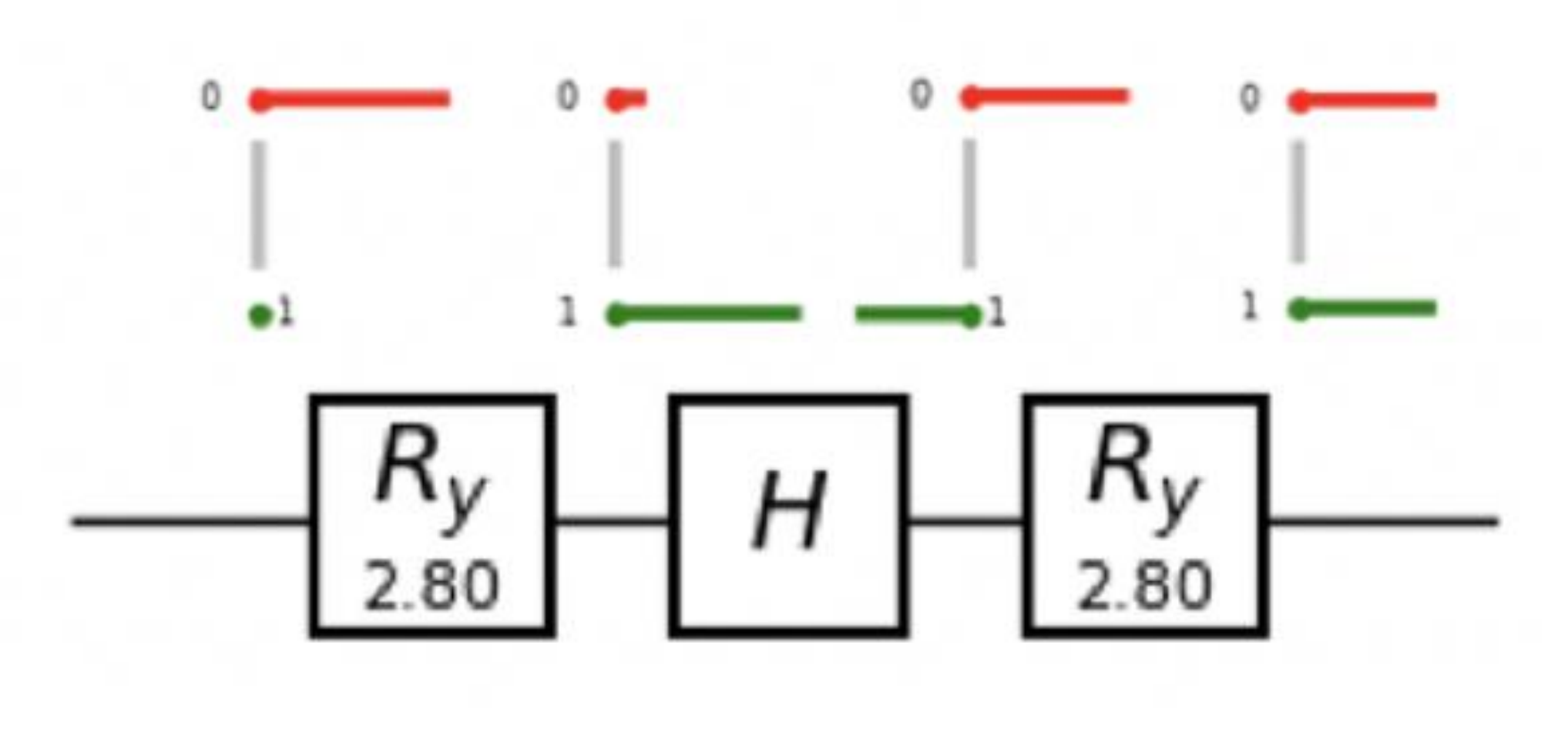}
\caption{\label{fig:26}A gate-by-gate arrow representation of the quantum state, showing how the two biased coin flips become unbiased with the inclusion of a Hadamard.}
\end{figure}

We can model two coin flips as two $R_Y$ gates in sequence, made fair by a Hadamard gate between them (Fig.~\ref{fig:25}).
To understand why this works, it may be helpful to look at the state before and after every gate in the circuit (Fig.~\ref{fig:26}).

\section{\label{sec:insights-on-quantum-algorithms}Insights on Quantum Algorithms}
 
\begin{figure}[htb]
\includegraphics[width=10cm]{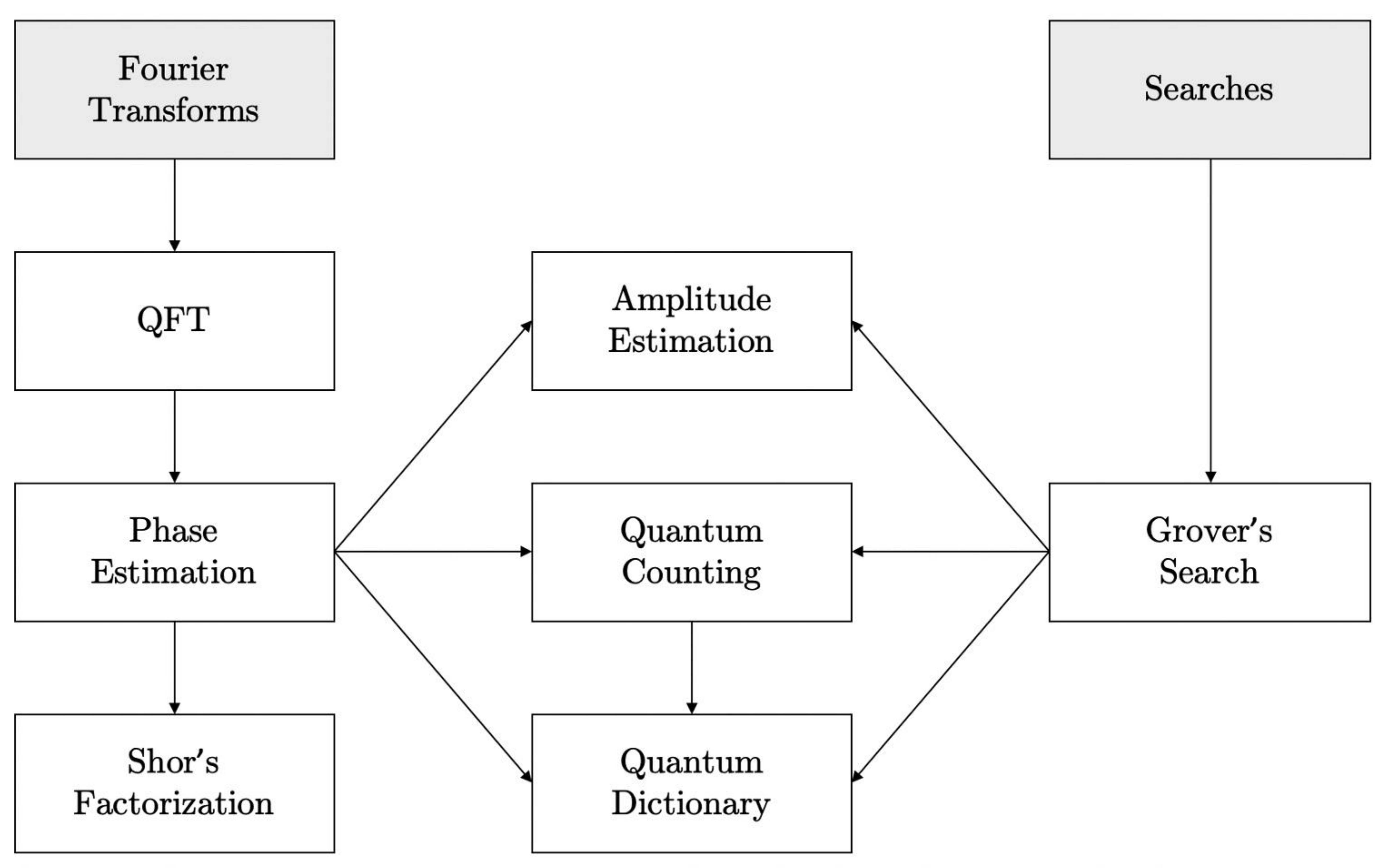}
\caption{\label{fig:27}The two main classes of quantum algorithms.}
\end{figure}

There are two main classes of algorithms where there is a potential quantum advantage---those that rely on Fourier transforms and those that perform searches~\cite{Shor1997,Grover1996,Chappell2011} (Fig.~\ref{fig:27}).

We will explore all the algorithms and patterns shown in Fig.~\ref{fig:27} except for Shor’s Factorization.

\subsection{\label{sec:classical-approx}Classical Approximations with Fourier Transforms}

\begin{figure}[htb]
\includegraphics[width=5cm]{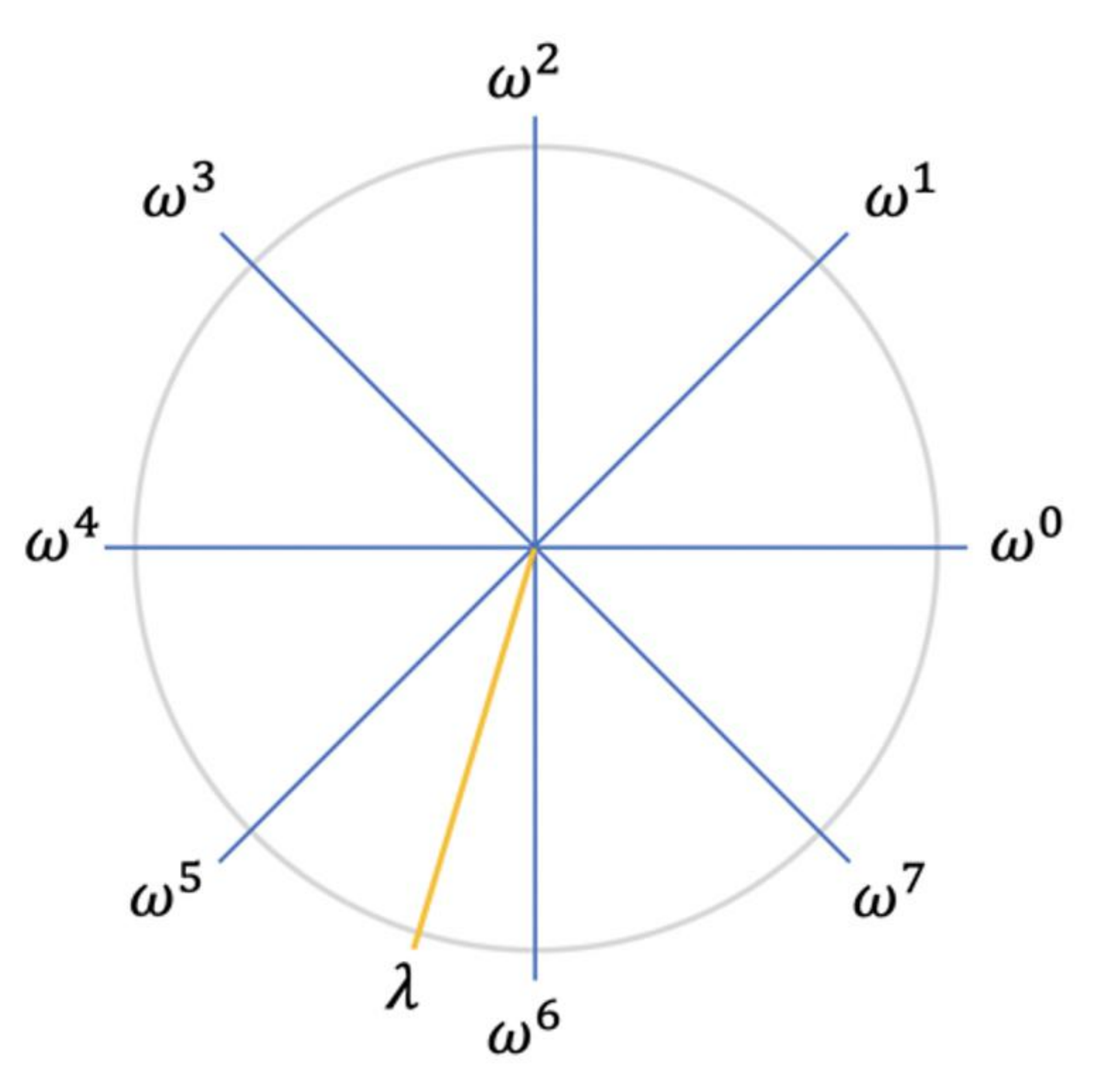}
\caption{\label{fig:28}The comparison of each $\omega^k$ for $N=8$ and a complex parameter $\lambda$.}
\end{figure}

For an unknown, unit complex parameter $\lambda=\cos\theta+i\sin\theta$, we want to approximate $\lambda$ with a root of unity $\omega^k$, where $\omega=\cos\frac{2\pi}{N}+i\sin\frac{2\pi}{N}$ and $k\in\{0,1,...,N-1\}$.
The best approximation is the closest root of unity, the one whose phase is the closest to the phase of $\lambda$ (Fig.~\ref{fig:28}).
The higher $N$ is, the better the approximation.
The inner product is a good similarity measure between two unit vectors, because it is the cosine of the angle (phase difference) between the vectors.

\begin{figure}[htb]
\includegraphics[width=15cm]{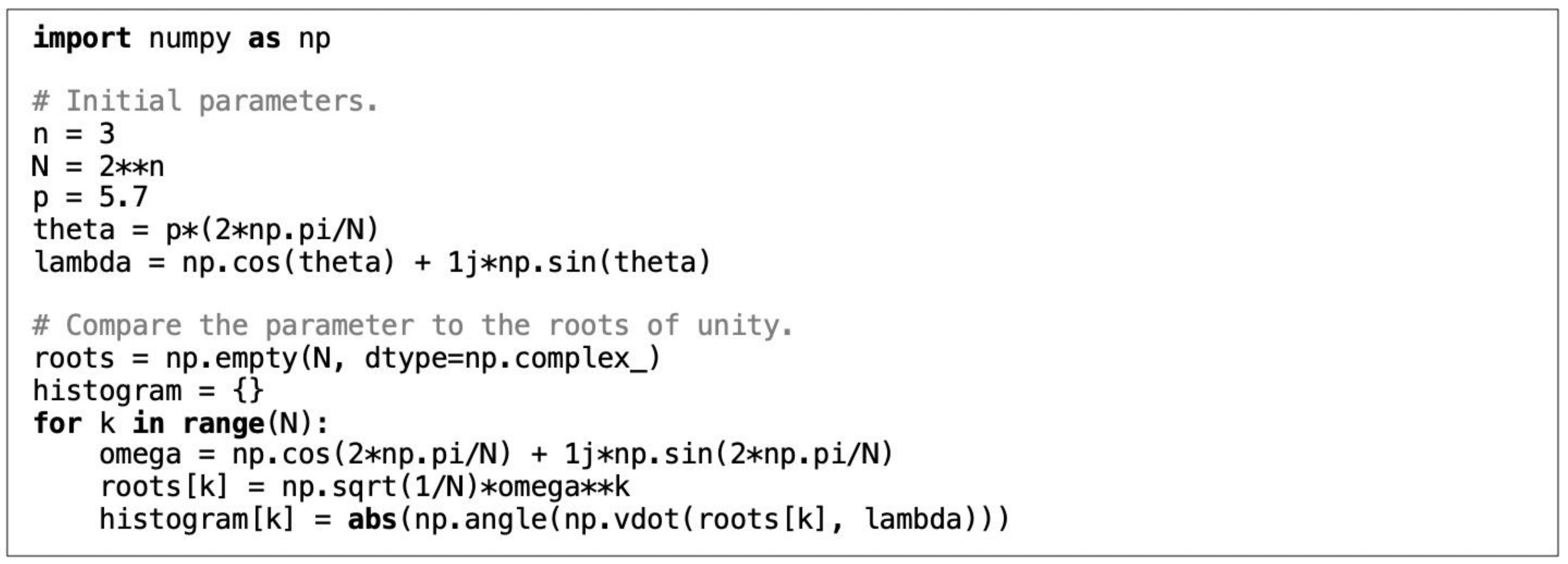}
\caption{\label{fig:29}A Python snippet that compares a complex parameter to the roots of unity using their angles.}
\end{figure}

\begin{figure}[htb]
\includegraphics[width=2cm]{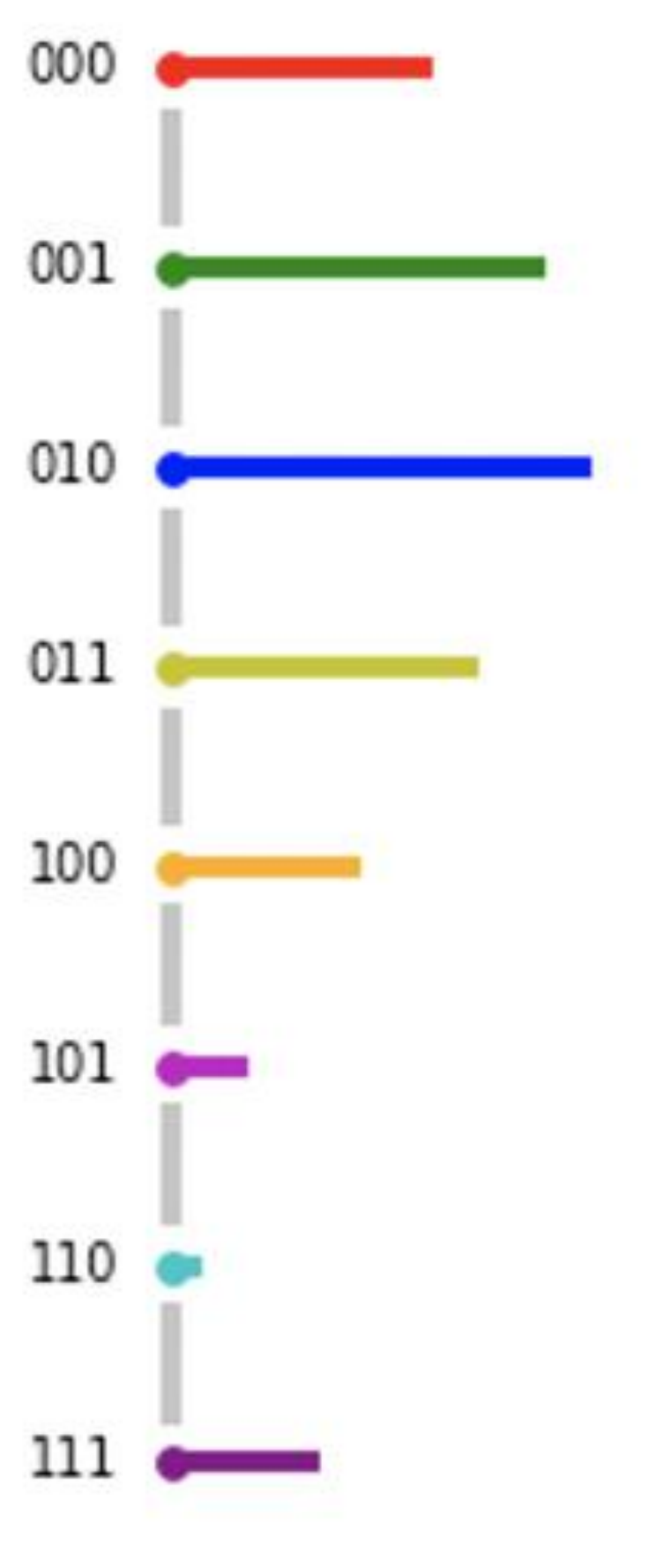}
\caption{\label{fig:30}The visual representation of the histogram variable in the code fragment above.}
\end{figure}

The phase $\theta$ of a unit complex number can be represented as a multiple $p$ of the phase $\frac{2\pi}{N}$ of the base root of unity $\omega$, i.e. $\theta=p\frac{2\pi}{N}$.
As an example, let’s look at encoding the unit complex number corresponding to $p=5.7$, and find the closest root of unity (Fig.~\ref{fig:29}).
In Fig.~\ref{fig:30}, $110$ ($6$) and $101$ ($5$) are the best approximations, as their inner products with their respective roots of unity have the smallest phase difference.

\begin{figure}[htb]
\includegraphics[width=12cm]{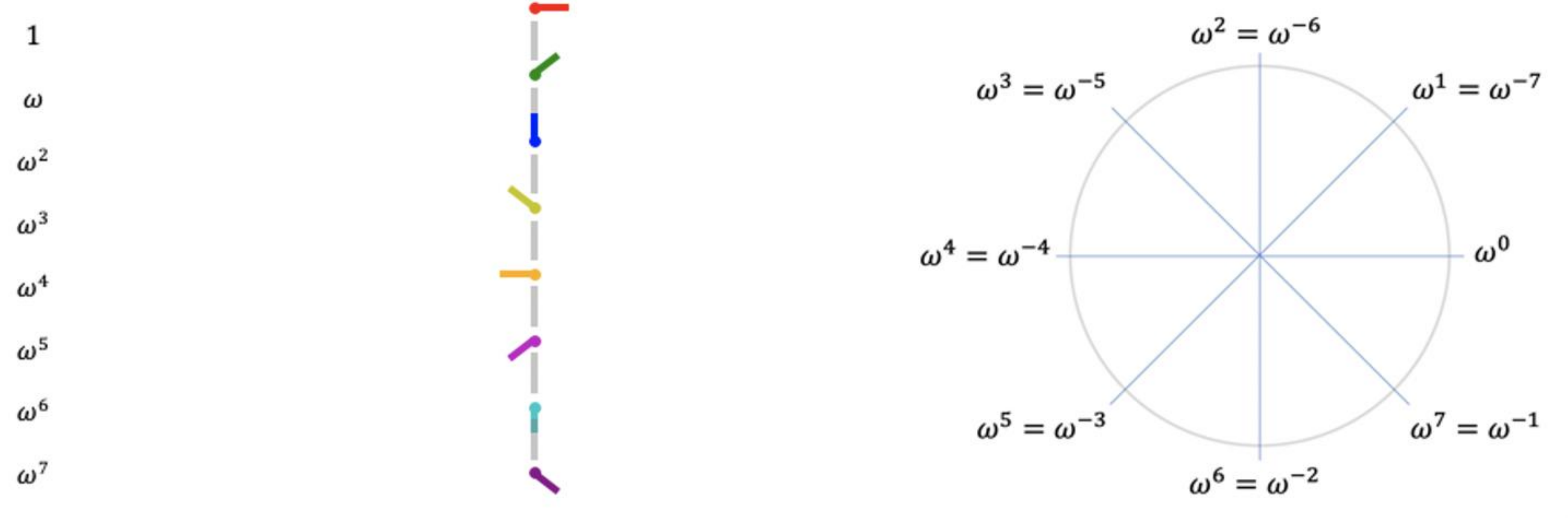}
\caption{\label{fig:31}A representation of the first non-trivial ($k=1$) Fourier basis for $N=8$ using a histogram and the angles.}
\end{figure}

This method does not directly work on a quantum system, as angles are not detected when measuring---only magnitudes.
Fourier transforms, described below, can help with converting differences in angle into magnitudes.
Note that we are describing the inverse Fourier transform in this section, which uses the positive powers of the roots of unity.
To apply the direct Fourier Transform use the negative powers instead.
Suppose we have a sequence of $N$ complex numbers, $[\lambda_0,\lambda_1,...,\lambda_{N-1}]$.
The Fourier basis consists of $N$ elements, with the element at index $k$ given by $[1, \omega^k, \omega^{2k},...,\omega^{(N-1)k}]$.
To calculate the Fourier transform of the given sequence, we take its inner product with each element of the Fourier basis (Fig.~\ref{fig:31}).
The result is a list of complex numbers that indicate how similar the given sequence is to elements of the Fourier basis.

\begin{figure}[htb]
\includegraphics[width=15cm]{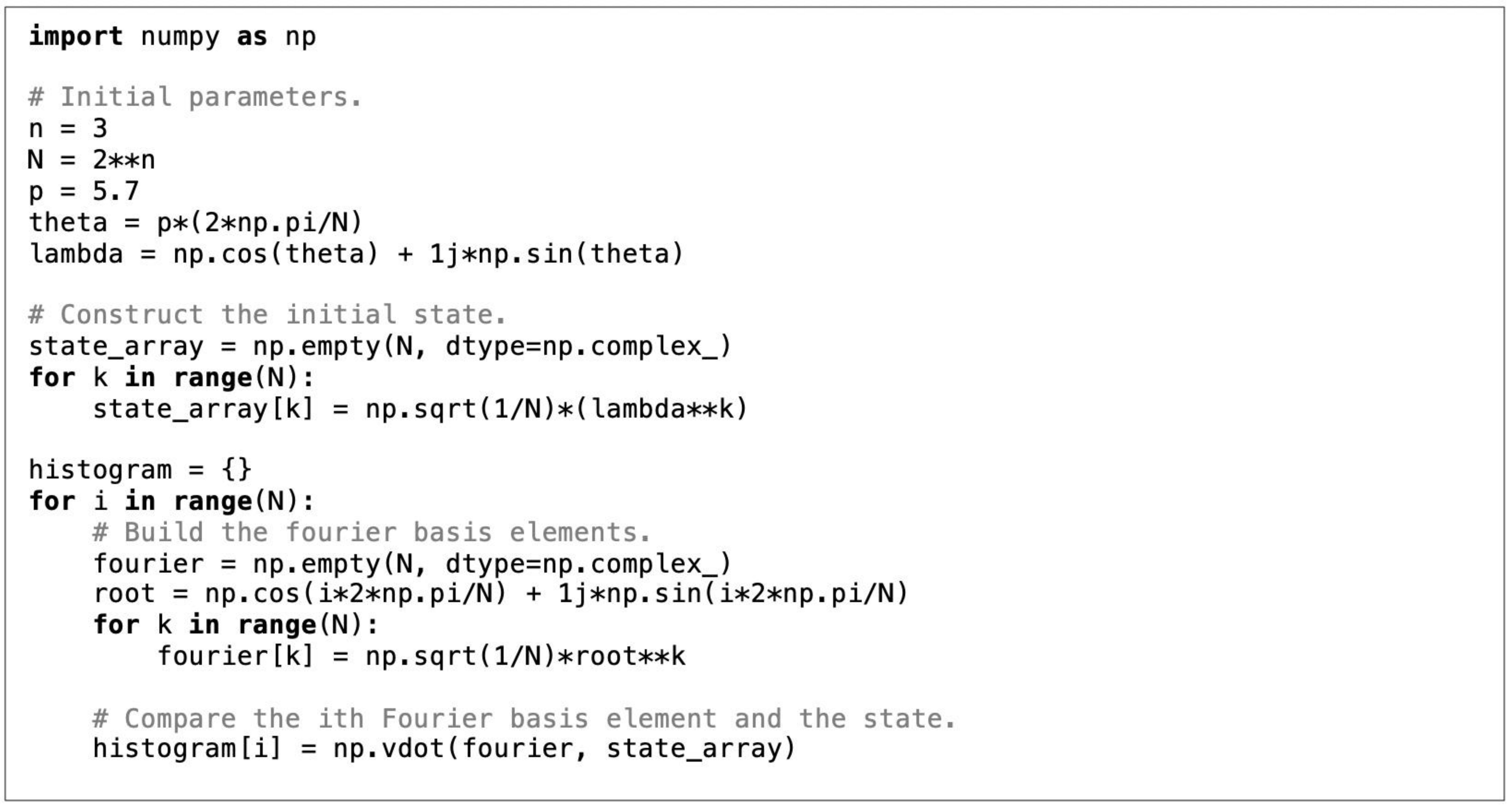}
\caption{\label{fig:32}A Python snippet that compares a complex parameter to the roots of unity using their magnitudes.}
\end{figure} 

\begin{figure}[htb]
\includegraphics[width=2cm]{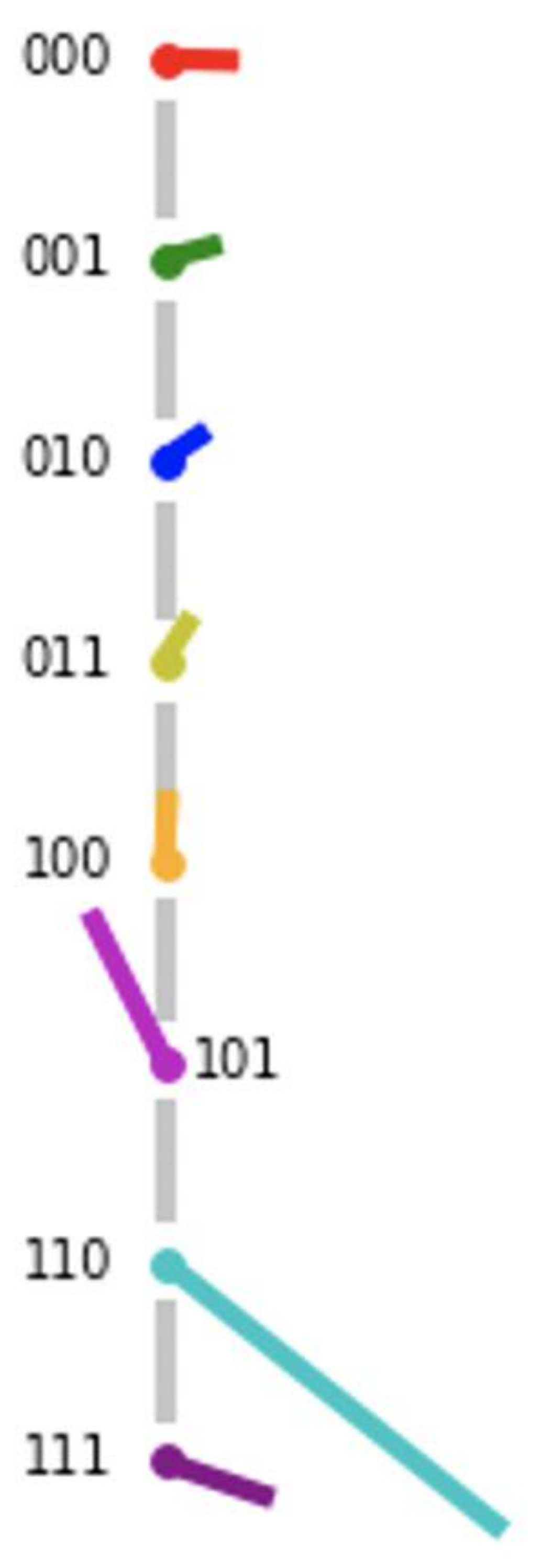}
\caption{\label{fig:33}The visual representation of the histogram variable in the code fragment above.}
\end{figure} 

\begin{figure}[htb]
\includegraphics[width=15cm]{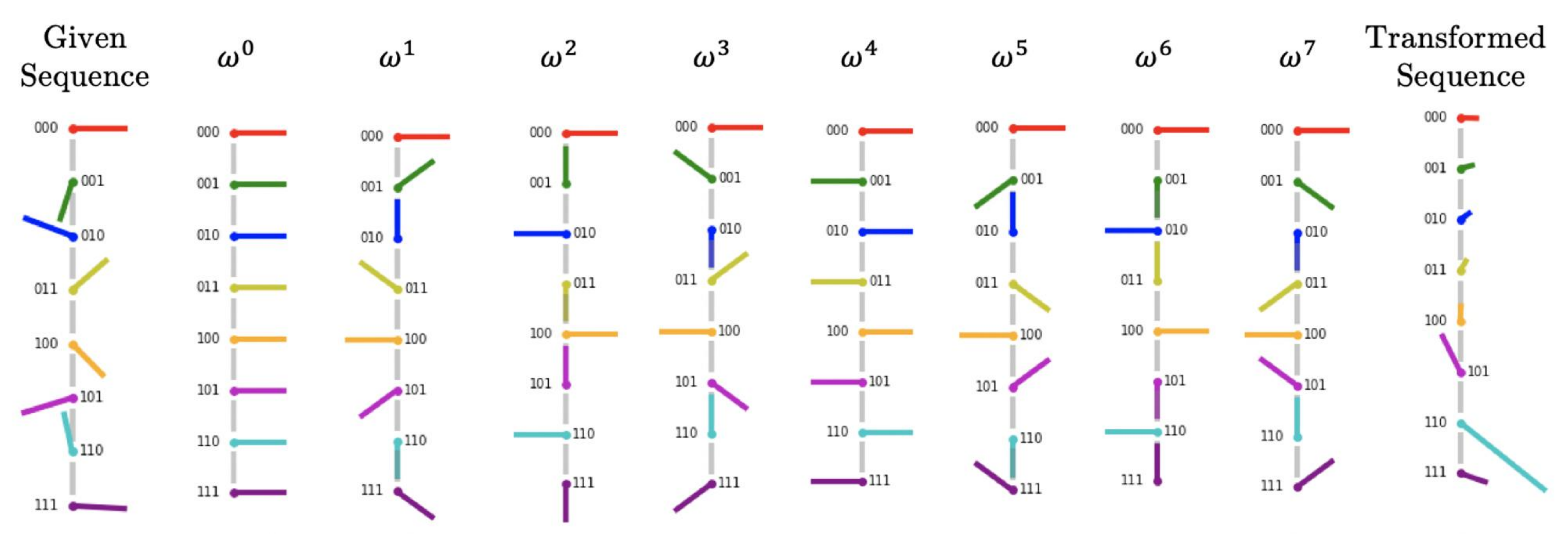}
\caption{\label{fig:34}A sequence of eight complex numbers, the Fourier basis elements, and the final state (left to right).}
\end{figure}

Note that the Fourier basis element at index $k$ consists of the geometric sequence of the $k$th power of $\omega$ ($\omega^k$).
Given a complex number $\lambda$ we can form the geometric sequence consisting of its powers $[1,\lambda,\lambda^{2},...,\lambda^{N-1}]$ (Fig.~\ref{fig:34}).
When we apply the Fourier transform to this sequence, the magnitude of each complex number in the resulting sequence correlates to the similarity of $\lambda$ with the corresponding power of $\omega^k$. 

\begin{figure}[htb]
\includegraphics[width=10cm]{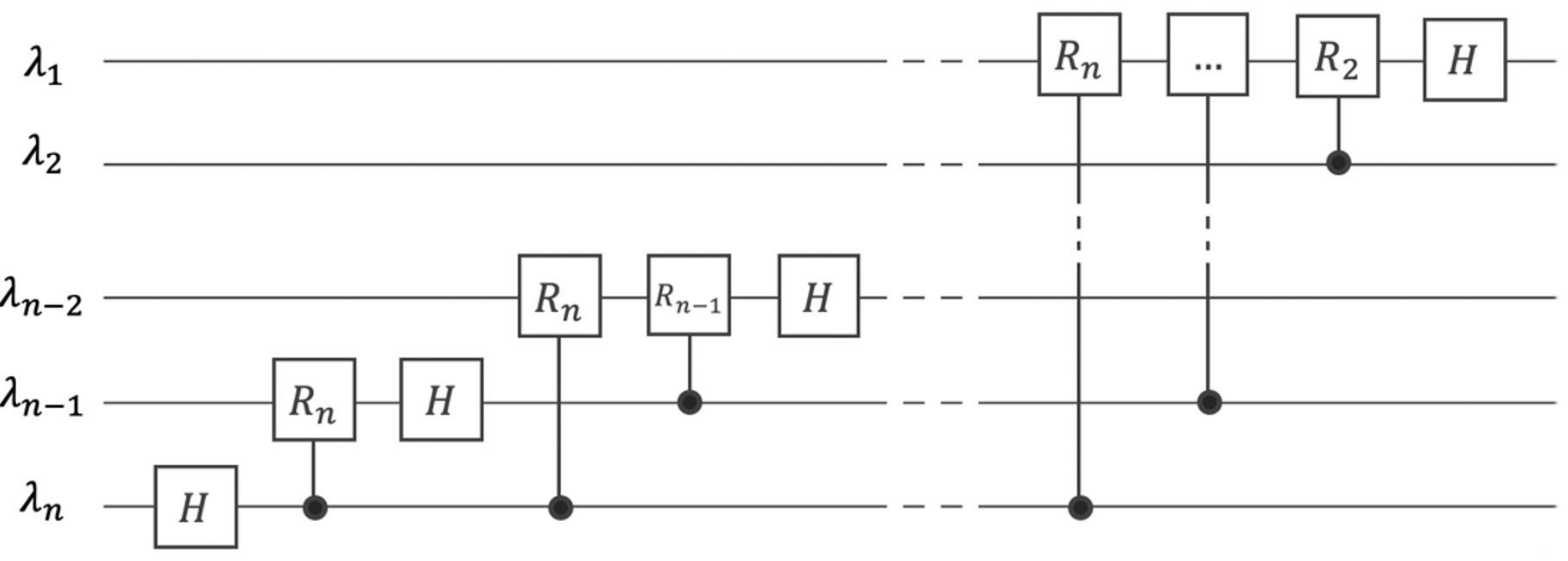}
\caption{\label{fig:35}A logical view of a circuit that applies the inverse $QFT$.}
\end{figure}

This classical computation has a quantum counterpart, called the Quantum Fourier Transform ($QFT$)~\cite{Shor1997}.
It is logically identical to the magnitude-based classical Fourier transform implemented in Fig.~\ref{fig:33}, but the quantum version is more efficient.
Compared to the classical fast Fourier transform that requires an exponential number of operations, $QFT$ requires a maximum of $n^2$ quantum gates, thus reducing the complexity from exponential to polynomial.
The logical circuit is shown in Fig.~\ref{fig:35}.

Note that unlike the classical Fourier transform (which returns all of the similarities) the $QFT$ returns the result with the highest probability---which is also the outcome that is most similar to the given input.
The $QFT$ is rarely useful on its own, and is instead used as a component in multiple efficient quantum algorithms.

\subsection{\label{sec:phase-estimation}Phase Estimation}

\begin{figure}[htb]
\includegraphics[width=4cm]{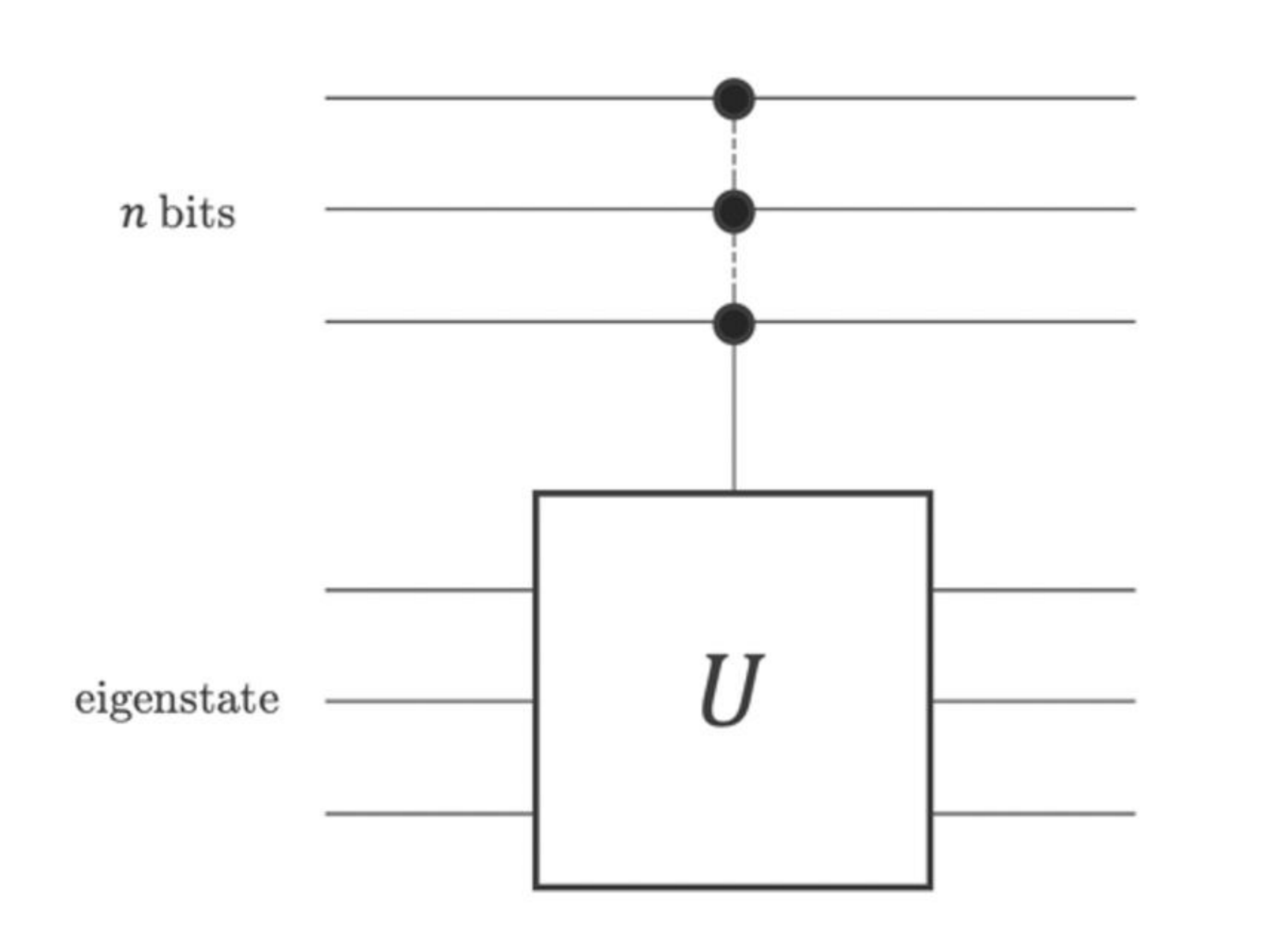}
\caption{\label{fig:36}Controlled-$U$ circuit.}
\end{figure}

Suppose we have an operator $U$ that multiplies a particular state  by a complex factor $\lambda=\cos\theta+i\sin\theta$, with $0 \le \theta < 2\pi$.
The angle $\theta$ can be expressed as a multiple of $\frac{2\pi}{N}$, with the factor $p$ being a real value with $0 \le p < N$.
Using an operator $U$ (Fig.~\ref{fig:36}), we can prepare the state $[1,\lambda,\lambda^{2},...,\lambda^{N-1}]$ (a geometric sequence) by applying $U$ $k$-times for each $k \in \{0,1,...,2^N-1\}$.

\begin{figure}[htb]
\includegraphics[width=6cm]{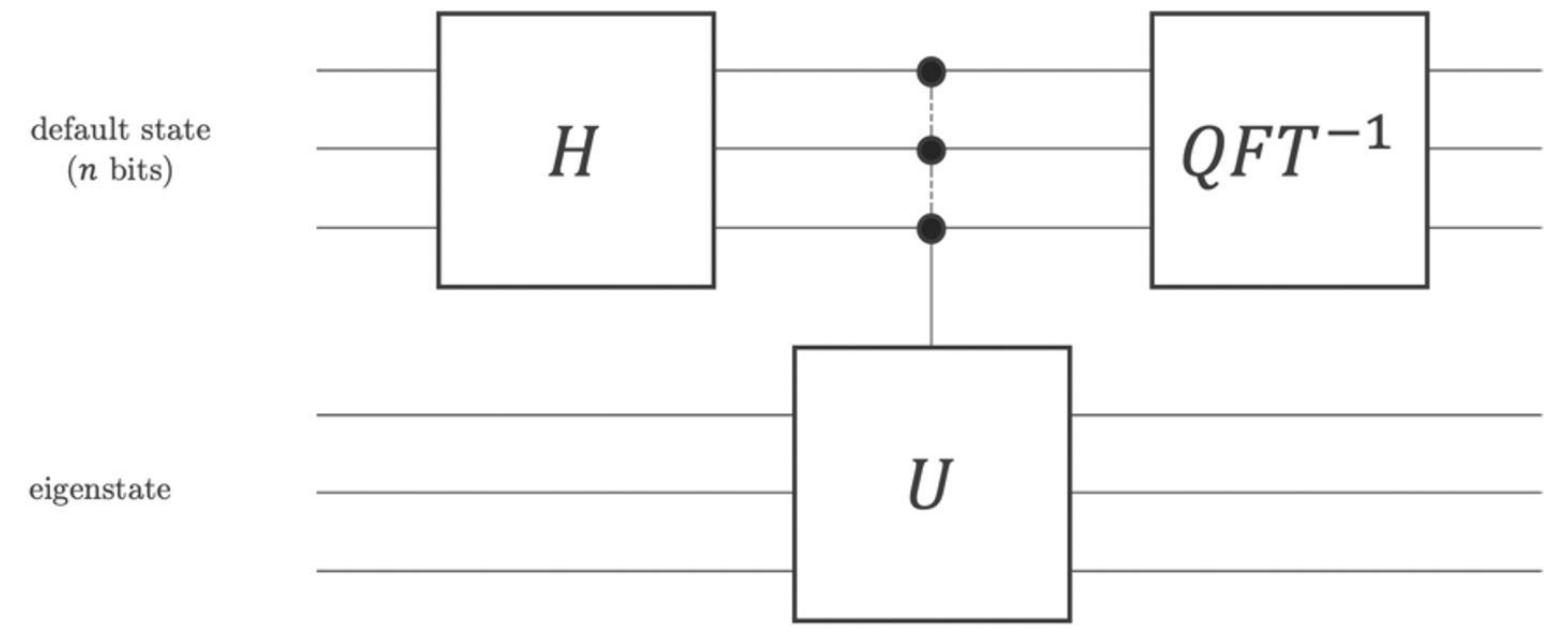}
\caption{\label{fig:37}A logical view of the Phase Estimation circuit.}
\end{figure}

The Phase Estimation algorithm uses two registers of qubits, the first of which is used for encoding the result, and the second the operator $U$ is applied to.
If the first register consists of $n$ qubits, then we have $N=2^n$ possible measurement outcomes of the first register.
The algorithm has two steps (Fig.~\ref{fig:37}):
\begin{enumerate}
    \item In the first register, prepare a state consisting of the geometric sequence of $\lambda$ as amplitudes as described above, where index $k$ is encoded by the superposition of the first register. Note that it is important to start with an initial state in the second register on which $U$ acts as a multiplication by $\lambda$.
    \item Apply inverse $QFT$ to the first register. The effect is that the magnitude of each possible output, and therefore its probability, will correlate to how close the corresponding output is to the factor $p$.
\end{enumerate}

\begin{figure}[htb]
\includegraphics[width=12cm]{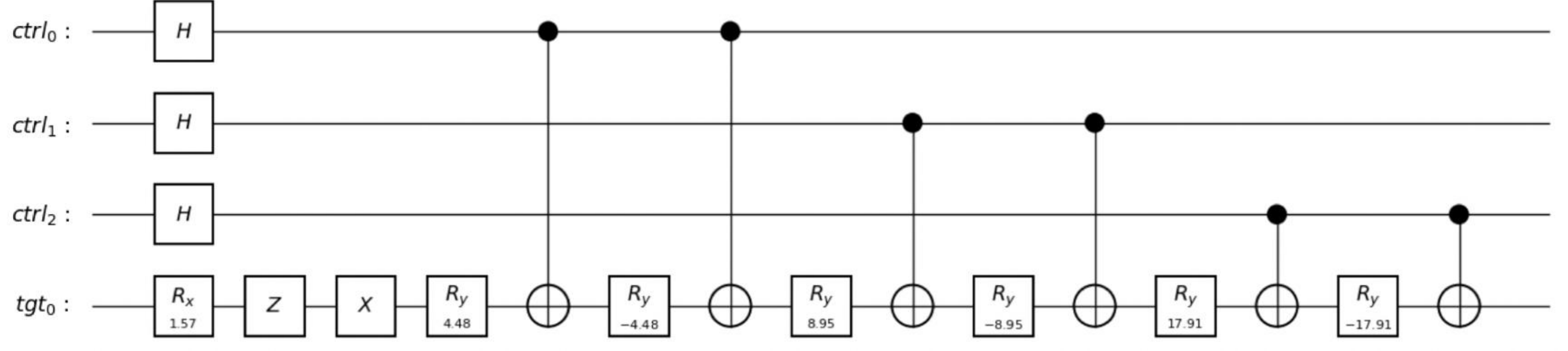}
\caption{\label{fig:38}A quantum circuit using $3$ qubits and an ancilla to prepare the sequence $[1,\lambda,\lambda^{2},...,\lambda^{N-1}]$ with $p=5.7$}
\end{figure}

\begin{figure}[htb]
\includegraphics[width=10cm]{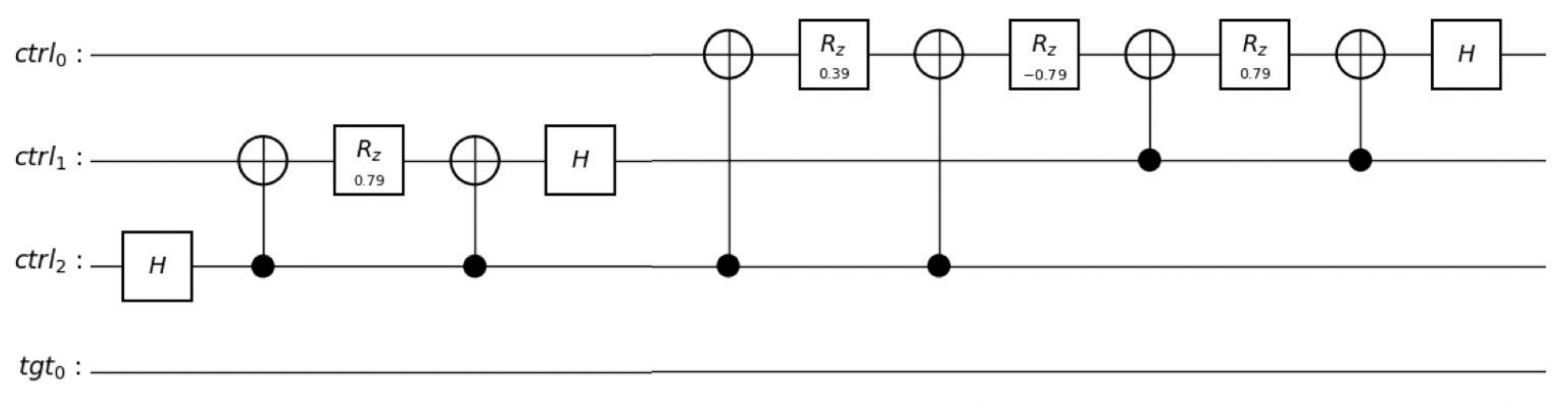}
\caption{\label{fig:39}A quantum circuit that applies the inverse $QFT$ after the circuit in Fig.~\ref{fig:38}.}
\end{figure}

As an example, Fig.~\ref{fig:38} and~\ref{fig:39} show the two steps of the algorithm (respectively) for $U=R_Y(2\theta)$, a single-qubit operator that acts as multiplication by $\lambda=\cos\theta+i\sin\theta$ on the initial state with amplitudes $a_0=\frac{i}{\sqrt{2}}$ and $a_1=\frac{1}{\sqrt{2}}$.

The name of the Phase Estimation algorithm points to estimating $\theta$ - the angle (or phase) of the eigenvalue of an operator.
In fact, we are estimating the factor $p$, and deriving $\theta$ from it.

\subsubsection{\label{subsec:example-estimating-unknown bias}Example – Estimating an Unknown Bias}

\begin{figure}[htb]
\includegraphics[width=6cm]{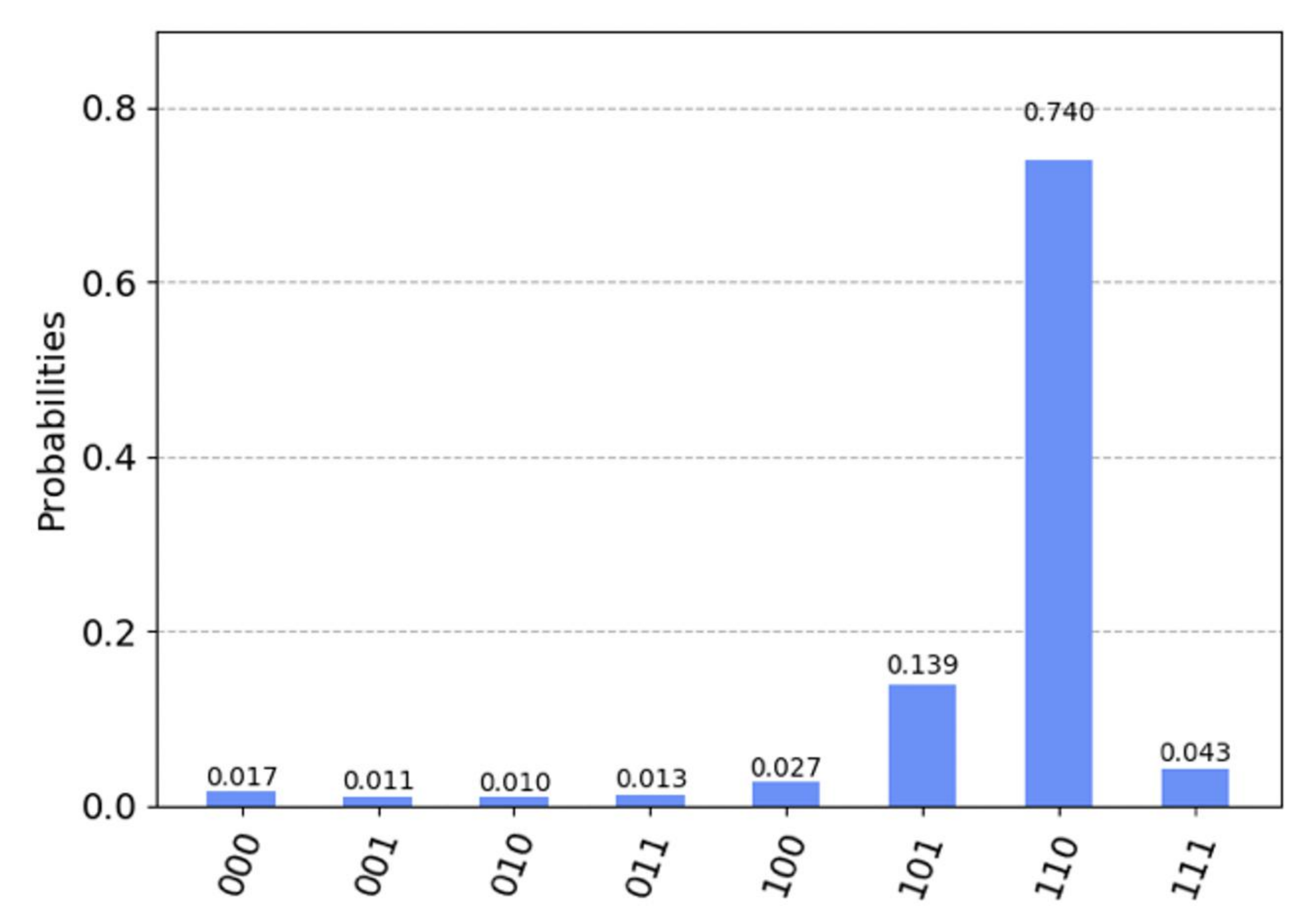}
\caption{\label{fig:40}The probabilities for each output in the range $[0,7]$ when the hidden factor is $5.7$ for $3$ qubits.}
\end{figure}

In Sec.~\ref{subsec:fair-results-example} we modelled a quantum coin using a single-qubit quantum system, and approximated its bias with $1000$ coin flips (shots).
We can estimate this bias using Phase Estimation instead.
Looking back to the definition of $R_Y$ in Sec.~\ref{sec:quantum-gates}, if we start with the default state $a_0=1$ and $a_1=0$, and follow with the application of $R_Y(\theta)$, we get a resulting state with amplitudes $b_0=\cos\frac{\theta}{2}$ and $b_1=\sin\frac{\theta}{2}$.
This means the bias is $\cos^2\frac{\theta}{2}$. 
As we discussed previously, $R_Y(\theta)$ has an eigenvector ($\lambda$) that is directly related to the angle $\theta$ and factor $p$ that the algorithm estimates.
The output that is the best approximation of $p$ will have the highest probability.
The algorithm succeeds with combined probability of the two integers (below and above the factor value) of at least $\frac{8}{\pi^2}$ (approximately $81\%$) with $1$ shot~\cite{Nielsen2011}.
This is sometimes considered a counterpart of the Monte Carlo simulation, as discussed by Woerner et. al.~\cite{Woerner2019Risk} in the context of Amplitude Estimation instead of Phase Estimation.
Amplitude Estimation can be applied to more general operators where it is not easy to identify an eigenvector, as needed by Phase Estimation. 
Since finding the bias of a quantum coin is equivalent to finding the factor $p=\theta\frac{N}{2\pi}$, we will show an example of its estimation.
As an example, in order to estimate $p=5.7$, we can apply the Phase Estimation algorithm using $R_Y(\theta)$---in essence the same example shown in Fig.~\ref{fig:34}.
Looking at the output probability histogram (Fig.~\ref{fig:40}), $5$ and $6$ have the highest probabilities.

\begin{figure}[htb]
\includegraphics[width=6cm]{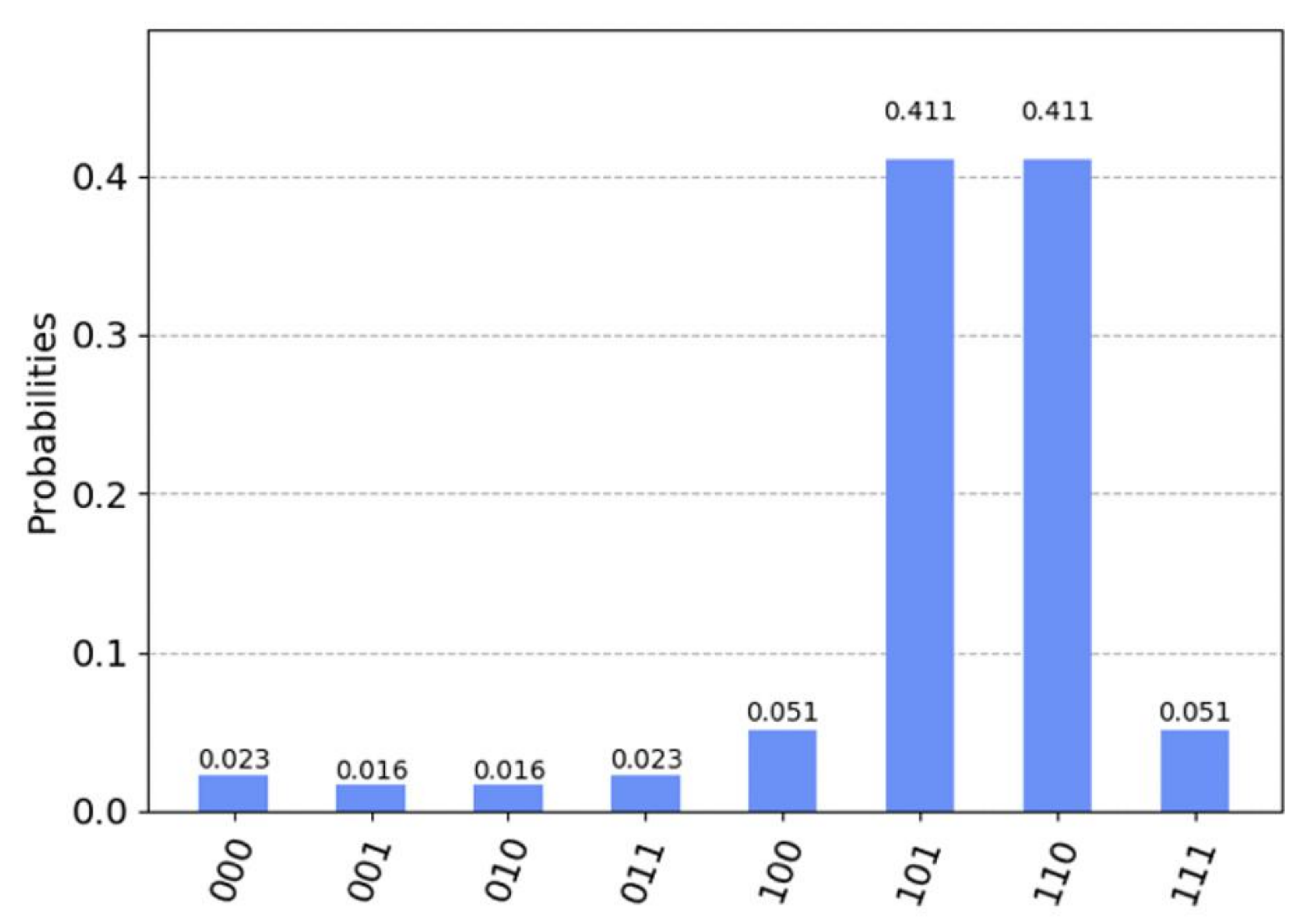}
\caption{\label{fig:41}The probabilities for each output in the range $[0,7]$ when the hidden parameter is $5.5$ for $3$ qubits}
\end{figure}

For a value of $5.5$ we get the same probability for $5$ and $6$, as the value of $p$ is halfway between the two measurement outputs (Fig.~\ref{fig:41}).

\begin{figure}[htb]
\includegraphics[width=7cm]{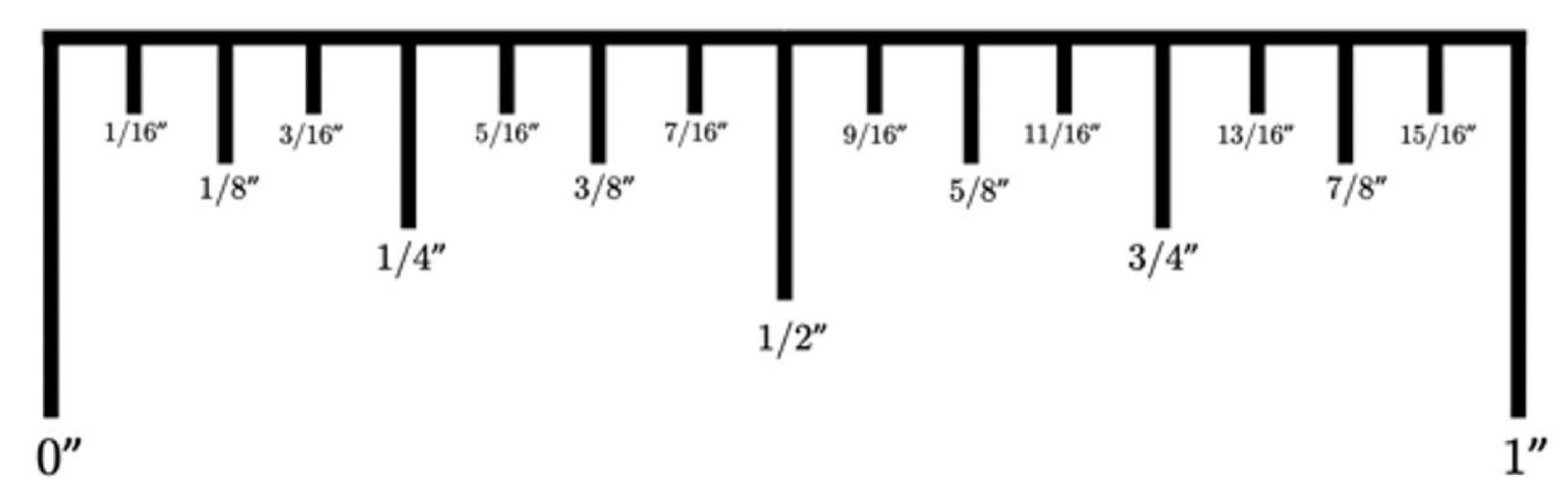}
\caption{\label{fig:42}A tape measure with precision to the 1/16th}
\end{figure}

If the parameter value is equal to one of the possible measurement outputs, e.g. $5$, that output will be measured $100\%$ of the time. 
One or two runs of the computation will be enough to provide an estimate of a given parameter with a desired precision.
This is the essence of efficient algorithms, and takes advantage of the fact that Fourier transforms are efficient on a quantum computer~\cite{Shor1997}.
Adding additional qubits to the first register in the Phase Estimation algorithm will provide a higher precision, but will also increases the cost of the computation.
It may be helpful to think about Phase Estimation as using a tape measure, which has finer precision as you add additional ticks (Fig.~\ref{fig:42}). 

\begin{figure}[htb]
\includegraphics[width=12cm]{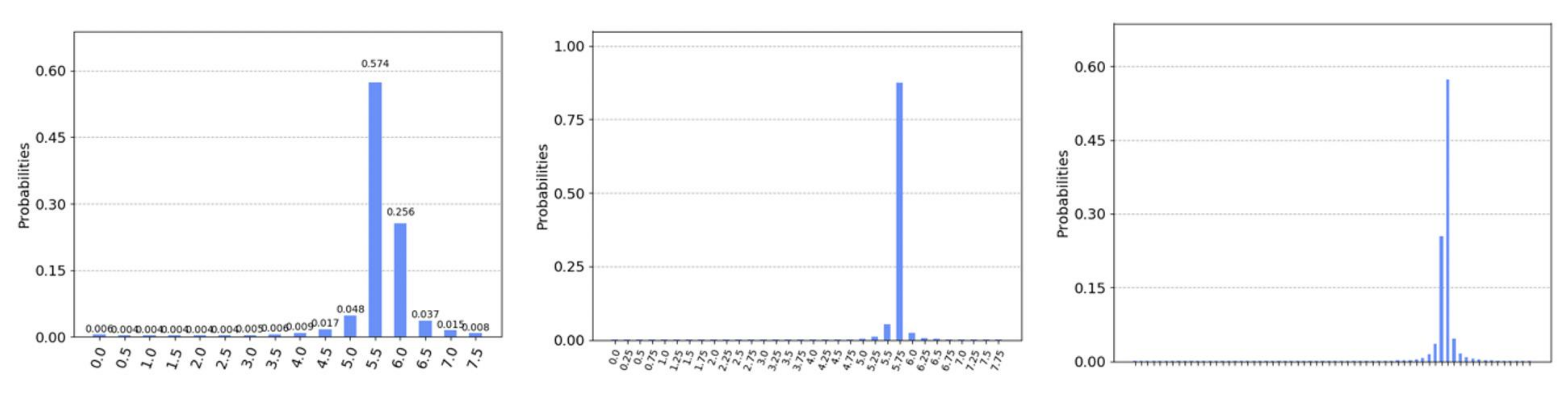}
\caption{\label{fig:43}The results of the Phase Estimation algorithm with $p=5.7$ for four, five, and six qubits (left to right).}
\end{figure}

As a comparison to Fig.~\ref{fig:40}, we can increase the precision using $4$, $5$, and $6$ qubits (Fig.~\ref{fig:43}). 

\begin{figure}[htb]
\includegraphics[width=12cm]{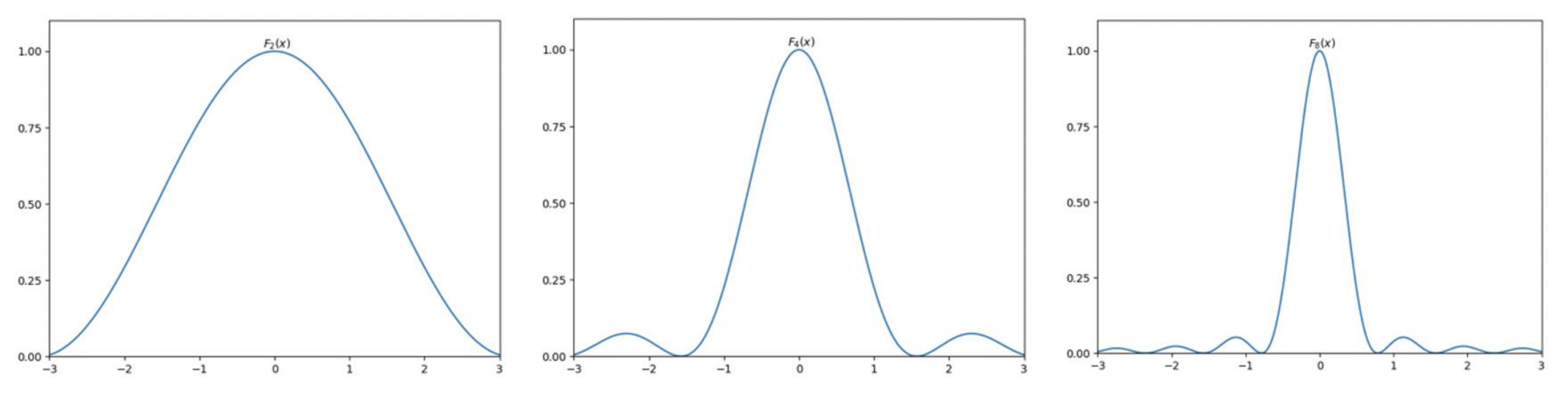}
\caption{\label{fig:44}A plot of the normalized Fejer kernel functions (from left to right) $F_2(x)$, $F_4(x)$, and $F_8(x)$.}
\end{figure}

The probability distributions of the estimates shown in Fig.~\ref{fig:43} turn out to match normalized Fejer kernels~\cite{Hoffman2007} (Fig.~\ref{fig:44}).
Although it is not widely-used in probability theory, this pattern is sometimes called the Fejer distribution~\cite{Janson2010} but this term's usage is rare, and was difficult to find in literature.
There is a closed expression for the probability distribution as the normalized Fejer kernels~\cite{Hoffman2007} seen below.

$$\frac{1}{N}F_N(x)=\frac{1}{N^2}\left(\frac{1-\cos(Nx)}{1-\cos(x)}\right)$$

\subsection{\label{subsec:single-slit-experiment}The Single Slit Experiment}

The following subsection connects Phase Estimation and the single slit experiment in an informal fashion. It is not required knowledge, and can be skipped.

\begin{figure}[htb]
\includegraphics[width=8cm]{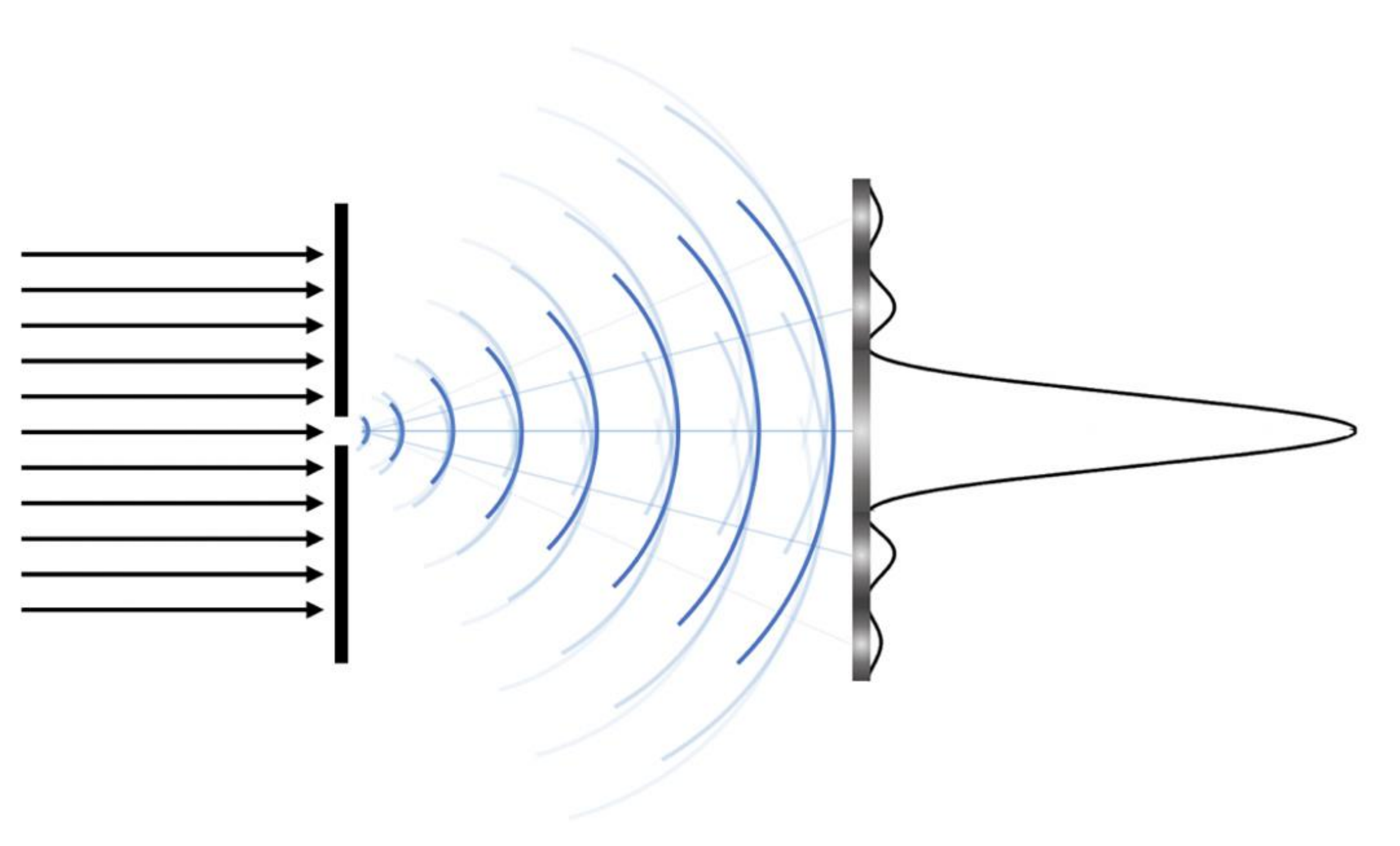}
\caption{\label{fig:45}A visual representation of the single slit experiment and the resulting intensity distribution.}
\end{figure}

The single slit experiment~\cite{Feynman2006}, well-known in the fields of optics and quantum mechanics, describes how light diffracts after passing through a slit of a width that is slightly wider than the wavelength of light. 
The distribution of the intensity of light, as observed on a panel far away from the slit, can be modeled by a squared sine cardinal (sinc) function, as seen in Fig.~\ref{fig:45}.

\begin{figure}[htb]
\includegraphics[width=6cm]{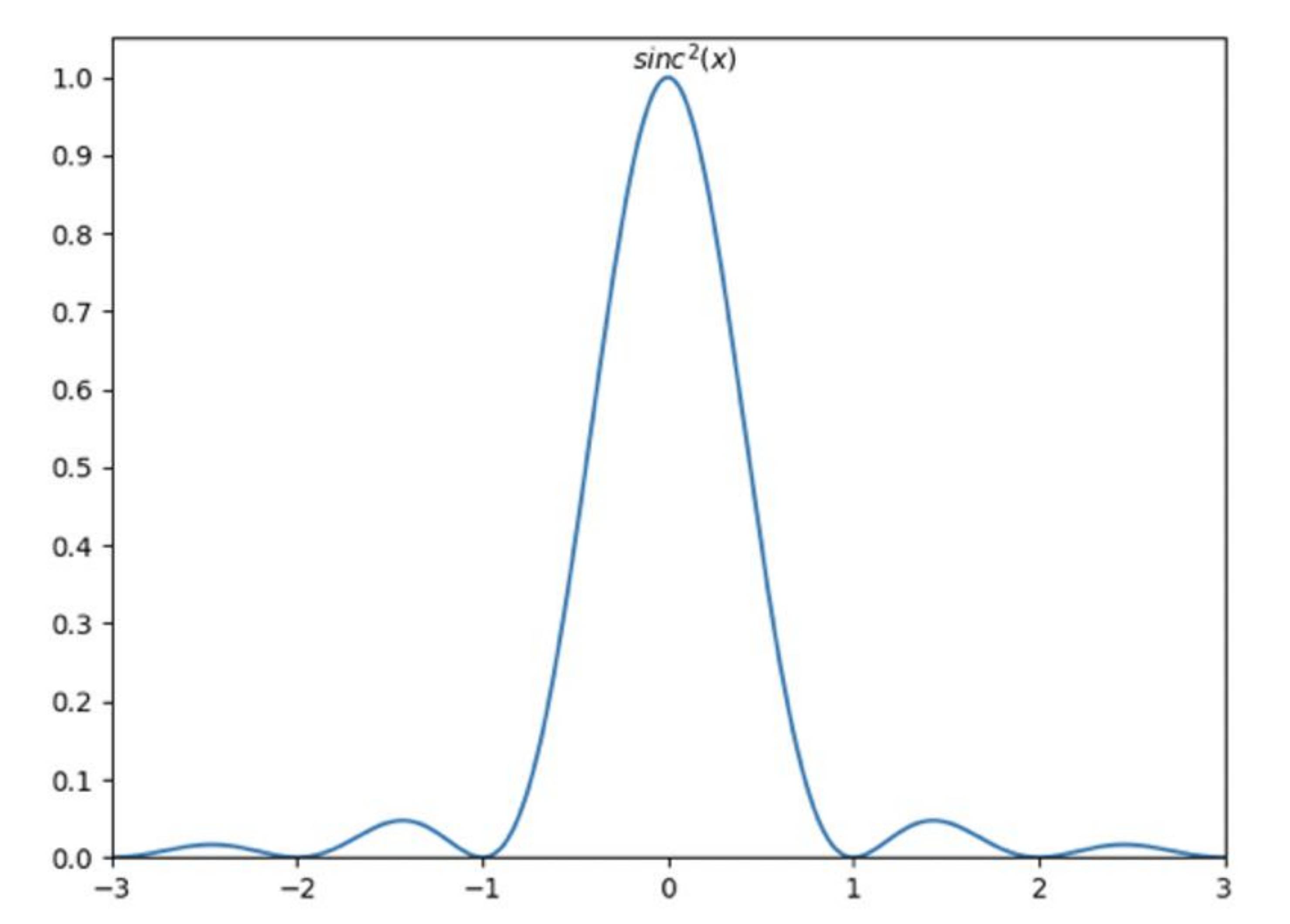}
\caption{\label{fig:46}A plot of a squared sinc function for $x \in [-3,3]$.}
\end{figure}

The pattern of light and dark areas in Fig.~\ref{fig:45} represents the probability that a particle will collide with that position. If this distribution (Fig.~\ref{fig:46}) looks familiar to the figures in Sec.~\ref{sec:classical-approx}, it is because the Phase Estimation algorithm works in a similar fashion to the single slit experiment---the Fejer distribution is the discrete version of the square sinc function.

\subsection{\label{sec:oracles}Oracles}

\begin{figure}[htb]
\includegraphics[width=12cm]{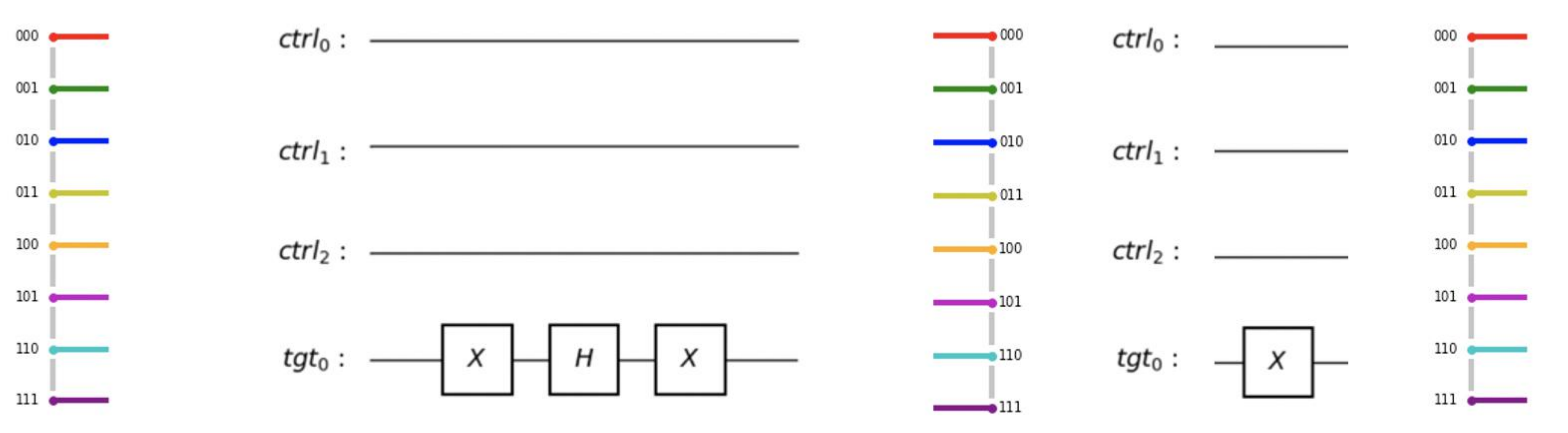}
\caption{\label{fig:47}A circuit that multiplies the state by $-1$ using a specially-prepared state, applied twice for illustration. The target qubit is not shown in the histograms.}
\end{figure}

\begin{figure}[htb]
\includegraphics[width=12cm]{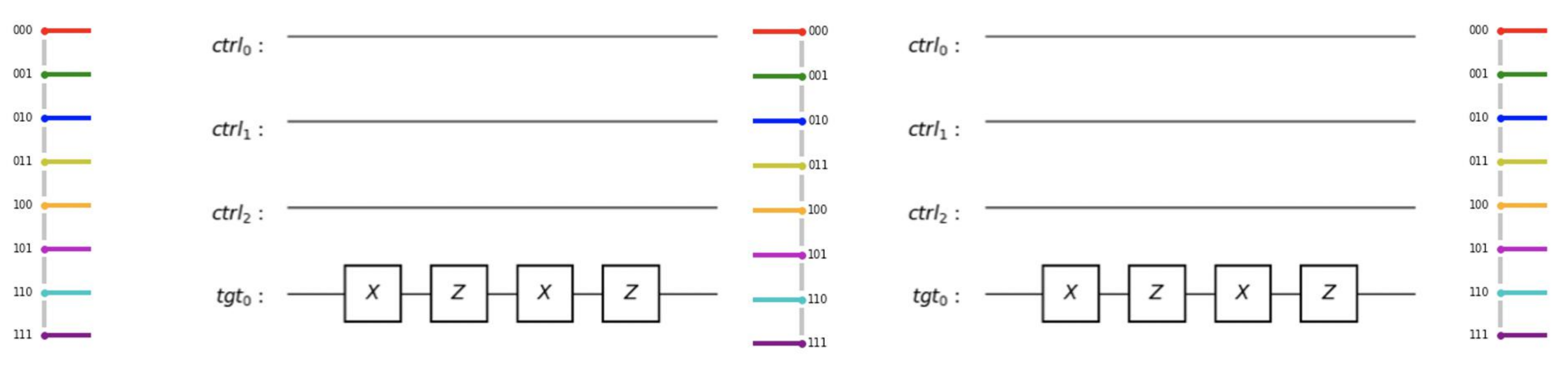}
\caption{\label{fig:48}A circuit using the $ZXZX$ gate sequence, which multiples any state by $-1$, applied twice for illustration. The target qubit is not show in the histograms.}
\end{figure}

We can think of an \textbf{oracle} as a black box with a contract---the oracle applies the contract when it recognizes an element of a specified subset of the basis states.
We call this subset of basis states the \textbf{good states}, and the rest of the states the \textbf{bad states}.
The most common contract is to conditionally multiply the amplitudes of the desired basis states by $-1$.
Many quantum algorithms rely on oracles to be efficient, and use a placeholder oracle in their definition.
Selecting an oracle depends on the nature of the problem that needs to be solved.
An application of an algorithm is only as efficient as the oracle it supplies. 
There are multiple ways to implement the contract of multiplying the amplitudes of the recognized states by $-1$:
\begin{enumerate}
    \item Most texts in quantum computing use a nice trick that efficiently performs the multiplication with a single qubit. Prepare the state of an ancillary qubit with amplitudes $a_0=\frac{1}{\sqrt{2}}$ and $a_0=-\frac{1}{\sqrt{2}}$, using an $HX$ gate sequence on the default state. We then apply an $X$ gate to that ancillary qubit, which flips the phase of control qubits (Fig.~\ref{fig:47}). Note that the trick only works on this specific state.
    \item Recall in Sec.~\ref{sec:quantum-gates} we discussed the $ZXZX$ gate sequence, which multiplies the amplitudes of a state by $-1$ (Fig.~\ref{fig:48}). This works on any state, unlike the trick above, but at the cost of an increased number of gates.
\end{enumerate}
	
\begin{figure}[htb]
\includegraphics[width=8cm]{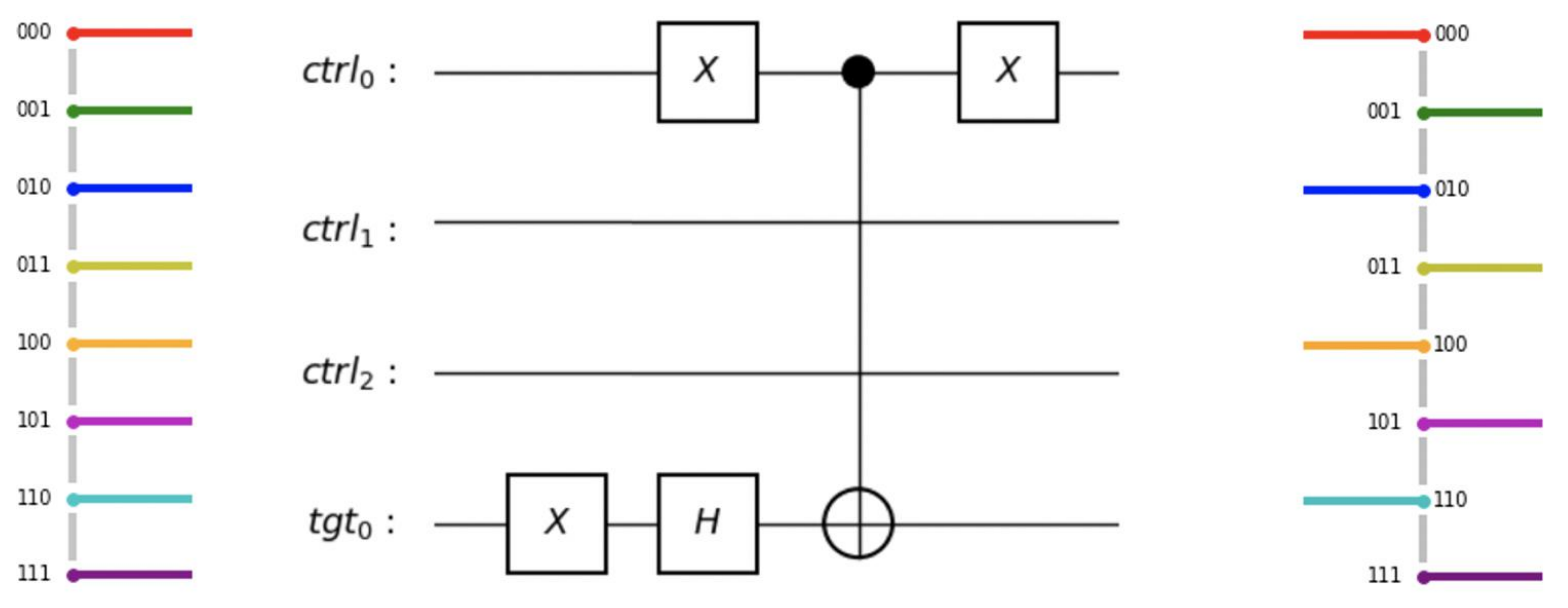}
\caption{\label{fig:49}An oracle that recognizes the even states and multiplies their amplitudes by $-1$ using the single-qubit trick.}
\end{figure}

\begin{figure}[htb]
\includegraphics[width=8cm]{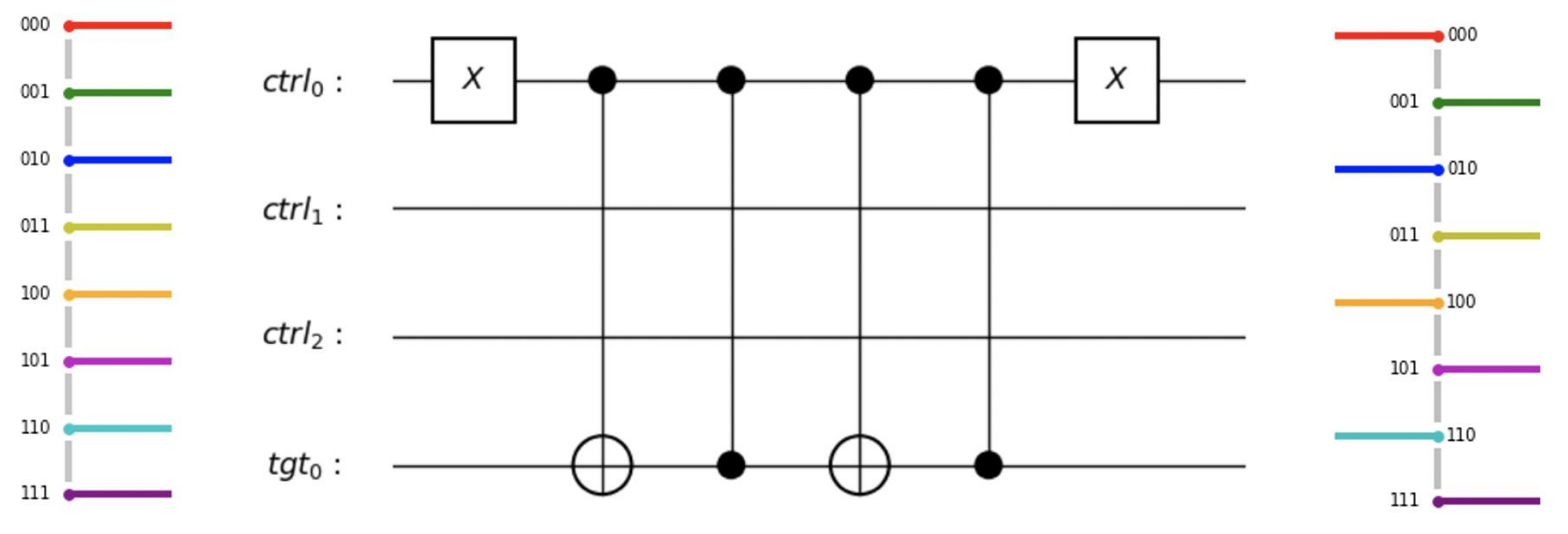}
\caption{\label{fig:50}An oracle that recognizes the even states and multiplies their amplitudes by $-1$ using the $ZXZX$ gate.}
\end{figure}

Let’s take a look at a simple oracle, recognizing the basis states (binary strings) representing even integers.
For a quantum system with three qubits, the oracle should negate the amplitudes of the basis states ending in $0$, e.g. $000$, $010$, $100$, and $110$ (Fig.~\ref{fig:49} and ~\ref{fig:50}).

\begin{figure}[htb]
\includegraphics[width=10cm]{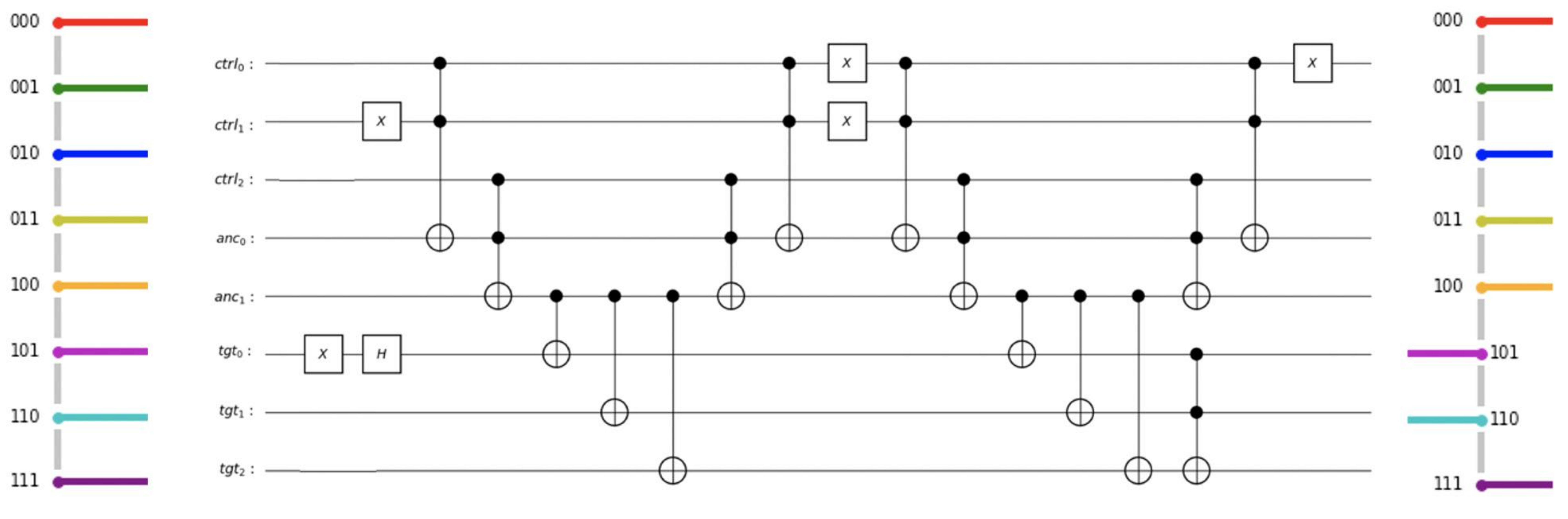}
\caption{\label{fig:51}An oracle that recognizes the states in a set, and multiplies their amplitudes by $-1$ using the single-qubit trick.}
\end{figure}

\begin{figure}[htb]
\includegraphics[width=10cm]{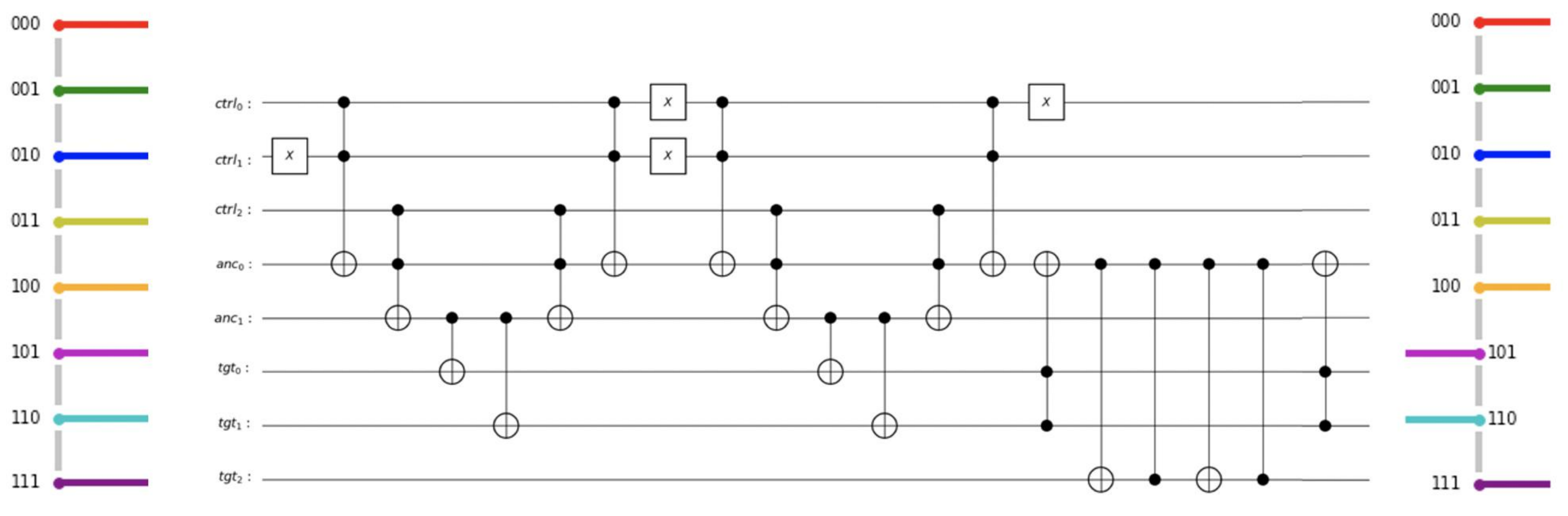}
\caption{\label{fig:52}An oracle that recognizes the states in a given set, and multiplies their amplitudes by $-1$ using the $ZXZX$ gate.}
\end{figure}

Note that we have to apply an $X$ gate to the qubit at index $j=0$.
Recall the control gates are applied when the control qubit is $1$.
In order to match $0$, we apply an $X$ to that qubit.
In the resulting state, the amplitudes of the even outputs have been multiplied by $-1$, and the amplitudes of the odd outputs remain unchanged.
Next, let’s look at an oracle that recognizes a particular set of basis states specified element by element, e.g. $\{101,110\}$.
In this example the oracle must examine all of the basis states, and multiply their amplitudes by $-1$ only if all qubits match an element in the given subset.
We will need to introduce $n-1$ ancillary qubits, used as the controls of the multiplication gate(s) on the target qubit (Fig.~\ref{fig:51} and ~\ref{fig:52}). 

While this is a useful oracle, it is not very efficient, as it requires a large number of controlled gates to match all bits in the recognized basis states. 
Oracles can also be used for satisfiability problems like $3$-SAT~\cite{Karp1972}, in which we create a black box that recognizes the basis states that satisfy a boolean expression.
While we won’t provide an in-depth discussion, there is one in~\cite{Nannicini2017}.
Another interesting oracle is one that recognizes prime numbers, which could potentially be built using Shor’s algorithm~\cite{Shor1997}, beyond the scope of this guide.

\subsection{\label{sec:grover-iteration-and-quantum-search}Grover Iteration \& Quantum Search}

\begin{figure}[htb]
\includegraphics[width=10cm]{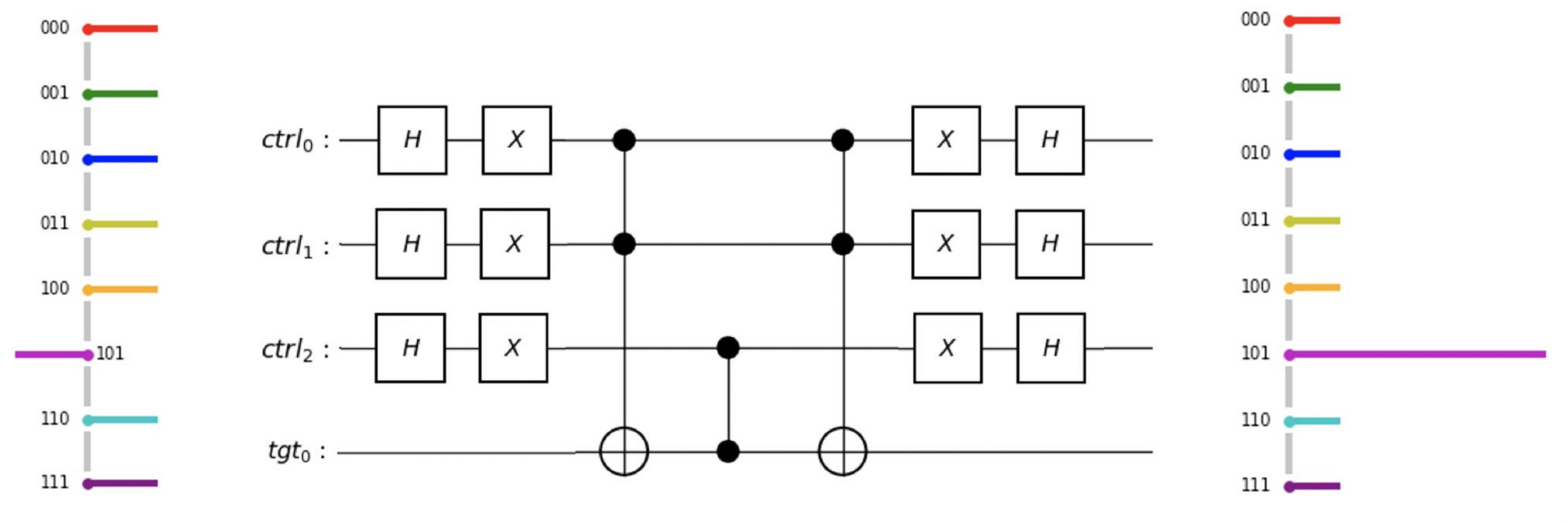}
\caption{\label{fig:53}A three-qubit state after a single pass of the oracle, which is then passed to the diffusion operator (multiplication of amplitudes by $-1$ not shown in the circuit).}
\end{figure}

In an efficient quantum computation we want to increase the probability of some desired outputs, and therefore decrease the number of times we need to repeat the computation.
We have already seen how Phase Estimation can help increase the amplitudes of the outputs that best approximate a parameter.
Grover’s algorithm~\cite{Grover1996} performs amplitude amplification by using an oracle that recognizes a single basis state.
The algorithm uses what is called the Grover iterate---consisting of a sequence of two steps (an oracle $O$ and the diffusion operator)---which is applied a specific number of times.
We discussed oracles in the previous subsection.
The diffusion operator $D$ has the net effect of inverting all amplitudes in the quantum state about their mean.
This causes all the amplitudes of the good states to be scaled by at least $\frac{1}{\sqrt{N}}$~\cite{Grover1996}, while the amplitudes of all other outcomes (the bad states) decrease.
If one thinks of the computation as a die, the probability of one of the faces is dramatically increased.
Note that the amplitudes remain real throughout the application of the oracle and the diffusion operator.
As an example, let’s examine a three-qubit quantum state with the desired outcome being $101$, using the set-based oracle described in the previous subsection (Fig.~\ref{fig:53}). 

\begin{figure}[htb]
\includegraphics[width=8cm]{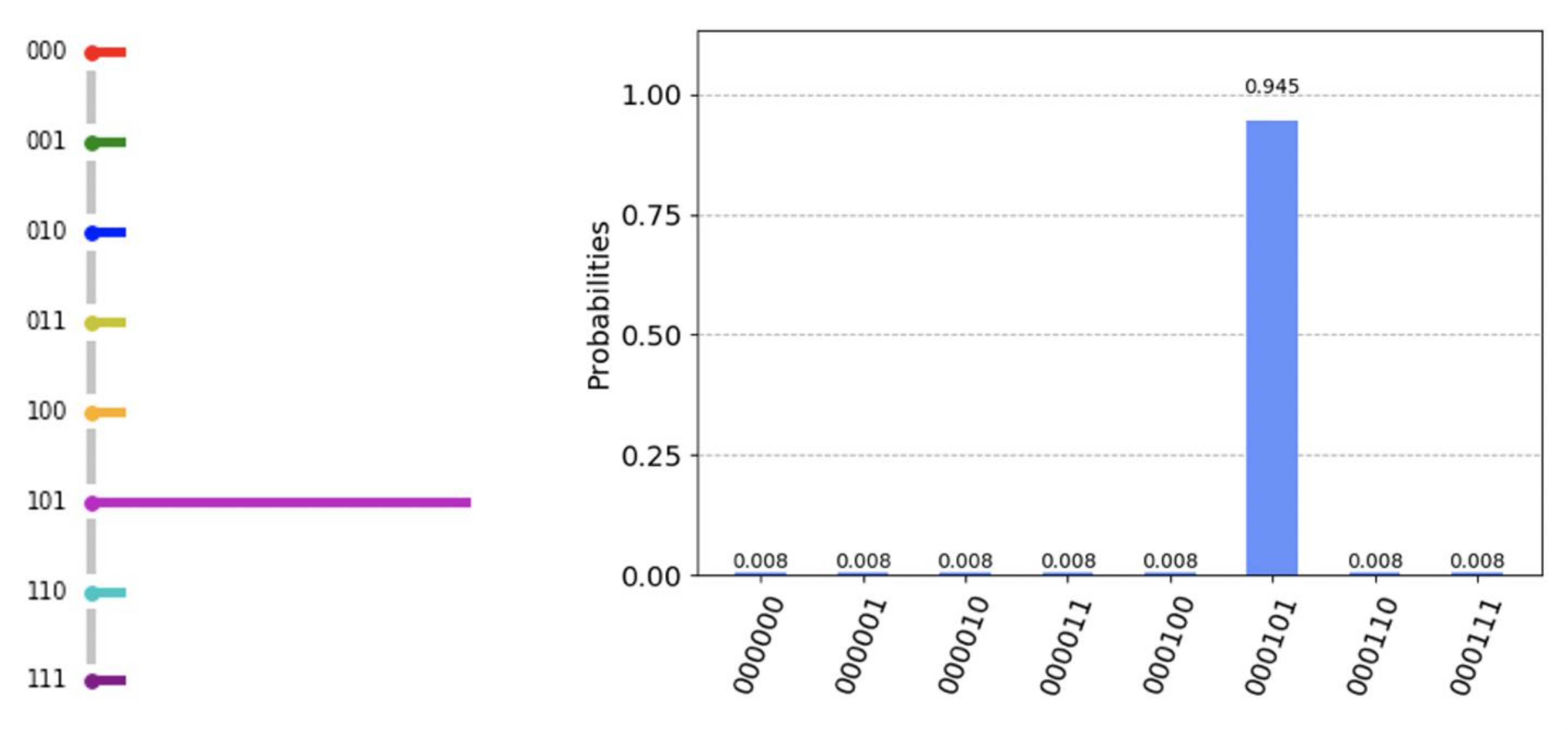}
\caption{\label{fig:54}The result of Grover’s algorithm after $\floor{\sqrt{8}}=2$ iterations of the oracle and diffusion operator.}
\end{figure}

\begin{figure}[htb]
\includegraphics[width=8cm]{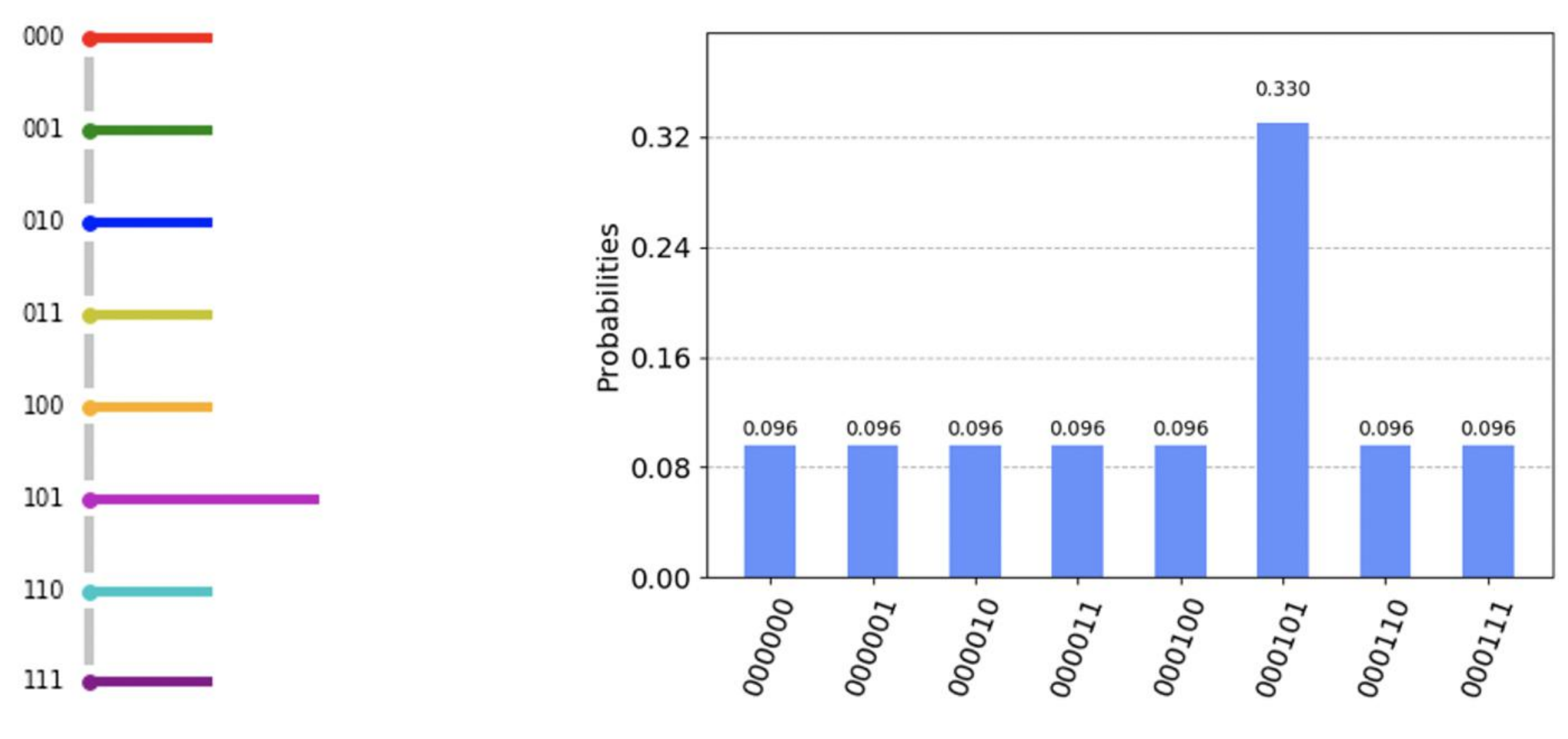}
\caption{\label{fig:55}The result of Grover’s algorithm after $\floor{\sqrt{8}}+1=3$ iterations of the oracle and diffusion operator.}
\end{figure}

We repeat the application of the oracle and the diffusion operator for $\floor{\sqrt{N}}$ iterations (Fig.~\ref{fig:54}).
Note that it is possible to over-iterate, which would decrease the magnitude of the desired output during the application of the diffusion operator (Fig.~\ref{fig:55}).

\subsection{\label{sec:basic-quantum-counting}Basic Quantum Counting}

\begin{figure}[htb]
\includegraphics[width=10cm]{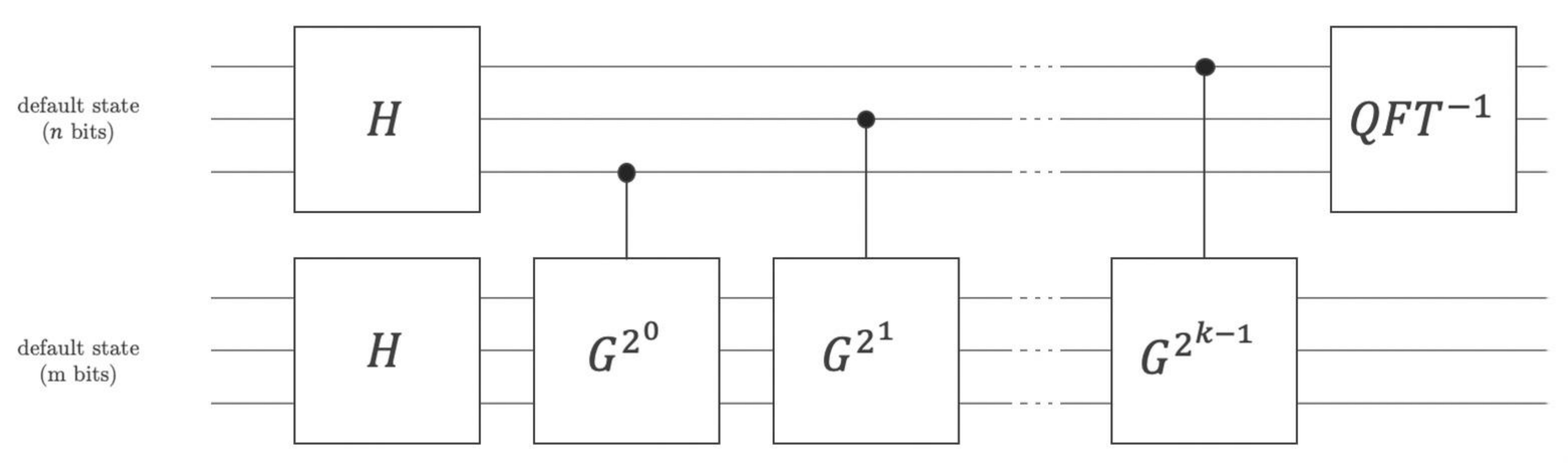}
\caption{\label{fig:56}A logical circuit for basic quantum counting.}
\end{figure}

Basic Quantum Counting uses the Grover iterator and the Phase Estimation algorithm to find how many basis states are recognized by a given oracle $O$, assuming no entanglement has been performed on the quantum state---i.e. how many good states are there?
The algorithm has three steps (Fig.~\ref{fig:56}):
\begin{enumerate}
    \item Initialize a control register with $n$ qubits and a target register with $m$ qubits, and put the basis states in superposition.
    \item Build the Grover iterator $G$, which performs the following operations:
        \begin{enumerate}
            \item Apply the oracle $O$.
	        \item Apply the diffusion operator $D$.
        \end{enumerate}
    \item Plug operator $G$ into the Phase Estimation algorithm, which applies the $G$ operator $2^k$ times to the target register~\cite{Mermin2007}. The most-probable measurement in the control bits will be translated into the count. As mentioned in Sec.~\ref{sec:phase-estimation}, the efficiency of the algorithm depends on how efficient applying powers of the Grover iterator is.
\end{enumerate}

We can think of the Grover iterator as creating a biased quantum die, and Phase Estimation tells us the bias.
Like the example in Sec.~\ref{subsec:entanglement-example}, we are effectively turning the quantum die into a coin – the bad states are seen as one side of the coin, and the good states the other.
The bias is the number of good states divided by the total possible outputs.
The count can be derived from the factor $p$ (the amount of base rotations $\frac{2\pi}{2^n}$) using the equation below.

$$count=2^m\cos^2\left(p\frac{\pi}{2^n}\right)$$

\begin{figure}[htb]
\includegraphics[width=1.25cm]{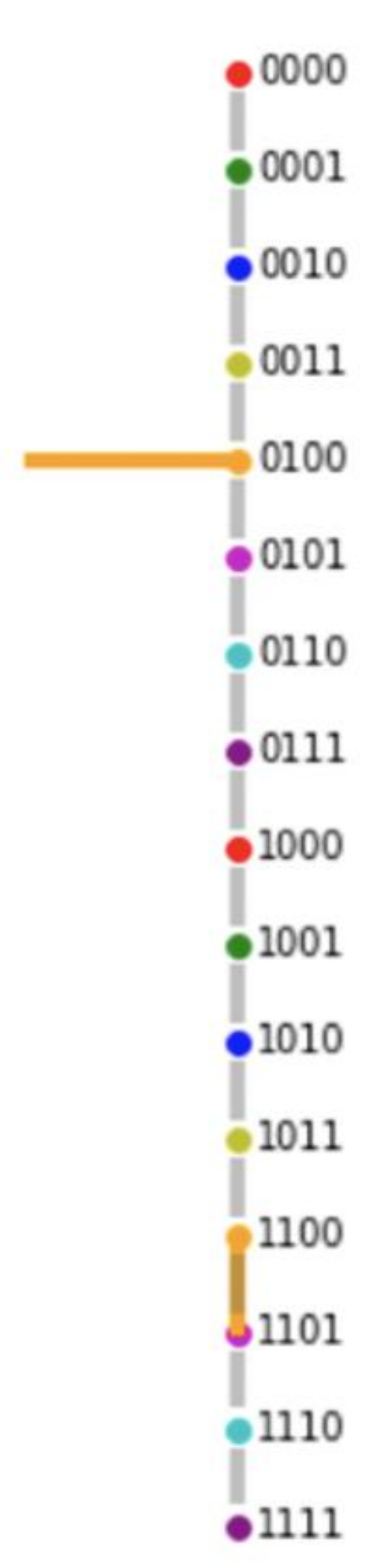}
\caption{\label{fig:57}The result of a quantum counting circuit for an oracle that recognizes even outputs.}
\end{figure}

As an example, let’s count the number of even outputs, using $n=4$ qubits for the control register and $m=3$ qubits for the target register.
Using the oracle from Sec.~\ref{sec:oracles}, we build the the Grover iterator $G$ and perform Phase Estimation.
In Fig.~\ref{fig:57}, the largest magnitudes are $0100$ ($4$) and $1100$ ($12$). 
Both outputs give us the correct result for a three-qubit state, as $\floor{2^3\cos^2\left(4\frac{\pi}{2^4}\right)}=4$ and $\floor{2^3\cos^2\left(12\frac{\pi}{2^4}\right)}=4$.
This is because $\cos^2\left(4\frac{\pi}{2^4}\right)$ and $\cos^2\left(12\frac{\pi}{2^4}\right)$ are multiples of the same angle.
Note that our implementation is slightly different from what is prescribed in~\cite{Brassard1998}, resulting in cosine being used instead of sine in the above equations. 

\begin{figure}[htb]
\includegraphics[width=1cm]{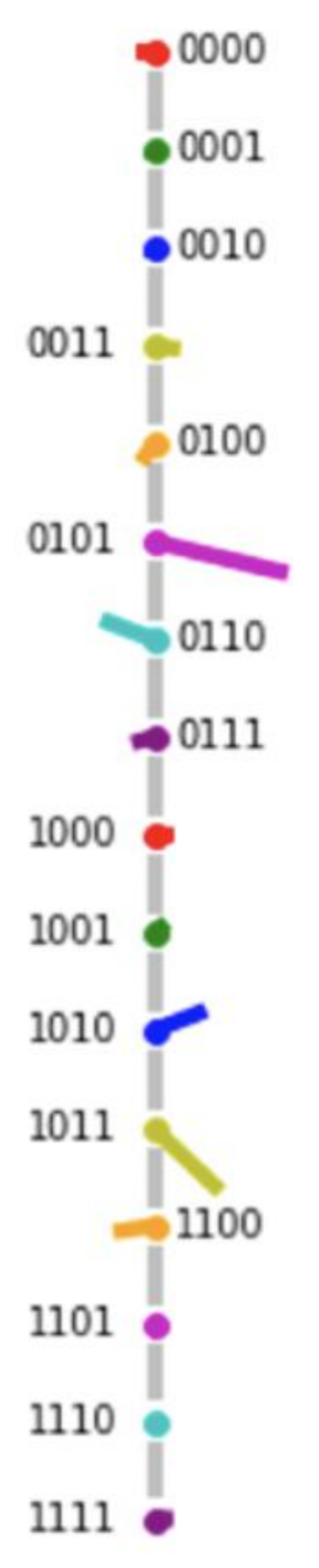}
\caption{\label{fig:58}The result of a quantum counting circuit for a set-based oracle.}
\end{figure}

Let’s look at a different example---counting the number of good states for a set-based oracle for $n=4$ and $m=3$.
We will use the oracle from Sec.~\ref{sec:oracles}, which recognizes $\{101,110\}$, and thus the algorithm should count two good states out of the eight possible.
As seen in Fig.~\ref{fig:58}, the output $0101$ has the largest magnitude.

The output corresponds to $p=5$ which corresponds to the number of good states $\floor{2^3\cos^2\left(5\frac{\pi}{2^4}\right)}=2$.

\subsection{\label{sec:generalized-quantum-computing}Generalized Quantum Counting \& Amplitude Estimation}
 
Generalized Quantum Counting allows entanglement on the quantum state, performed by an operator $A$, but still requires an equal superposition of the possible outputs. 
Amplitude Estimation has the same form of generalized Quantum Counting without assuming equal superposition of the possible outputs, thus giving an estimate of the probability of good states as a whole~\cite{Brassard2000}.
Going forward, we will focus on generalized Quantum Counting.

The algorithm for generalized Quantum Counting consists of three steps:
\begin{enumerate}
    \item Initialize a control register and a target register, and put the basis states in superposition.
    \item (Optional) If useful, prepare an eigenstate of operator $A$ in the qubits the operator is applied to.
    \item Build the Grover iterator $G$, which performs the following operations:
        \begin{enumerate}
            \item Prepare the desired state by applying the operator $A$.
            \item Apply the oracle $O$.
            \item Reverse the state preparation in Step 3a.
            \item Apply the diffusion operator $D$.
        \end{enumerate}
	\item Plug operator $G$ into the Phase Estimation algorithm, as in simple Quantum Counting.
	\item (Optional) Reverse the preparation of the state described in Step 2.
\end{enumerate}

The next section contains more details about the implementation and usage of generalized Quantum Counting.
The use of the Amplitude Estimation algorithm in the context of option pricing is presented in~\cite{Stamatopoulos2019}, including a simple example using $A=R_Y(\theta)$, which implies $G=R_Y(2\theta)$, as explained in the paper. 
Examples of other operators $A$ are used in~\cite{Woerner2019Risk}.
 
\section{\label{sec:quantum-dictionary} Quantum Dictionary}
 
Many problems in quantum computing require the encoding of some logic or constraints into the quantum state.
In a classical computer we do this by using expressions, conditional statements, etc.
In a quantum system we can use entanglement, superposition, and interference to achieve the same goals.
The following pattern uses a combination of algorithms discussed in this paper, and can be applied across large classes of business problems.
 
\subsection{\label{sec:definition}Definition}

Software developers use data structures---such as lists, associative arrays, dictionaries,  time series, etc.---to model problems of interest.
In particular, a dictionary (a mapping of key-value pairs) is a very flexible construct, that is well understood by developers across various languages, and is often used to encode (partial) functions.
We introduce a \textbf{quantum dictionary} as a quantum counterpart to the similar concept in classical computing, and show how it can be used to model problems in quantum computing. 

Two quantum registers---one for keys and one for values---are used to represent the dictionary by taking advantage of quantum entanglement.
In a way, we can think of the entanglement of registers in this context as a substitute for using pointers in classical computers.
If a measurement is performed, the output will contain a key paired with its corresponding value.
Sometimes we are interested in the value corresponding to a key, other times we are interested in the information about the values only. 

The population of the dictionary with the entangled key-value pairs requires a quantum procedure that we call an \textbf{encoding operator}.
It is both an art and a science to define such an efficient operator, which is highly-dependent of the problem at hand.
In this section, we will focus on a class of operators inspired by the techniques used in the Phase Estimation algorithm.
Recall that in Phase Estimation we encode the eigenvalue of an operator into a quantum state, represented by the geometric sequence of that eigenvalue.
The rotation gate $R_Y(\theta)$ for a particular angle $\theta \in [0,2\pi]$ has the nice property that its eigenvectors don’t depend on $\theta$.
One eigenvector of $R_Y(2\theta)$ is $\left(\frac{i}{\sqrt{2}},\frac{1}{\sqrt{2}}\right)$, with corresponding eigenvalue $\cos\theta+i\sin\theta$.
Encoding the eigenvalue implicitly provides an encoding for the angle $\theta$.

\begin{figure}[htb]
\includegraphics[width=5cm]{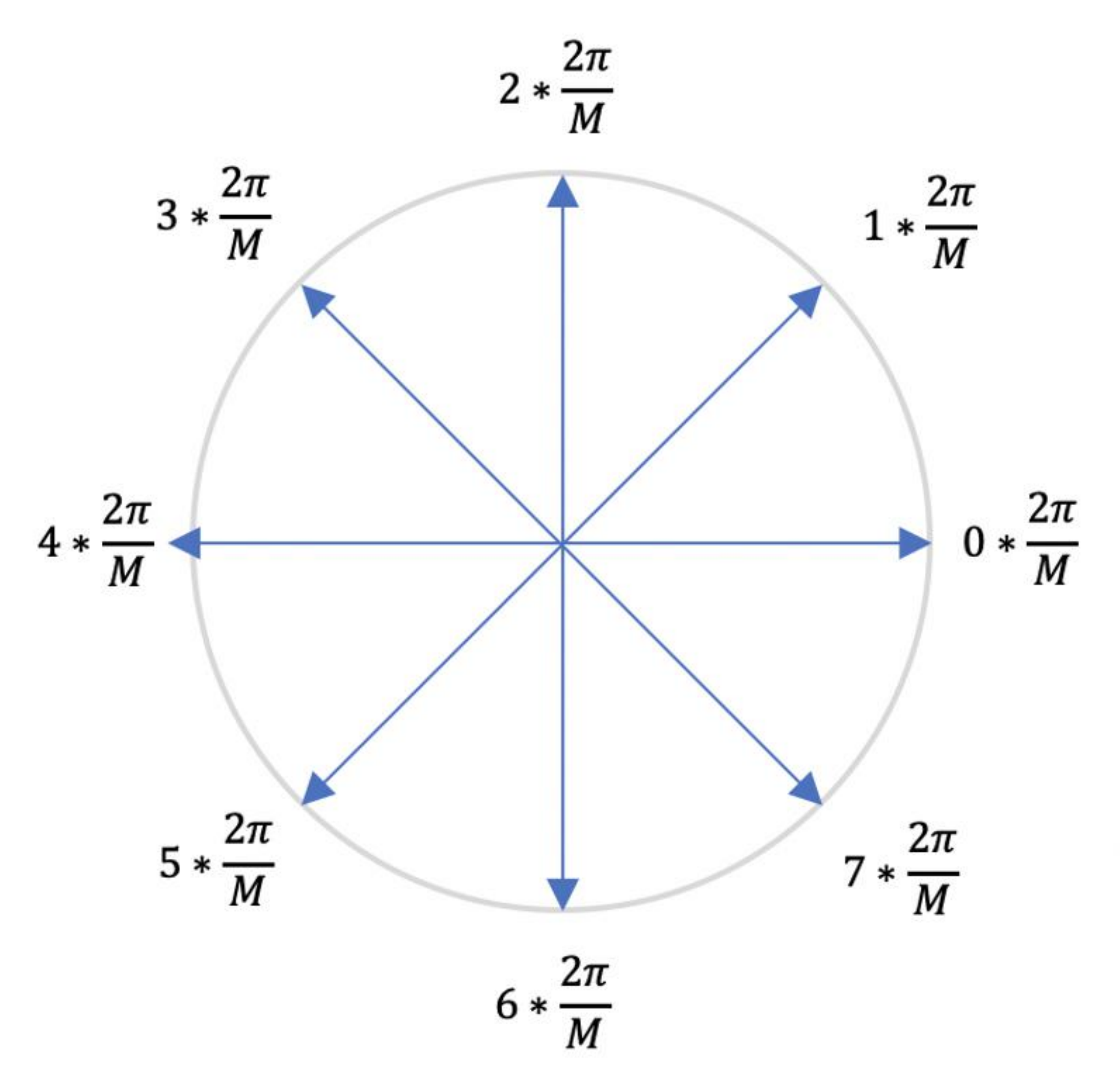}
\caption{\label{fig:59}A unit circle labelled by multiples of a base angle. $M=2^m$, where $m$ is the number of value qubits.}
\end{figure}

We can exploit this idea to encode and manipulate numbers, similar to what we discussed in Sec.~\ref{subsec:example-estimating-unknown bias}, when we estimated a real parameter.
We pointed out that if the parameter is an integer, the estimation is exact.
Building on this idea, we can map numbers to angles, but we can do even better.
We choose a base angle and a number will be mapped to the corresponding multiple of the base angle.
For example, restricting ourselves to integers, $1$ will be represented by the base angle, $5$ by five times the base angle, $-3$ by three times the negative of the base angle, and so on.
This way numbers will be represented as rotations.
For a value register consisting of $m$ qubits, it allows the representation of integers $\{0,1,2,...,M-1\}$ (where $M=2^m$) when the register is measured.
In this context, we can choose $\frac{2\pi}{M}$ as the base angle, and the numbers will correspond to unit vectors, as shown in Fig.~\ref{fig:59}.

\begin{figure}[htb]
\includegraphics[width=7cm]{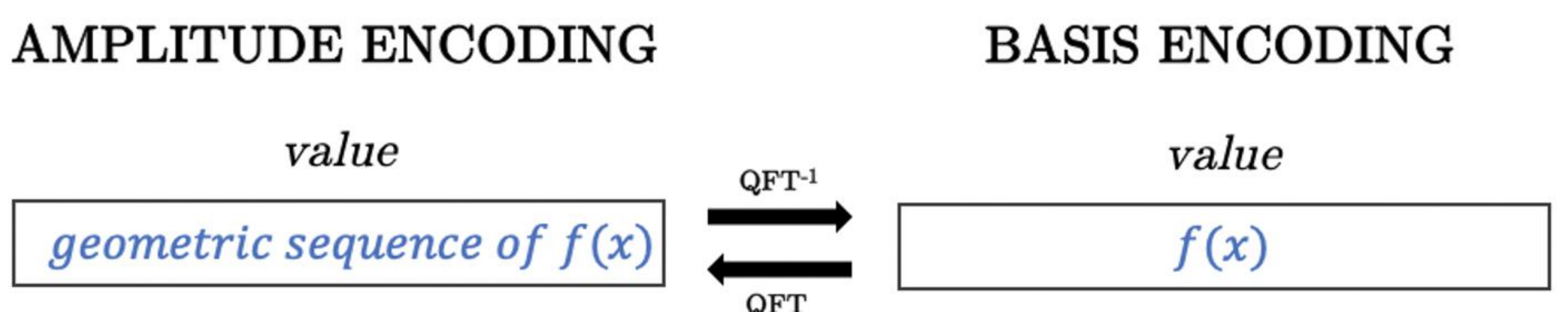}
\caption{\label{fig:60}A visual representation of the transition between amplitude and basis encoding.}
\end{figure}

By convention, counter-clockwise rotations represent positive numbers, and clockwise rotations represent negative numbers.
An integer will fall on a “tick” (which is one of the exact multiples of the base angle) and non-integers will fall between.
Integers are therefore measured precisely, while non-integers are approximated by their closest “tick”.
Once we have the angle representing a number, we can use it to create an \textbf{amplitude encoding} for it, as the geometric sequence of the unit vector for that angle, as prescribed by the Phase Estimation algorithm.
The inverse $QFT$ will convert the amplitude encoding into a \textbf{basis encoding} of the given number. The power of $QFT$ and Phase Estimation comes from the ability to serve as a bridge between basis and amplitude encoding (Fig.~\ref{fig:60}). 

\begin{figure}[htb]
\includegraphics[width=6cm]{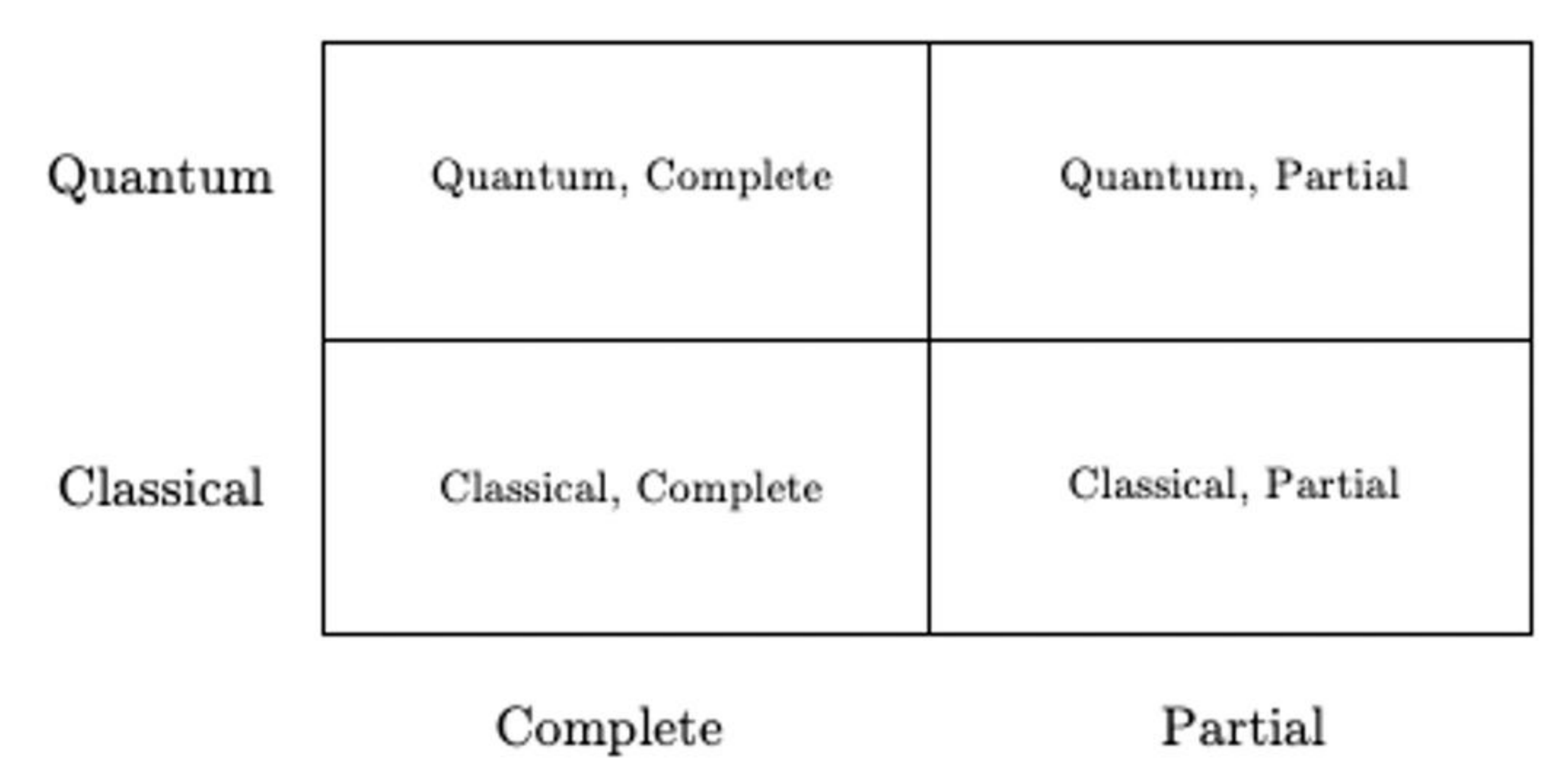}
\caption{\label{fig:61}A visual representation of the different kinds of encoding operators.}
\end{figure}

Four classes of encoding operators are described by the four quadrants shown in Fig.~\ref{fig:61}.
A \textbf{complete encoding} maps every key in the dictionary to a value, similar to a total function in mathematics.
A \textbf{partial encoding} maps a specific subset of keys to corresponding values, and the rest of the values are “filled in” automatically.
A \textbf{classically-defined encoding} uses a classical system to compute values in advance, and a \textbf{quantumly-defined encoding} computes values as part of a quantum computation.

In summary, given a function $f: \{0,1,...2^n-1\}\rightarrow\{0,1,...,2^m-1\}$ for positive integers $n$ and $m$, the steps for populating a quantum dictionary are as follows: 
\begin{enumerate}
    \item Create two registers in the quantum system---a key register with $n$ qubits (with total number of outputs $N=2^n$) and a value register with $m$ qubits (with total number of outputs $M=2^m$). In many cases, it is useful to put these registers in an equal superposition.
    \item (Optional) Prepare an eigenstate of the encoding operator (for example, multiple applications of $R_Y$) in the qubits the operator is applied to.
    \item Apply the encoding operator to entangle keys and values.
    \item Apply the inverse $QFT$ to the value register.
    \item (Optional) Reverse Step 2 to simplify measurements.
\end{enumerate}

\subsection{\label{sec:classical-function-encoding}Classically-Defined Function Encoding}

\begin{figure}[htb]
\includegraphics[width=4cm]{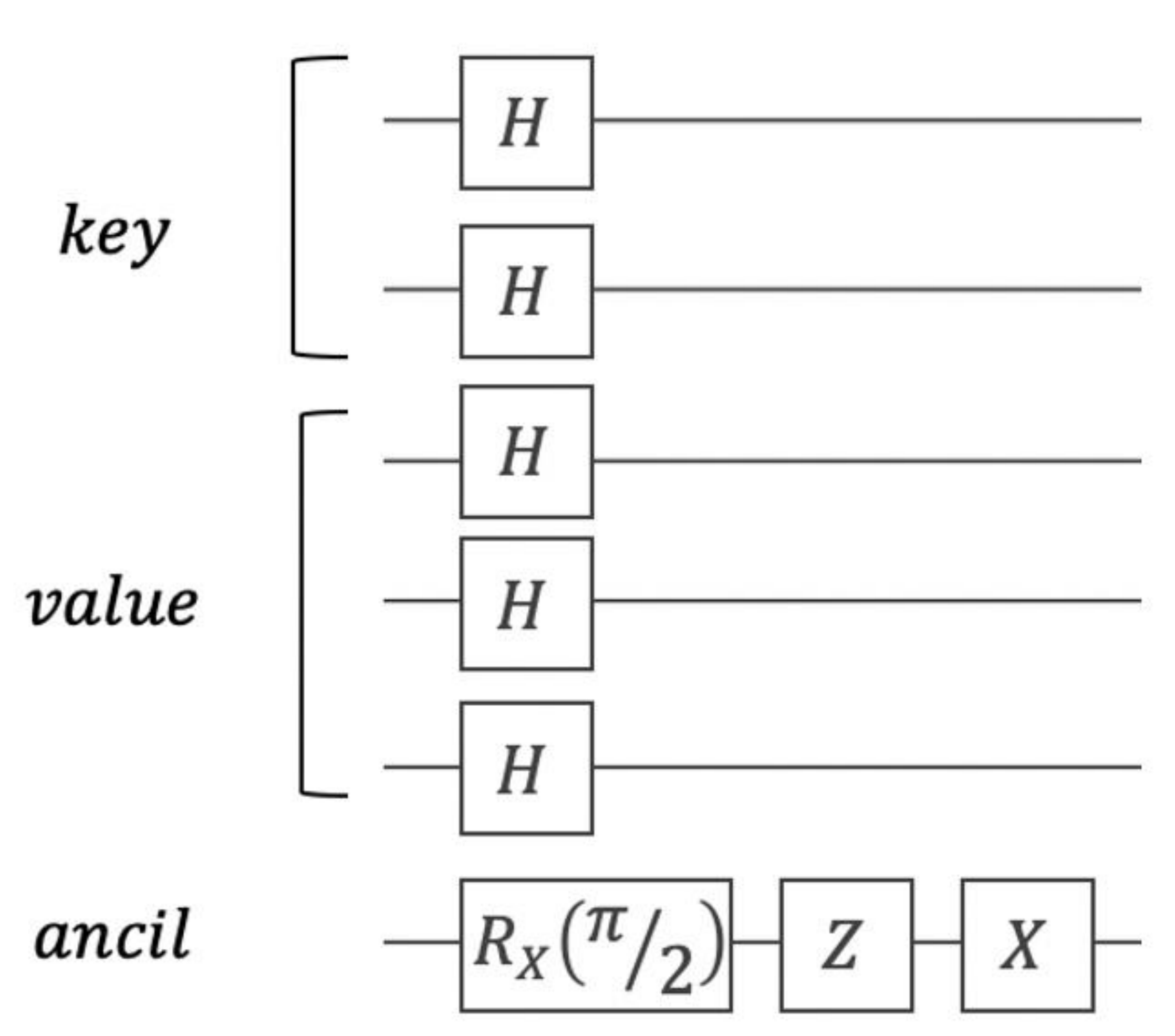}
\caption{\label{fig:62}A quantum circuit preparing an equal superposition for the key and value, and an eigenstate in the ancilla.}
\end{figure}

As an example of a quantum dictionary with a classically-defined encoding operator, let’s encode the function $f:\{0,1,2,3\}\rightarrow\{0,1,2,3,4,5,6,7\}$ defined by:
\begin{equation*}
  f(x) =
  \begin{cases}
    5 & \text{if $x=0$} \\
    3 & \text{if $x=1$} \\
    1 & \text{if $x=2$} \\
    7 & \text{if $x=3$}
  \end{cases}
\end{equation*}
into a quantum state.
In order to do that, we can use a key register with $2$ qubits, a value register with $3$ qubits, and $1$ ancilla qubit.
The function can be represented classically as a list $[5,3,1,7]$.
Steps $1$ and $2$ in the procedure described above are shown in Fig.~\ref{fig:62}.

\begin{figure}[htb]
\includegraphics[width=10cm]{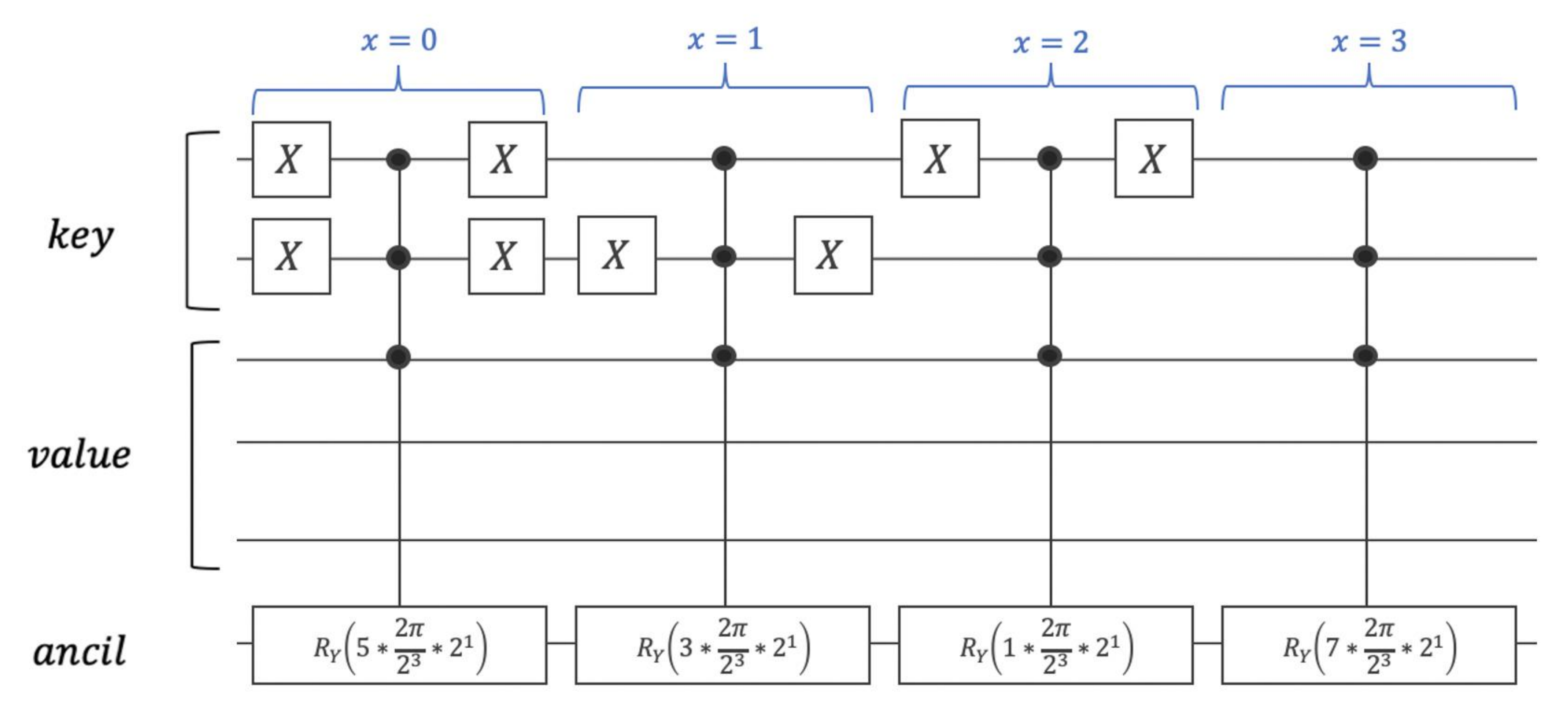}
\caption{\label{fig:63}A circuit representing an encoding operator fragment corresponding to the top value qubit.}
\end{figure}

Step $3$ consists of a sequence of rotations applied to the ancilla and controlled by individual value qubit and groups of key qubits.
Fig.~\ref{fig:63} shows the circuit corresponding to one of the value qubits.

\begin{figure}[htb]
\includegraphics[width=12cm]{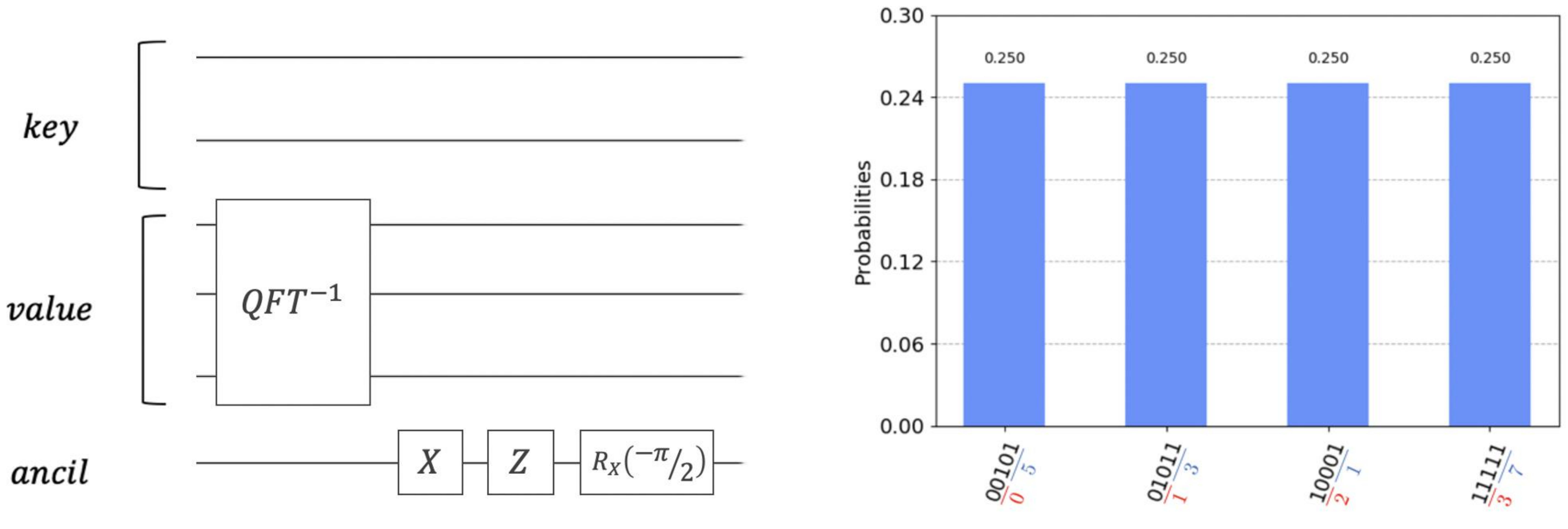}
\caption{\label{fig:64}The circuit and result of applying an inverse $QFT$ gate to the value register after encoding the function.}
\end{figure}

Fig.~\ref{fig:64} shows Steps $4$ and $5$.
There are only four possible outputs, which are the key-value pairs that define the function.

\subsection{\label{sec:quantum-function-encoding}Quantumly-Defined Function Encoding}

\begin{figure}[htb]
\includegraphics[width=10cm]{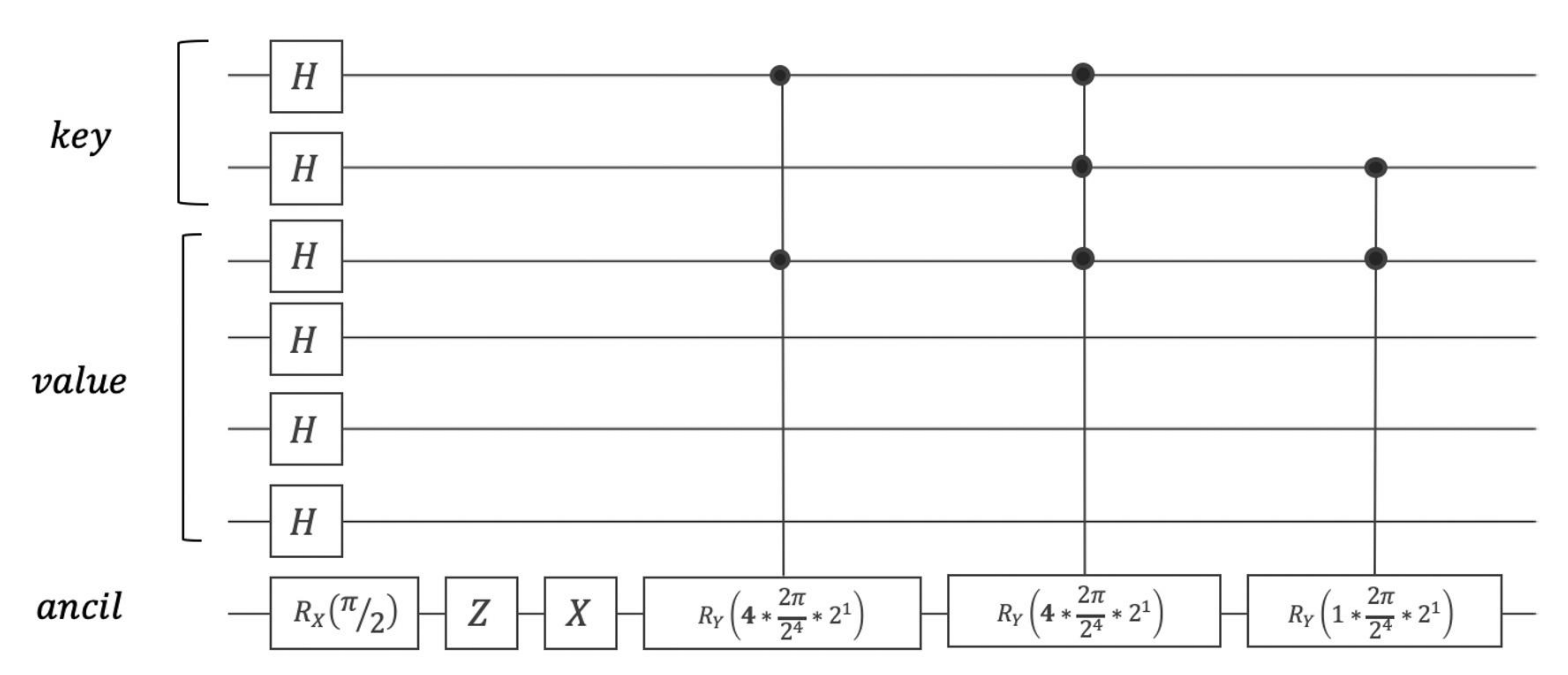}
\caption{\label{fig:65}The first section of a circuit that encodes the function $x^2$ for a two-qubit system. Here we show the initial preparation followed by the controlled-$R_Y$ gates for $i=0$.}
\end{figure}

As an example of a quantumly-defined encoding, let’s revisit the calculating squares example from Sec.~\ref{subsec:composition-example}.
As discussed in~\cite{Woerner2019Risk}, we can define x as $x=2x_1+x_0$ for a two-qubit register, and for a qubit $x_i$ (a binary variable), we know that $x_i^2=x_i$.
Therefore, we can derive $x^2=4x_1+4x_1x_0+x_0$, which is the formula we want to encode. 
For each value qubit $i$ we apply a controlled-$R_Y$ gate for each non-zero key, which correspond to the three terms in the function for $x^2$ (we rotate by qubit, which is efficient). 
The presence of a qubit in each term determines the control bits (e.g. the first term controls on $x_1$, the second on $x_1$ and $x_0$, etc.).
We show the beginning of the circuit in Fig.~\ref{fig:65}, which prepares the initial state followed by the $cR_Y$ gates for $i=0$.

\begin{figure}[htb]
\includegraphics[width=6cm]{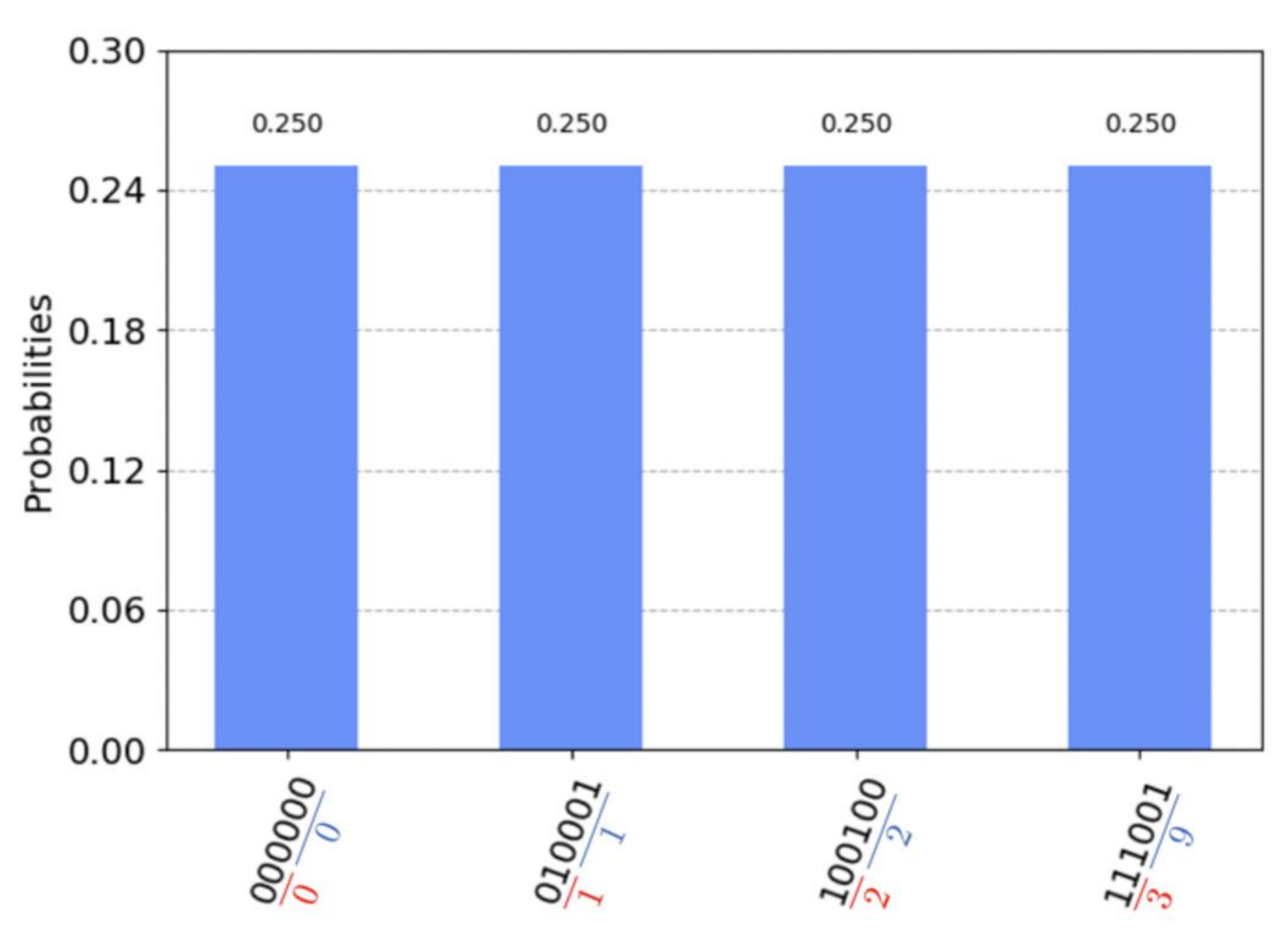}
\caption{\label{fig:66}The result of applying an inverse $QFT$ gate to the value register after encoding the square function.}
\end{figure}

After applying the inverse $QFT$ and setting the ancilla qubit to $0$, we get the results in Fig.~\ref{fig:66}.

Note the difference from the example in previous subsection---we do not classically calculate $f(x)$ in advance and encode the value into the circuit.
Instead, we encode the calculation directly. 
We can use quantumly-defined encoding for a number of hard problems, such as the Quadratic Unconstrained Binary Optimization (QUBO) model~\cite{Glover2019}.
The goal of QUBO is to minimize a quadratic polynomial 
$$f(x_0,x_1,...,x_{n-1})=\sum_{0 \le i < n}{l_ix_i} + \sum_{0 \le i < j < n}{q_{ij}x_ix_j}$$
where $\{x_0,x_1,...,x_{n-1}\}$ is a set of binary variables and $l_i$ and $q_{ij}$ are real coefficients.

\begin{figure}[htb]
\includegraphics[width=6cm]{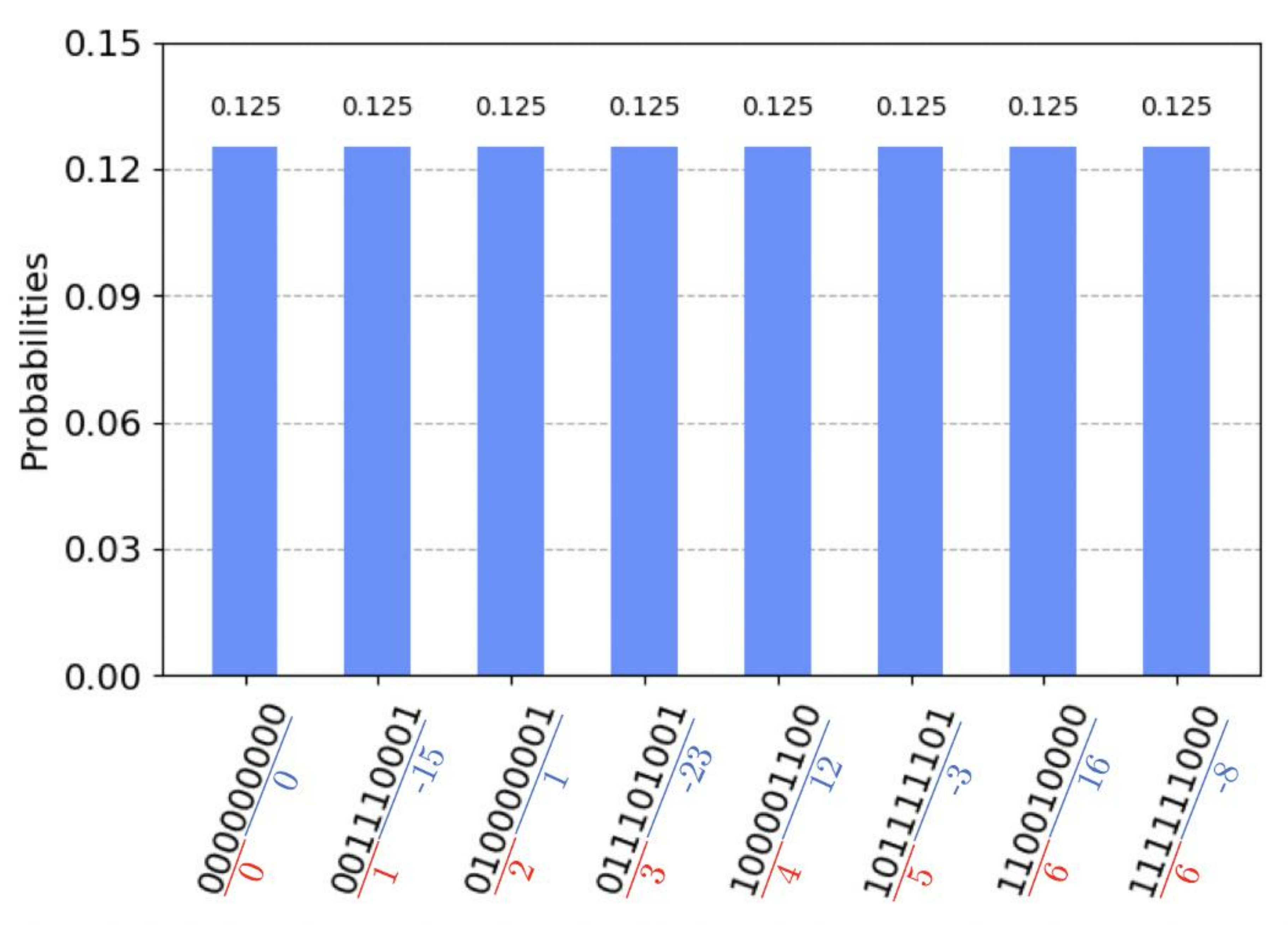}
\caption{\label{fig:67}The possible values of a function encoded by a QUBO quantum dictionary.}
\end{figure}

As an example, let’s look at the function $f(x_0,x_1,x_2 )=12x_0+x_1-15x_2+3x_0x_1-9x_1x_2$ and try to get information about its minimum.
As with encoding the squares (Fig.~\ref{fig:65}), we apply rotations controlled on a single key qubit for all linear components, and on pairs for the quadratic ones.
Using $3$ key qubits (the number of binary variables) and $6$ qubits to encode the value, the populated dictionary contains all eight possible values the function can take (Fig.~\ref{fig:67}).

Recall that the representation of numbers by rotations implicitly handles negative numbers using Two’s Complement~\cite{vonNeumann1993}.
For example, while index $1$ ($001$) seems to give a possible solution of $49$ ($110001$), classically we derive:
$$(12*\mathbf{0})+(1*\mathbf{0})-(15*\mathbf{1})+(3*\mathbf{0}*\mathbf{0})-(9*\mathbf{0}*\mathbf{1})=-15$$

We recognize a value as negative if the left-most bit in the value register is set to $1$, and get its Two’s Complement by inverting all bits and adding one.
In this case $49$ ($110001$) is the complement of $-15$ ($001111$), and thus the computation is correct.
We will continue this example in Sec.~\ref{sec:counting-values}, in which we will look at finding the minimum value of the function.

\subsection{\label{sec:probability-distributions}Probability Distributions}

\begin{figure}[htb]
\includegraphics[width=6cm]{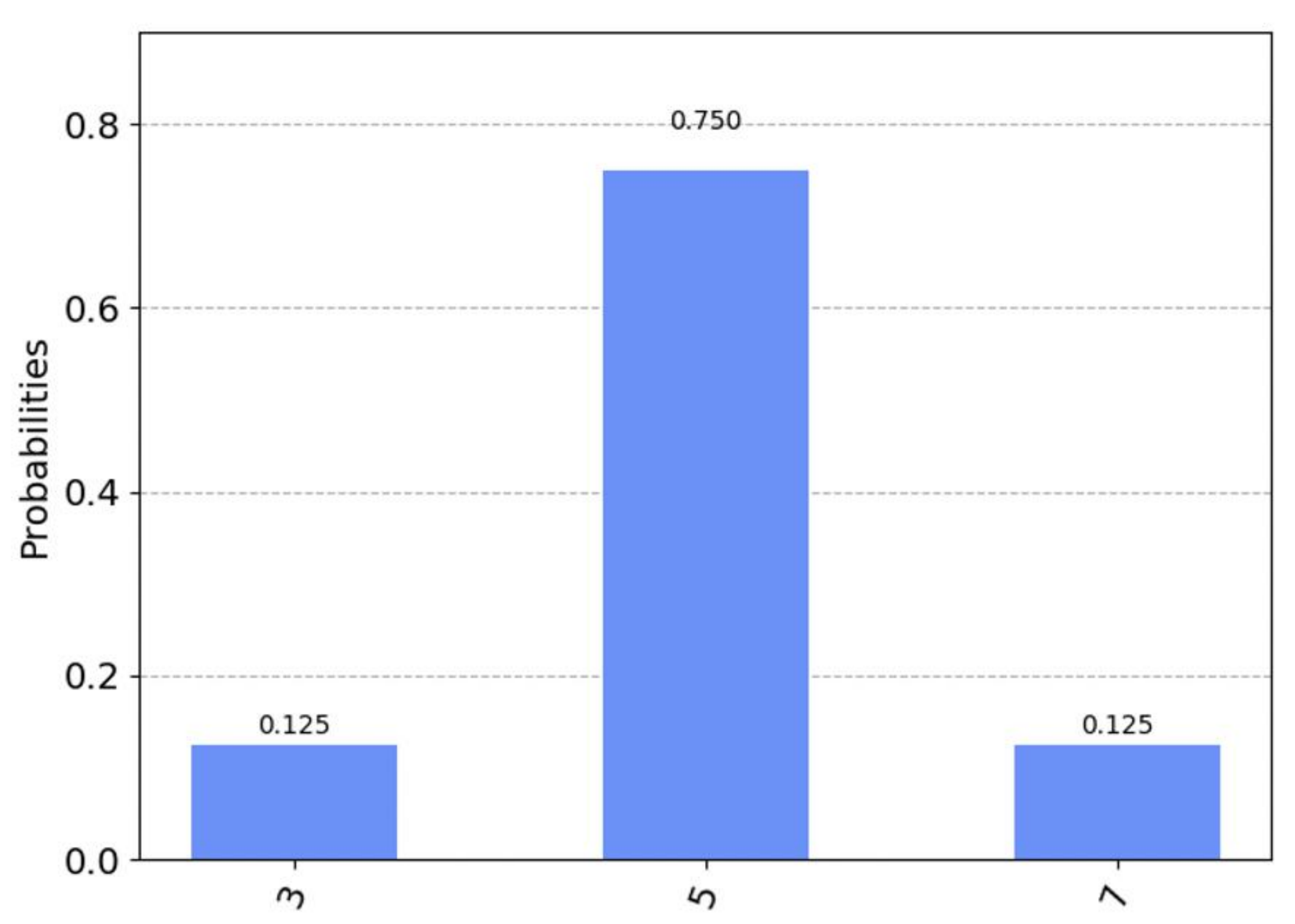}
\caption{\label{fig:68}A probability distribution created from encoding a function in the quantum state.}
\end{figure}

The ability to encode a function also allows us to create any probability distribution as the result of a quantum computation.
Given a function, the repetition of an output increases the probability of it being measured in the value register.
As we control the mapping of the inputs to outputs, we can manipulate the outputs to create a specific probability distribution. For example, let’s consider the function $f:{0,1,...,7}\rightarrow{0,1,...,7}$ defined as:
\begin{equation*}
  f(x) =
  \begin{cases}
    3 & \text{if $x=0$} \\
    5 & \text{if $0<x<7$} \\
    7 & \text{if $x=7$}
  \end{cases}
\end{equation*}
After populating a quantum dictionary using the function above, we can measure the key and value registers to reveal a simple probability distribution (Fig.~\ref{fig:68}).

\begin{figure}[htb]
\includegraphics[width=5cm]{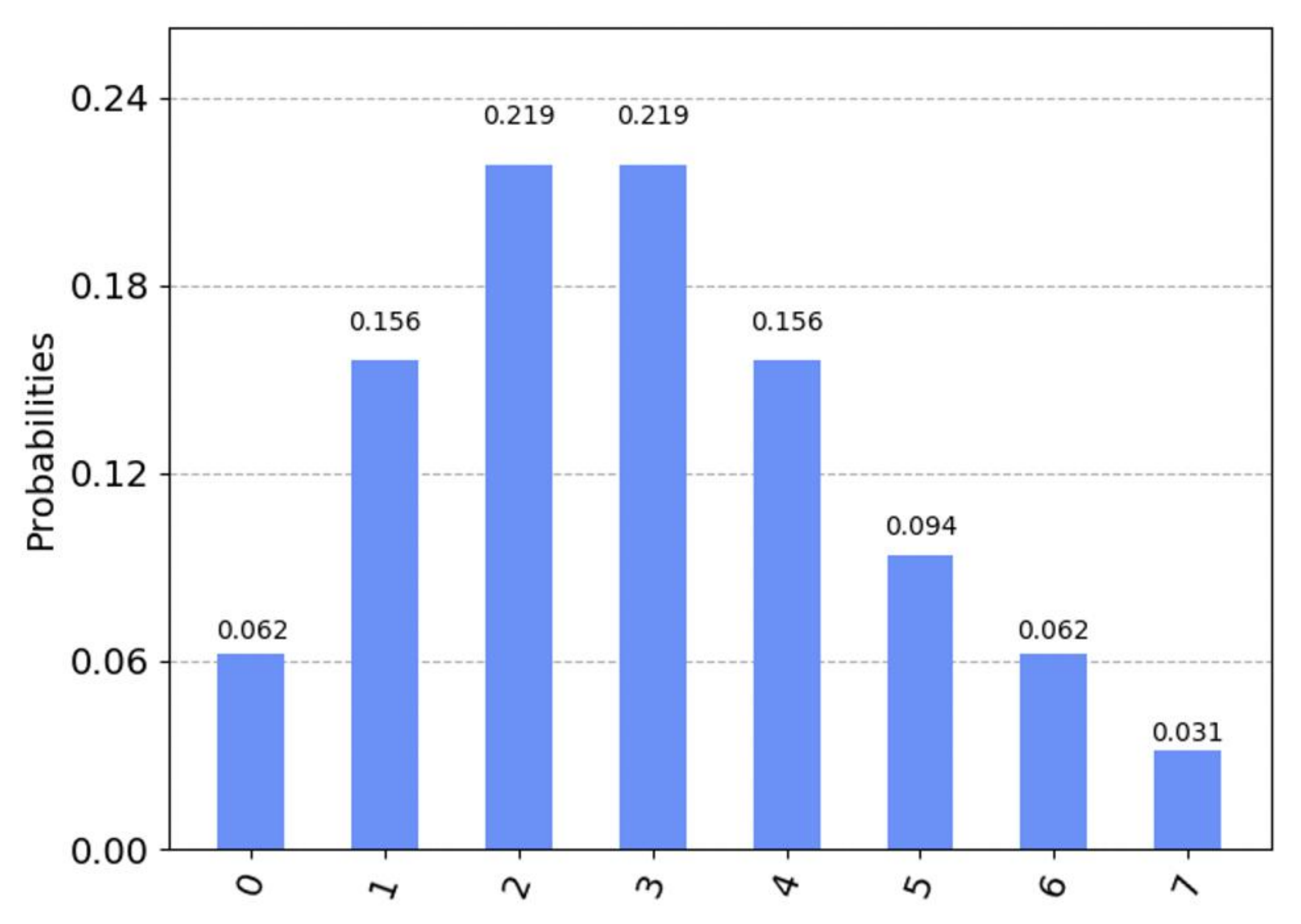}
\caption{\label{fig:69}A Poisson distribution ($\lambda=3$) created using a quantum dictionary.}
\end{figure}

Note that this example uses a complete-encoding operator, as we had to know the values in advance to create the distribution.
The value added by utilizing a quantum system comes from sampling, as the output of a quantum state measurement is truly random, as opposed to the pseudo-randomness of a classical system.
Now let’s consider the Poisson distribution~\cite{Katti1968}, which is used to model the number of times an event may occur in a given interval of time.
To encode this distribution, we use the probability mass function for the Poisson distribution $P(v)=e^{-\lambda}\frac{\lambda^v}{v!}$ with a given rate $\lambda$ and a value $v \in \{0,1,...,N-1\}$.
The encoding of the probability is done by assigning each value to a number of keys that matches the value frequency, specified by the probability mass function of the distribution (Fig.~\ref{fig:69}).

\subsection{\label{sec:partially-encoded-functions}Partially-Encoded Functions}

\subsubsection{\label{subsec:addition}Addition}

\begin{figure}[htb]
\includegraphics[width=10cm]{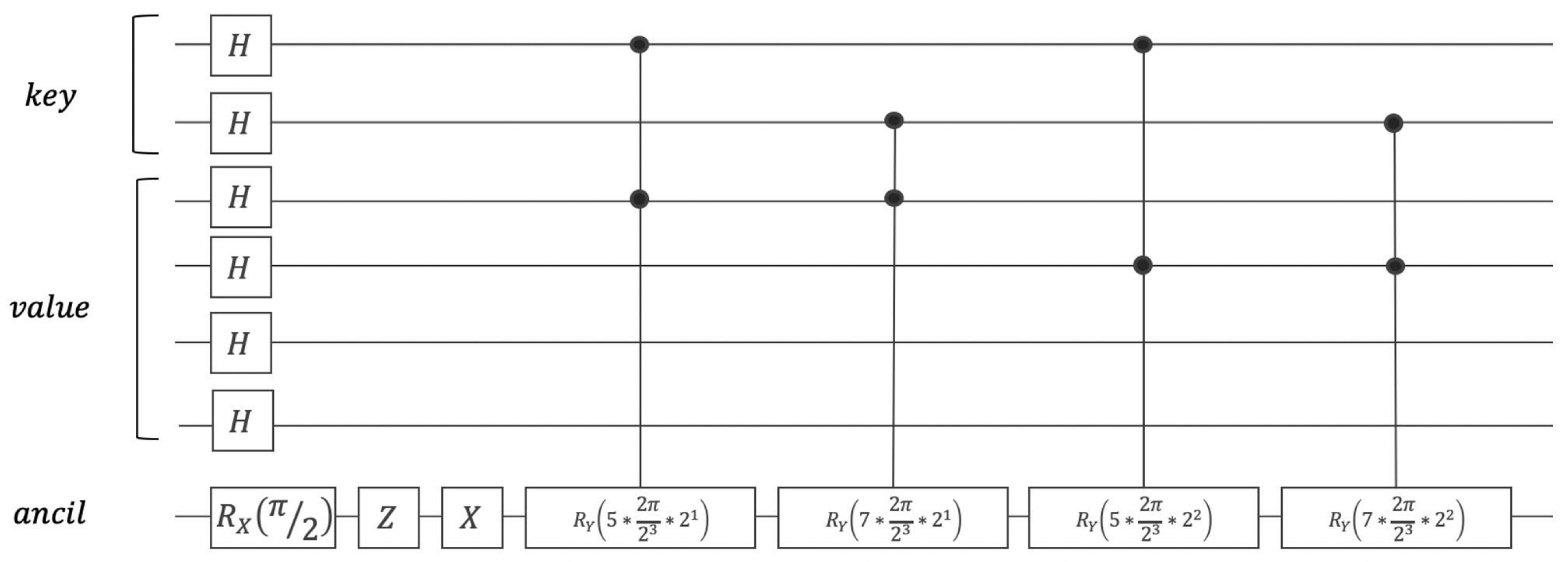}
\caption{\label{fig:70}The first section of an addition circuit. Here we show the initial preparation followed by the controlled-$R_Y$ gates for $i=0$ and $i=1$.}
\end{figure}

As an example of a partially-encoding operator, let’s model the addition of two numbers.
Given two inputs $x_1=5$ and $x_2=7$, we encode them into the keys represented by $01$ and $10$ (Fig.~\ref{fig:70}).
We don’t encode a value for the keys $00$ or $11$.

\begin{figure}[htb]
\includegraphics[width=6cm]{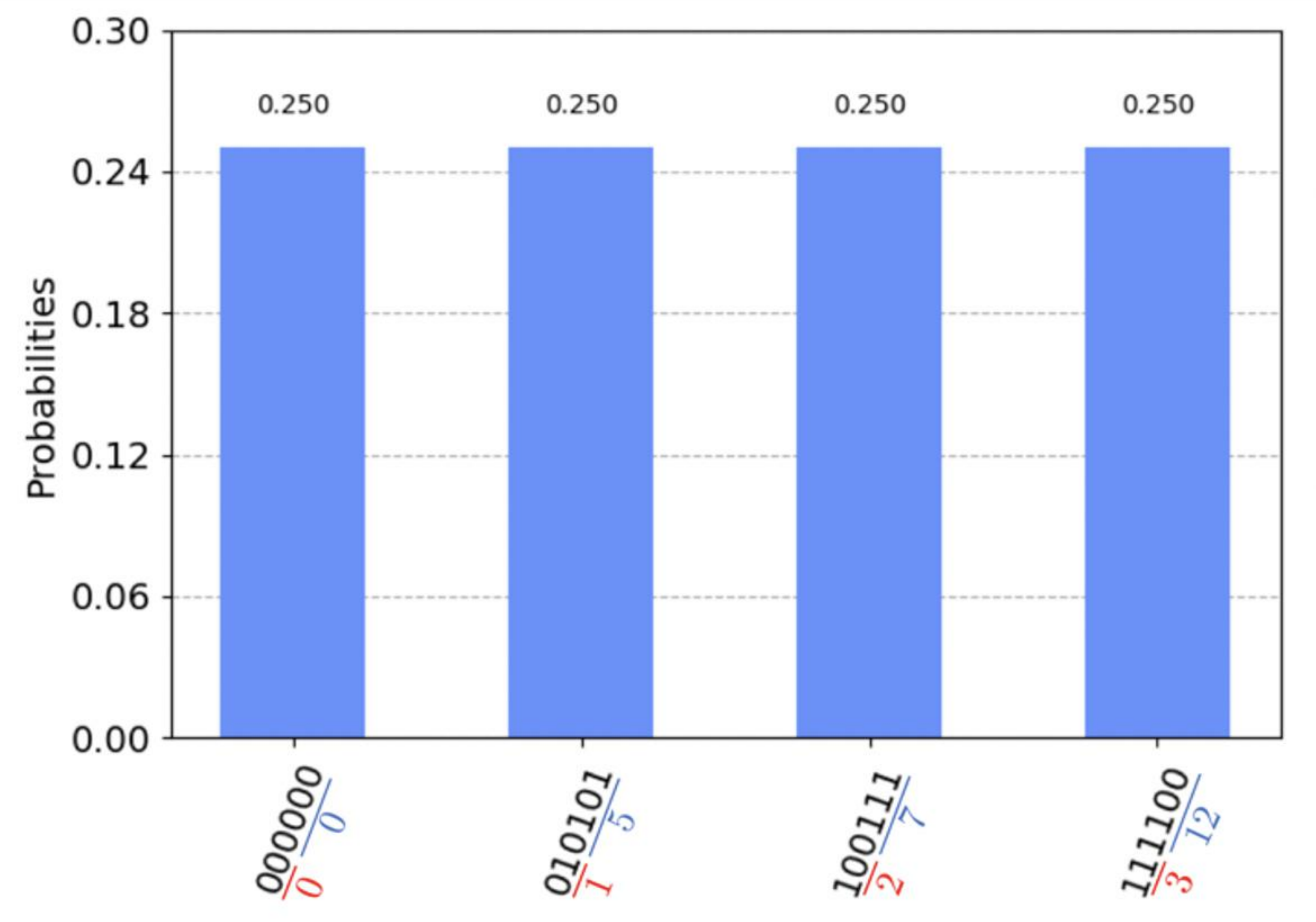}
\caption{\label{fig:71}The results of applying an inverse QFT gate to the value register after encoding the inputs.}
\end{figure}

Note also that we don’t explicitly encode an addition function in this example---we only encode the value $5$ to the key $01$, and value $7$ to $10$.
The results of this circuit are shown in Fig.~\ref{fig:71}, in which the value for key $11$ is $12$---the sum of the inputs.

This circuit creates a state that represents all possible sums of the inputs.
The presence of an input in each sum is denoted by the corresponding bit in the key register, as seen below.
\begin{eqnarray*}
\mathbf{00}:   \mathbf{0}*7+\mathbf{0}*5 &= 0 \\
\mathbf{01}:   \mathbf{0}*7+\mathbf{1}*5 &= 5 \\
\mathbf{10}:   \mathbf{1}*7+\mathbf{0}*5 &= 7 \\
\mathbf{11}:   \mathbf{1}*7+\mathbf{1}*5 &= 12
\end{eqnarray*}

This can be used to add any number of inputs e.g. given three inputs $(c_0,c_1,c_2)$, the state will contain all partial sums as well as the total sum with key $111$.

\begin{eqnarray*}
\mathbf{000}:   \mathbf{0}*c_0+\mathbf{0}*c_1+\mathbf{0}*c_2 \\
\mathbf{001}:   \mathbf{0}*c_0+\mathbf{0}*c_1+\mathbf{1}*c_2 \\
\mathbf{010}:   \mathbf{0}*c_0+\mathbf{1}*c_1+\mathbf{0}*c_2 \\
\mathbf{011}:   \mathbf{0}*c_0+\mathbf{1}*c_1+\mathbf{1}*c_2 \\
\mathbf{100}:   \mathbf{1}*c_0+\mathbf{0}*c_1+\mathbf{0}*c_2 \\
\mathbf{101}:   \mathbf{1}*c_0+\mathbf{0}*c_1+\mathbf{1}*c_2 \\
\mathbf{110}:   \mathbf{1}*c_0+\mathbf{1}*c_1+\mathbf{0}*c_2 \\
\mathbf{111}:   \mathbf{1}*c_0+\mathbf{1}*c_1+\mathbf{1}*c_2
\end{eqnarray*}

\subsubsection{\label{subsec:multiplication}Multiplication}

\begin{figure}[htb]
\includegraphics[width=5.5cm]{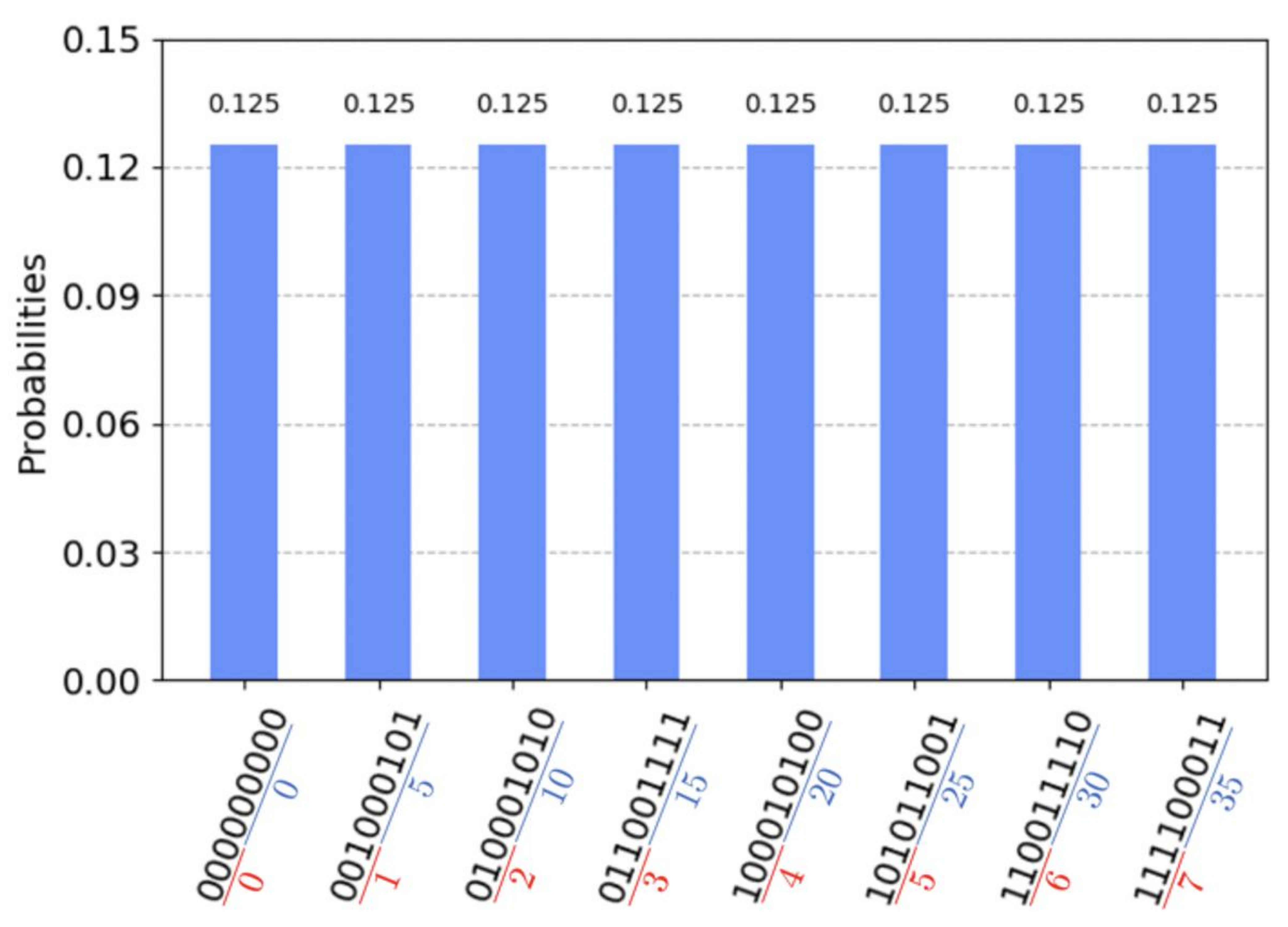}
\caption{\label{fig:72}The results of applying an inverse $QFT$ gate to the value register after encoding the multiplication function.}
\end{figure}

Using the same circuit above, we can also create something akin to a multiplication table for a given integer $x_0$.
The input of the circuit is defined as the product of $x_0$ with the powers of $2$ represented by $n$ qubits, i.e. $x=\{2^0x_0, 2^1x_0,...,2^{n-1}x_0\}$.
As an example, let’s define $x_0=5$ and $n=3$, thus $x=\{5,10,20\}$. The results of the computation are shown in Fig.~\ref{fig:72}.

Even though we only provided the products of $x_0$ with the powers of $2$ to the circuit, the resulting state is encoded with all multiples of $x_0$ represented by $n+m$ qubits.
The key register controls the precision for the multiplier, and the value register controls the precision for the product---i.e. we can create a larger multiplication table by scaling $n$ and $m$.
It’s important to mention that multiplying by powers of $2$ in a binary representation is as easy as appending the corresponding number of zeros to the right-hand side of the input binary string, similar to what we do for multiplying decimal values by powers of $10$ ($10$, $100$, etc.). 
The explanation for this circuit’s result is similar to the previous example:
\begin{eqnarray*}
\mathbf{000}:   \mathbf{0}(5*2^2)+\mathbf{0}(5*2^1)+\mathbf{0}(5*2^0) &= 0 \\
\mathbf{001}:   \mathbf{0}(5*2^2)+\mathbf{0}(5*2^1)+\mathbf{1}(5*2^0) &= 5 \\
\mathbf{010}:   \mathbf{0}(5*2^2)+\mathbf{1}(5*2^1)+\mathbf{0}(5*2^0) &= 10 \\ 
\mathbf{011}:   \mathbf{0}(5*2^2)+\mathbf{1}(5*2^1)+\mathbf{1}(5*2^0) &= 15 \\ 
\mathbf{100}:   \mathbf{1}(5*2^2)+\mathbf{0}(5*2^1)+\mathbf{0}(5*2^0) &= 20 \\
\mathbf{101}:   \mathbf{1}(5*2^2)+\mathbf{0}(5*2^1)+\mathbf{1}(5*2^0) &= 25 \\
\mathbf{110}:   \mathbf{1}(5*2^2)+\mathbf{1}(5*2^1)+\mathbf{0}(5*2^0) &= 30 \\
\mathbf{111}:   \mathbf{1}(5*2^2)+\mathbf{1}(5*2^1)+\mathbf{1}(5*2^0) &= 35
\end{eqnarray*}

Note that we can use this technique to explore the greatest common divisor of two integers. In particular, finding if two integers are relatively-prime. 

\subsubsection{\label{subsec:prob-dist}Probability Distributions}

\begin{figure}[htb]
\includegraphics[width=5.5cm]{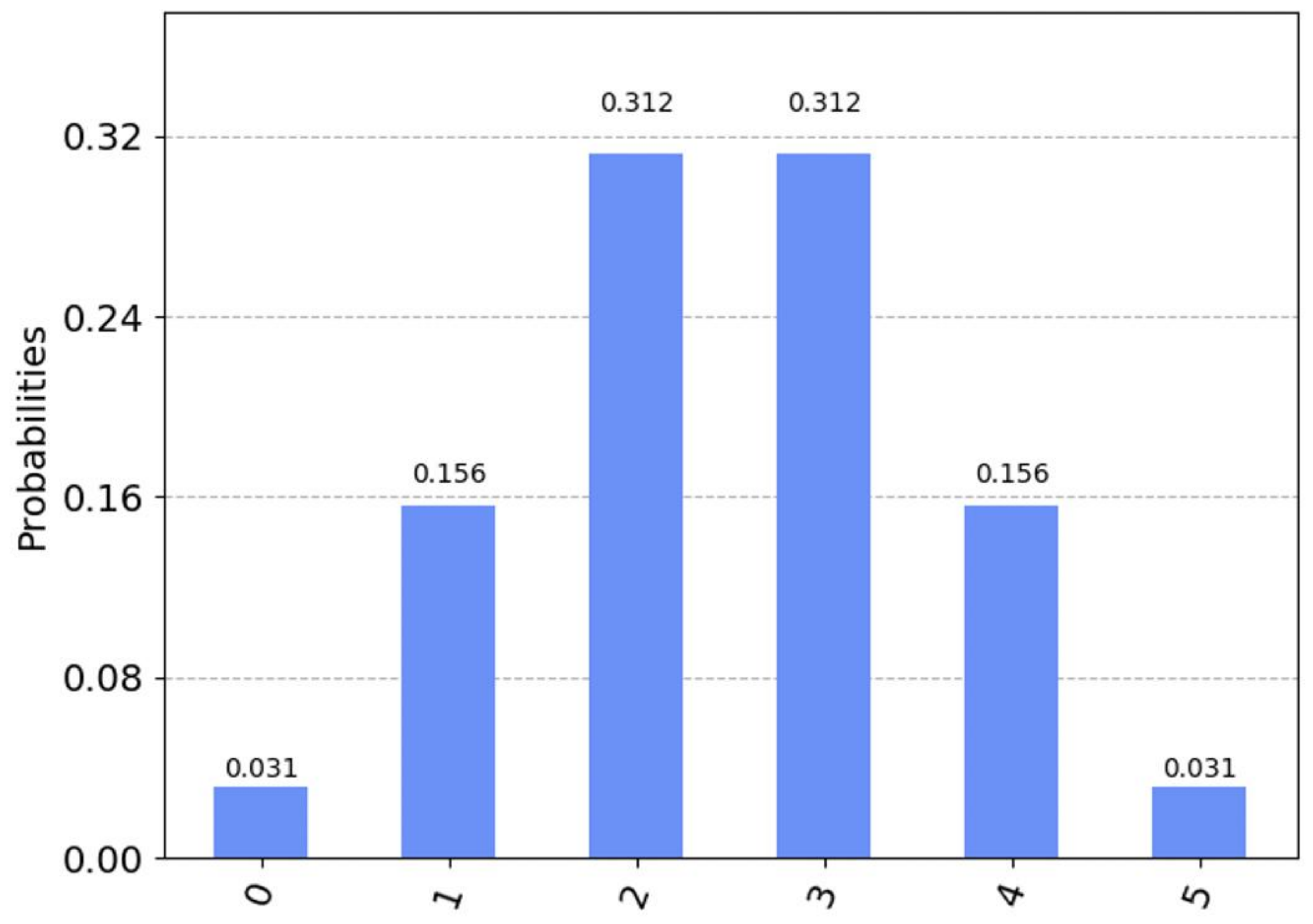}
\caption{\label{fig:73}A binomial probability distribution created from partially-encoding a function in the quantum state.}
\end{figure}

We can also use partial-encoding operators to efficiently encode certain probability distributions, such as a binomial distribution.
This can be achieved by encoding the value $1$ to each of the powers of $2$, and leaving all other inputs alone.
For $n=5$ key qubits and $m=3$ value qubits, we can create the binomial distribution in Fig.~\ref{fig:73}.

The distribution is derived from the implicit sums discussed previously.
In this example we have less value qubits than key qubits, and thus values are repeated across the $2^n$ inputs.
When the probabilities for each value are summed together the result is a binomial distribution. 

\subsection{\label{sec:value-lookup-by-key}Value Lookup by Key}

\begin{figure}[htb]
\includegraphics[width=6cm]{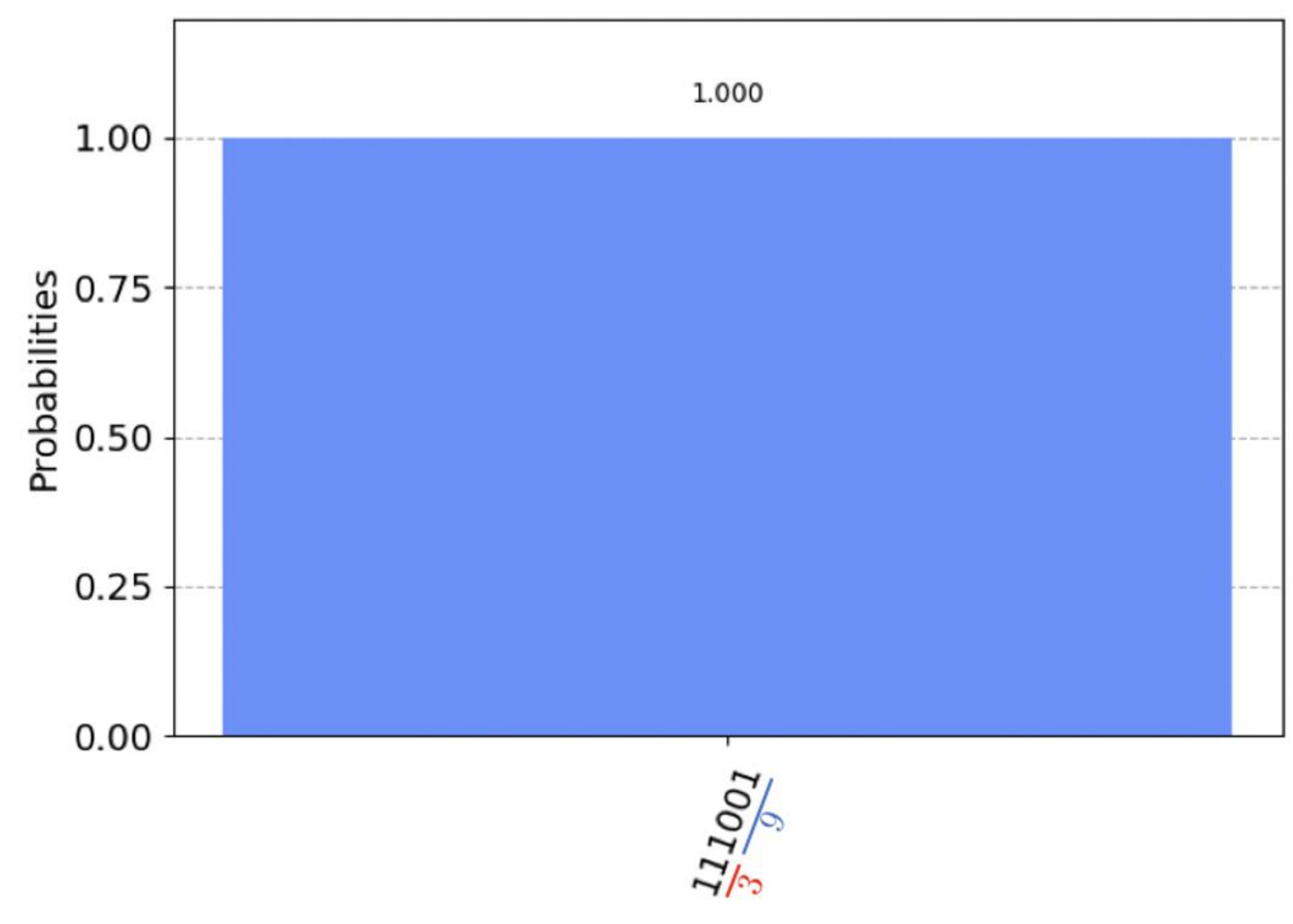}
\caption{\label{fig:74}The result of using Grover’s algorithm on key $11$ before encoding $x^2$ into the quantum state.}
\end{figure}

When using a partial-encoding operator, we only provide some of the values encoded in the quantum state.
We can implement a lookup capability for quantum dictionary using Grover’s Search algorithm, which increases the probability of measuring a specific output.
In particular, this pattern shows the true power of Grover's Search---we can match on just part of output.
With the lookup operator we can revisit the calculating squares example defined in the introduction.
As discussed in Sec.~\ref{sec:quantum-systems}, all results are equally likely to be measured, and thus we’d have to repeat the computation to get all the squares (especially if we get duplicate outputs).
If we use Grover’s algorithm to increase the probability of measuring a specific key, we simultaneously increase the probability of measuring the corresponding value (Fig.~\ref{fig:74}).

\begin{figure}[htb]
\includegraphics[width=6cm]{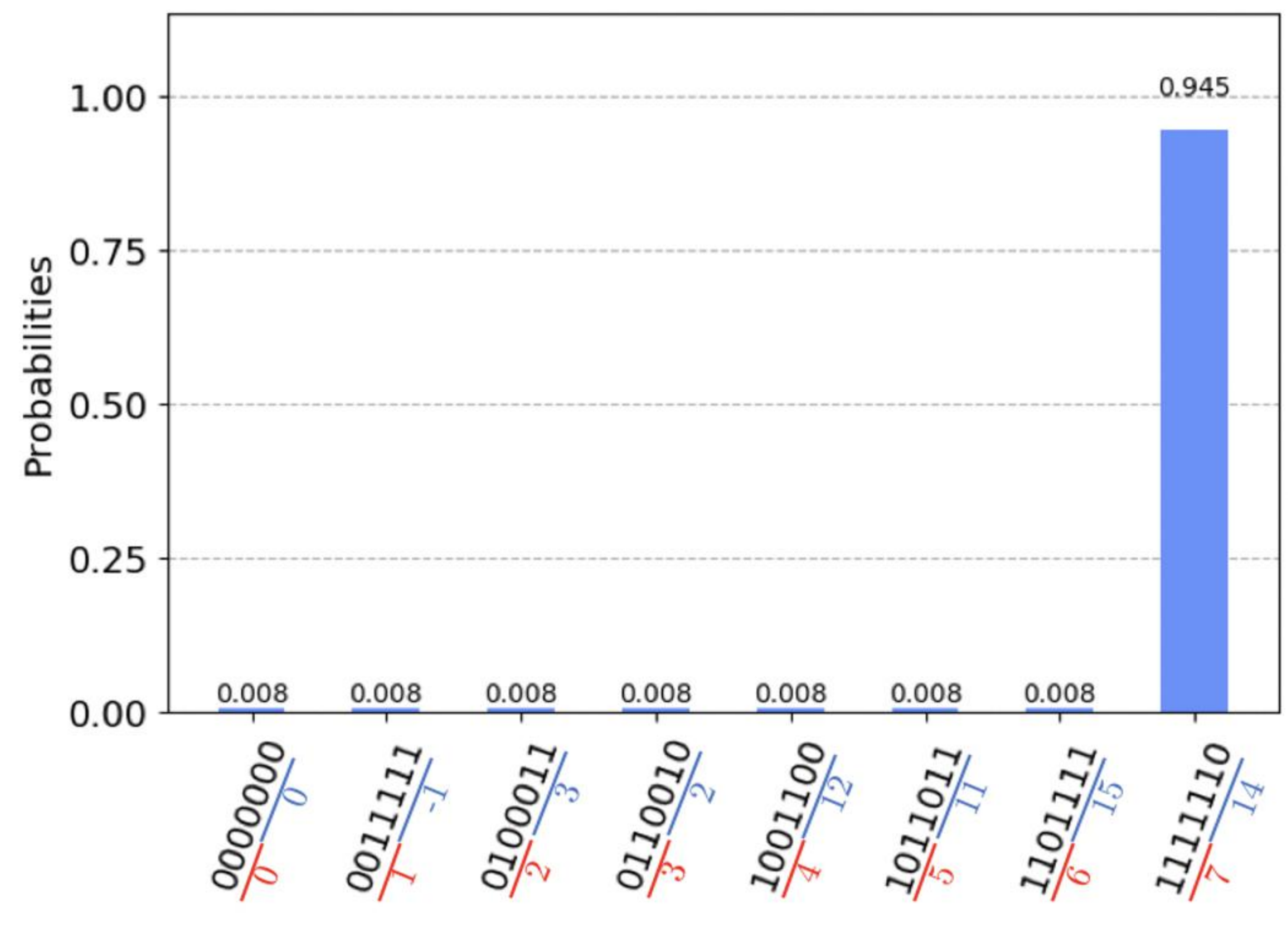}
\caption{\label{fig:75}The result of using Grover’s algorithm to recover the total sum of three inputs.}
\end{figure}

The lookup operator also helps retrieve the result of adding numbers. For example, let’s say we want to calculate the sum of three integers $x_1=12$, $x_2=3$, and $x_3=-1$.
The result should be $x_1+x_2+x_3=12+3-1=14$.
As discussed previously, the quantum state will include all partial sums and the total sum as values.
Using the lookup feature, we can find the total sum ($14$) paired with the key with all $1$s (Fig.~\ref{fig:75}).

\subsection{\label{sec:counting-values}Counting Values}

\subsubsection{\label{subsec:equality-value-matching}Equality-Based Value Matching}

Using a quantum dictionary and Quantum Counting, we can count the number of times a specific value appears, i.e. the number of keys that have that same value associated to them.
If we have an operator $A$ that populates the quantum dictionary and an oracle $O$ that recognizes a given value in the value register, we can use Quantum Counting with Grover iterate $DA^{-1}OA$ to count, where $D$ is the diffusion operator introduced in Sec.~\ref{sec:grover-iteration-and-quantum-search}.

\begin{figure}[htb]
\includegraphics[width=2cm]{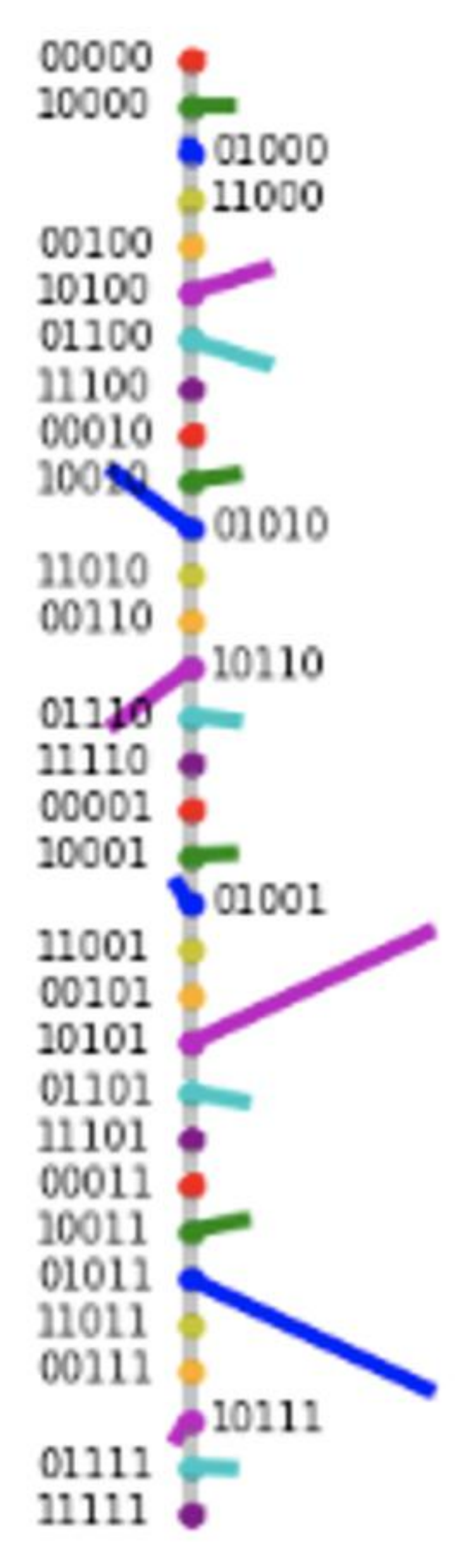}
\caption{\label{fig:76}The quantum state after the application of the zero-sum Quantum Counting circuit.}
\end{figure}

As an example let’s look at the \textbf{subset sum problem}~\cite{Horowitz1974}.
We want to find how many subsets of a set of integers, for example $\{1,0,2,-1\}$, have a sum of $0$.
An exhaustive check shows there are four subsets with zero sum: $\{\}$, $\{0\}$, $\{1,-1\}$, and $\{0,1,-1\}$. 
While for small sets an exhaustive check is reasonable, for large sets the computation becomes prohibitive.
The quantum version will avoid an exhaustive search, as is the case with all Grover’s algorithm applications.
In order to apply the quantum dictionary pattern to this problem, we encode the set of integers as a list $[1,0,2,-1]$ into the keys whose integer representation is a power of two $\{1,2,4,8\}$.
The values for the other keys are all the sums of all the subsets of the given set.
Using value counting, we can get the number of subsets with a sum of $0$ (Fig.~\ref{fig:76}).

In Fig.~\ref{fig:76} the most-probable outputs are $21$ ($10101$) and $11$ ($01011$).
Recall that a post-processing step is needed where the outputs are converted into corresponding counts.
Using the equation from~\ref{sec:basic-quantum-counting}, we derive:
$$\ceil{2^4\cos^2\left(11\frac{\pi}{2^5}\right)} = \ceil{2^4\cos^2\left(21\frac{\pi}{2^5}\right)} = 4$$

Note that knowing the count of a value allows us to retrieve an output that contains that value and one of the keys mapped to it, by choosing the right number of applications of the Grover iterate~\cite{Brassard2000}.

\subsubsection{\label{subsec:inequality-value-matching}Inequality-Based Value Matching (Comparison)}

\begin{figure}[htb]
\includegraphics[width=1.5cm]{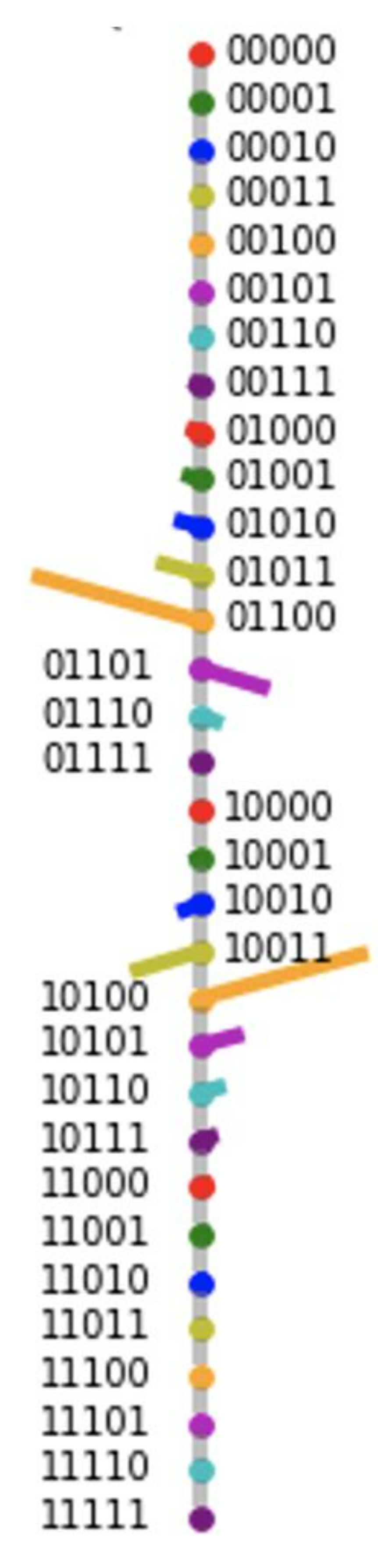}
\caption{\label{fig:77}The quantum state after the application of the negative-sum Quantum Counting circuit.}
\end{figure}

In addition to counting the values that are equal to a given number, we can also count the values that are less than a given number.
In order to do that, we can use an oracle that recognizes negative values.
This is not difficult to implement---as we pointed out in previous subsections, negative values are easy to recognize because they have a $1$ in the left-most qubit.
Let’s revisit the previous example, which counted the zero sums of the subset sum problem for the given set $\{1,0,2,-1\}$.
Now we want to know how many of the subset sums are negative---using an exhaustive check we determine there are two: $\{-1\}$ and $\{0,-1\}$.
We can use the less-than operator to find how many of the solutions are negative (Fig.~\ref{fig:77}).
The most-probable outputs ($12$ and $20$) correspond to the expected count:
$$\ceil{2^3\cos^2\left(12\frac{\pi}{2^5}\right)} = \ceil{2^3\cos^2\left(20\frac{\pi}{2^5}\right)} = 2$$

\subsection{\label{sec:qubo}QUBO and Other NP-Hard Problems}

\subsubsection{\label{subsec:qubo}QUBO}

\begin{figure}[htb]
\includegraphics[width=1.5cm]{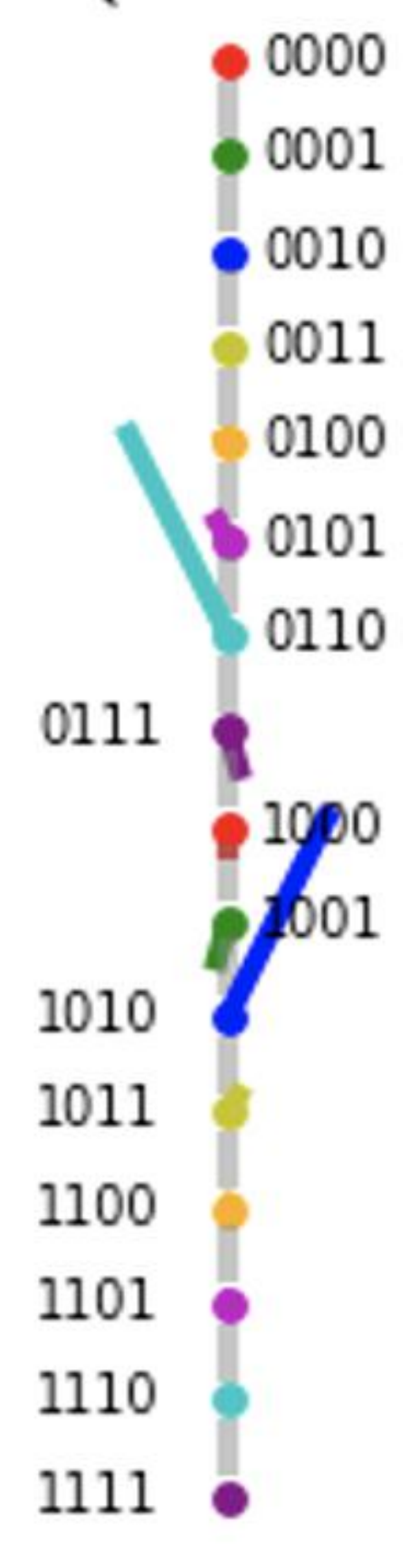}
\caption{\label{fig:78}The quantum state after the application of a Quantum Counting circuit that counts negative QUBO solutions.}
\end{figure}

Returning to QUBO, we can’t directly find the minimum with a quantum dictionary, but we can use the inequality value matching capability to gain information about the minimum in an iterative manner.
This is similar to variational approaches to quantum computing, where a computation is repeated with adaptive parameters.
Let’s revisit the example from a previous subsection:
$$f(x_0,x_1,x_2)=12x_0+x_1-15x_2+3x_0x_1-9x_1x_2$$
Measuring the state represented in Fig.~\ref{fig:67} returned $-15$.
Using the comparative capability of the quantum dictionary, we can find out how many function values are less than $-15$ (Fig.~\ref{fig:78}).

The most-probable outputs are $6$ and $10$, which represent the count:
$$\floor{2^3\cos^2\left(6\frac{\pi}{2^4}\right)} = \floor{2^3\cos^2\left(10\frac{\pi}{2^4}\right)} = 1$$
which indicates there is a lower value than $-15$, which has to be the minimum of the function.
An iterative process will reveal that the minimum is $-23$.

\subsubsection{\label{subsec:fibonacci}Fibonacci Numbers}

For a given length $n$, the count of binary strings without consecutive ones is the $(n+2)$nd Fibonacci number~\cite{Malesevic2004}.
We can fit this into a quadratic optimization context (QUBO) by creating a quadratic equation with $0$ coefficients except for the neighboring binary variables, which are $1$.

\begin{figure}[htb]
\includegraphics[width=1.5cm]{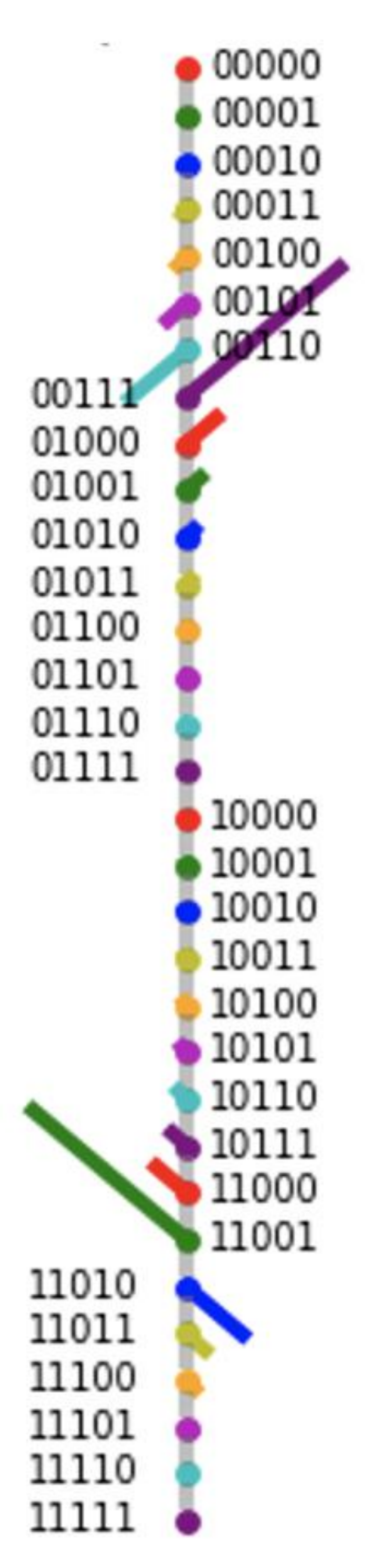}
\caption{\label{fig:79}The quantum state after the application of a zero-sum Quantum Counting circuit.}
\end{figure}

For example, if we have $n=3$ key qubits we can calculate the $5$th Fibonacci number, which is $5$.
To achieve this we encode a function $f(x_1,x_2,x_3 )=x_1x_2+x_2x_3$ using the QUBO quantum dictionary.
The keys without consecutive ones will have a value of $0$, and thus we can use the equality value matching capability of the quantum dictionary to look for all solutions with this value, similar to the subset-sum problem (Fig.~\ref{fig:79}).
 
The most-probable outputs in Fig.~\ref{fig:79} are $25$ and $7$, which represent a count of:
$$\ceil{2^3\cos^2\left(7\frac{\pi}{2^5}\right)} = \ceil{2^3\cos^2\left(25\frac{\pi}{2^5}\right)} = 5$$

\section{\label{sec:conclusion}Conclusion}
Finding efficient operators and oracles is not easy, in fact there are very few examples we are familiar with.
Algorithms assume their existence, and they are usually left as an exercise to the reader.
We hope that this paper helps with that difficult task.
We have found a geometric approach very useful, and some of the insights presented in the paper are inspired by Geometric Algebra concepts.
 
The Quantum Dictionary is a standalone pattern that incorporates a number of ingredients in a well-defined interface.
We have shown how it can be used to explore QUBO (Quadratic Unconstrained Binary Optimization) problems using the advantages of standard algorithms, but many of the more recent developments in the optimization space, like QAOA, can be applied to the dictionary as well.
There are other classes of problems that can be explored in the same context, especially those that require looking at “all possible combinations” of inputs.
Many hard problems fall into that category, with a large number of applications to business problems.
It is important to distinguish between the various levels involved in solving problems, including technical and business aspects.
Take, for example, the problems of integer factorization and encryption, related to Shor’s algorithm.
Breaking certain types of encryption is easy using integer factorization, which is a much harder problem.
However, many people confuse the two.
Similarly, sampling, or search are hard problems in themselves, with obvious  applications to an enormous number of business problems.
To use an example from this paper, Grover’s algorithm can be applied to lookup a value in a dictionary by key, and not necessarily as a database search tool, as it has been sometimes presented, in a rush to get closer to a business application.
We barely touched on exploring probabilistic spaces with a Quantum Dictionary.
The outputs of a quantum computation can be seen as either simple outcomes or events (sets of outcomes) of a probabilistic space.
Richard Feynman introduced the concept of “paths” leading to a certain event, and his famous sum and product rules in the same context.
“Good states” recognized by oracles can be regarded as events as well.
We have more insights and patterns related to this concept that we are currently exploring.

\section{Acknowledgments}
 
We would like to thank Amarendra Mishra and Nikitas Stamatopoulos for reading the draft copies, and for their useful comments and suggestions

\section{Disclaimers}
 
This material is for informational purposes only and is not the product of JPMorgan Chase \& Co.’s Research Department.  This material is not intended as research, a recommendation, advice, offer or solicitation for the purchase or sale of any financial product or service, and is not a research report and is not intended as such. This material is not intended to represent any position or opinion of JPMorgan Chase \& Co.  JPMorgan Chase \& Co. disclaims any responsibility or liability whatsoever for the quality, accuracy or completeness of the information herein, and for any reliance on, or use of this material in any way. \copyright 2020 JPMorgan Chase \& Co.

IBM, IBM Q, Qiskit are trademarks of International Business Machines Corporation, registered in many jurisdictions worldwide. Other product or service names may be trademarks or service marks of IBM or other companies.

\bibliography{main}

\clearpage

\appendix

\section{\label{sec:posulates}The Postulates of Quantum Mechanics}

“The postulates of quantum mechanics were derived after a long process of trial and (mostly) error, which involved a considerable amount of guessing and fumbling by the originators of the theory."~\cite{Nielsen2011}

\subsection*{State Space}

A quantum system is completely described by its state, which is an assignment of an arrow to each possible outcome.

\begin{figure}[htb]
\includegraphics[width=5cm]{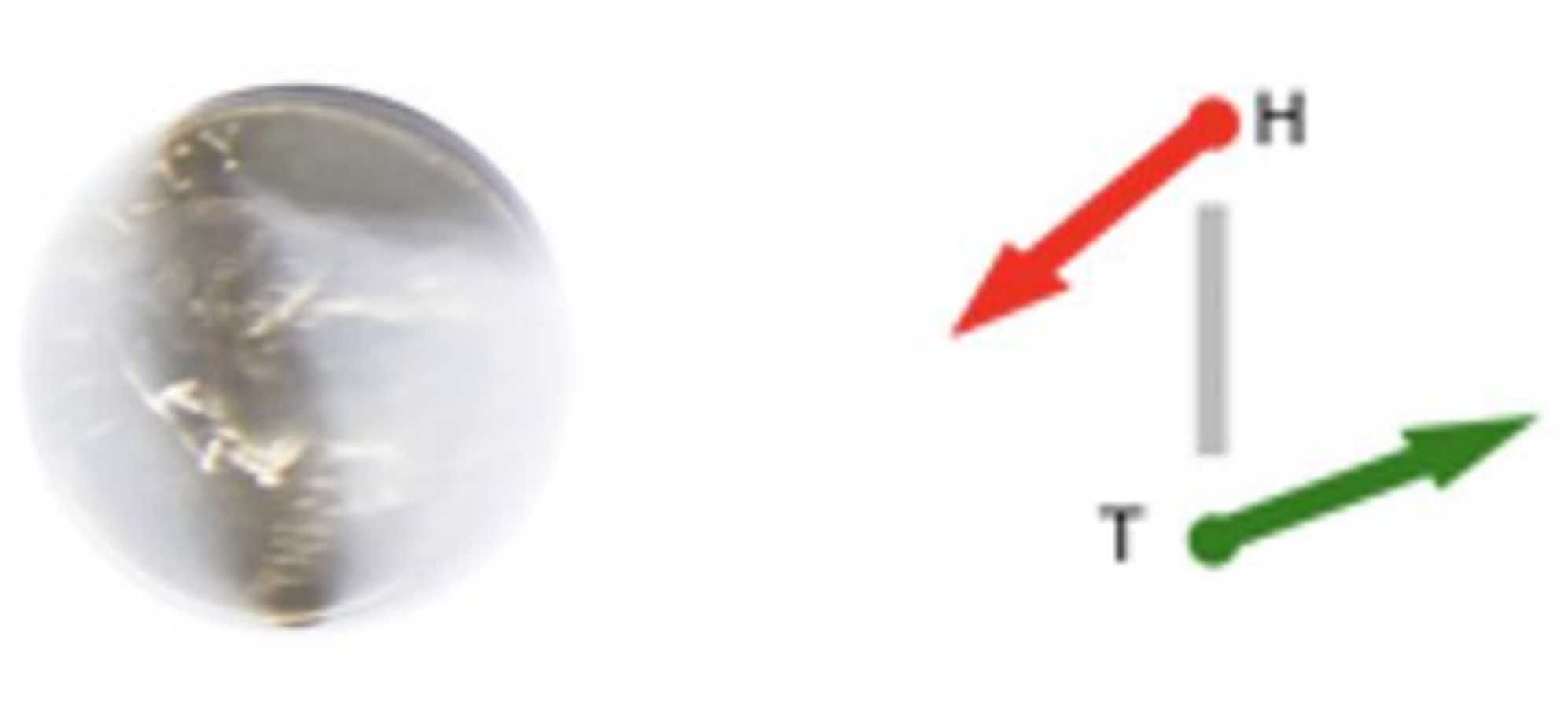}
\end{figure} 

\subsection*{Composition}

The state space of a composite physical system is the tensor product of component states.

\begin{figure}[htb]
\includegraphics[width=7cm]{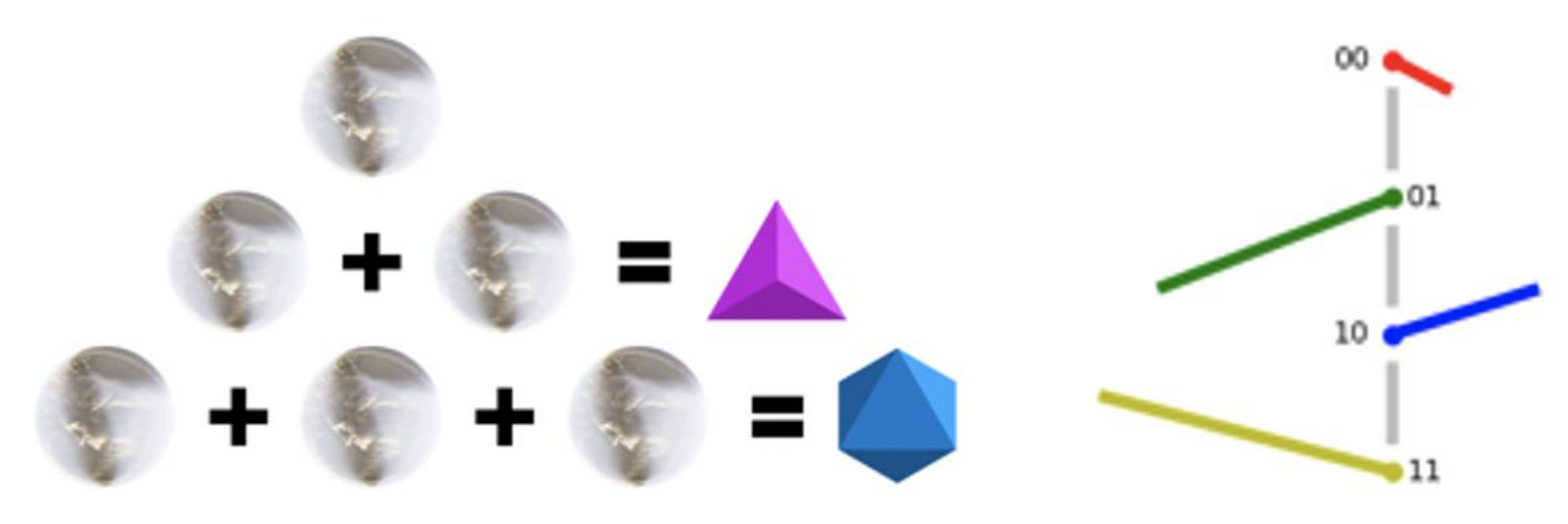}
\end{figure} 

\subsection*{Evolution}

States at two different times are related by a unitary operator.

\begin{figure}[htb]
\includegraphics[width=7cm]{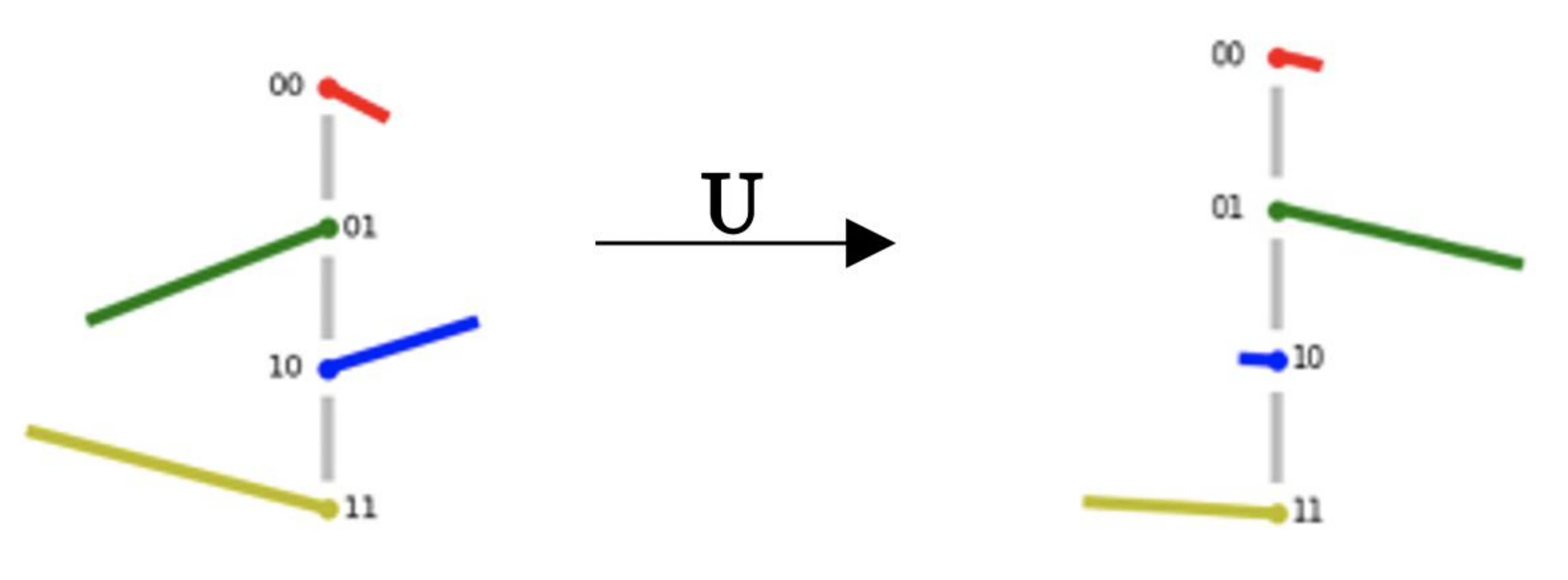}
\end{figure} 

\subsection*{Measurement}

One possible outcome occurs randomly when a computation is repeated. The probability of an outcome is the square of the length of the amplitude.

\begin{figure}[htb]
\includegraphics[width=5cm]{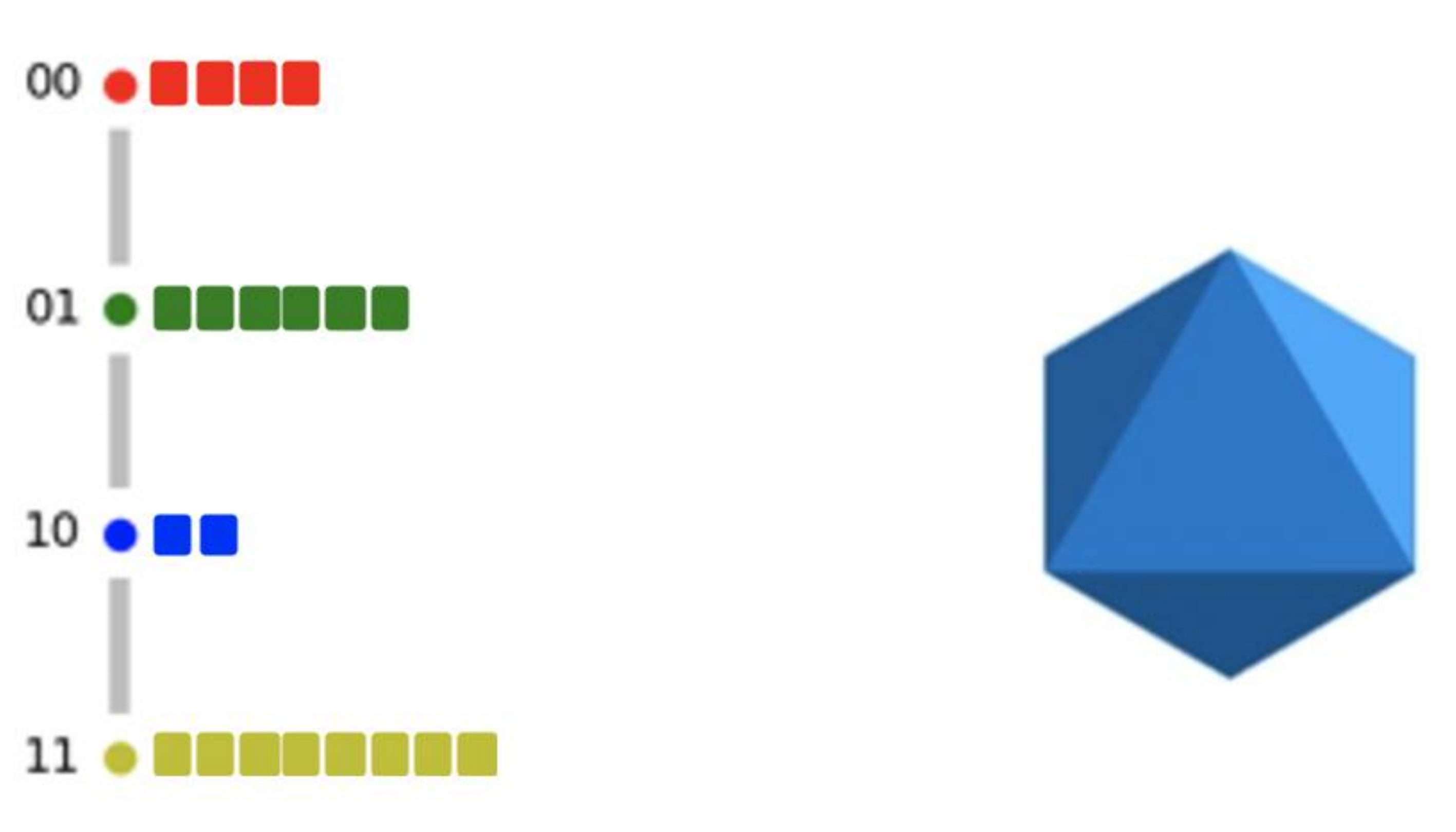}
\end{figure} 

\end{document}